\newcommand{\be}{\begin{equation}}
\newcommand{\ee}{\end{equation}}
\newcommand{\bea}{\begin{eqnarray}}
\newcommand{\eea}{\end{eqnarray}}
\newcommand{\nn}{\nonumber}
\newcommand{\Appendix}[1]%
    {\renewcommand{\thesection}{Appendix~\Alph{section}:}%
         \section{#1}}%
\catcode`@=11
\long\def\@makecaption#1#2{
   \vskip 10pt
   \setbox\@tempboxa\hbox{{\small\bf #1.} \ {\small #2}}
   \ifdim \wd\@tempboxa >\hsize       % IF longer than one line:
   {\small\bf #1.} \ {\small #2}\par  % THEN set as ordinary paragraph.
   \else                              %   ELSE  center.
        \hbox to\hsize{\hfil\box\@tempboxa\hfil}
   \fi}
\catcode`@=12

\catcode`@=11
\def\secteqno{\@addtoreset{equation}{section}%
\def\theequation{\thesection.\arabic{equation}}}
\def\endsecteqno{\def\theequation{\@ifundefined{chapter}%
{\arabic{equation}}{\thechapter.\arabic{equation}}}}
\newcounter{subequation}
\def\thesubequation{\alph{subequation}}
\def\sneqnarray{\stepcounter{equation}\let\@currentlabel=\theequation
\setcounter{subequation}{1}
\def\@eqnnum{{\rm (\theequation\thesubequation)}}
\global\@eqcnt\z@\tabskip\@centering\let\\=\@eqncr\let\@@eqncr=\@@sneqncr
$$\halign to \displaywidth\bgroup\@eqnsel\hskip\@centering
 $\displaystyle\tabskip\z@{##}$&\global\@eqcnt\@ne
 \hskip 2\arraycolsep \hfil${##}$\hfil
 &\global\@eqcnt\tw@ \hskip 2\arraycolsep
$\displaystyle\tabskip\z@{##}$\hfil
tabskip\@centering&\llap{##}\tabskip\z@\cr}
\def\endsneqnarray{\@@sneqncr\egroup $$\global\@ignoretrue}
\def\@@sneqncr{\let\@tempa\relax
   \ifcase\@eqcnt \def\@tempa{& & &}\or \def\@tempa{& &}
   \else \def\@tempa{&}\fi
     \@tempa \if@eqnsw\@eqnnum\stepcounter{subequation}\fi
     \global\@eqnswtrue\global\@eqcnt\z@\cr}
\def\nobiblabels{\def\@lbibitem[##1]##2{\@bibitem{##2}}}
\catcode`@=12

\def\beq{\begin{equation}}
\def\eeq{\end{equation}}
\def\bea{\begin{eqnarray}}
\def\eea{\end{eqnarray}}
\def\nn{\nonumber}

    \def\lapp{\lambda^{\prime\prime}}
\def\la{\lambda} \def\lap{\lambda^{\prime}} \def\pa{\partial}   \def\dag{\dagger}

\def\lQ{\Lambda_{\rm QCD}}
\documentclass[aps,10pt,prd,showpacs,amsmath,amssymb,preprintnumbers,superscriptaddress,nofootinbib,showkeys]{revtex4-1}

\usepackage[german,english]{babel}
\usepackage{hyperref}
\usepackage{amssymb}
\usepackage{amsmath}
\usepackage{bm}% bold math   
\usepackage{braket}
\usepackage{slashed}
\usepackage{multirow}
\usepackage{epsfig}
\usepackage{epstopdf}
\usepackage{bbm}
\usepackage{bbold}

\begin{document}

\title{Effective field theory for double heavy baryons at strong coupling}

\author{Joan Soto}
\email{joan.soto@ub.edu}
\affiliation{Departament de F\'\i sica Qu\`antica i Astrof\'isica and Institut de Ci\`encies del Cosmos, Universitat de Barcelona, Mart\'\i$\,$ i Franqu\`es 1, 08028 Barcelona, Catalonia, Spain}

\author{Jaume Tarr\'us Castell\`a}
\email{jtarrus@ifae.es}
\affiliation{Grup de F\'\i sica Te\`orica, Dept. F\'\i sica and IFAE-BIST, Universitat Aut\`onoma de Barcelona,\\ 
E-08193 Bellaterra (Barcelona), Catalonia, Spain}

\date{\today}

\begin{abstract}
We present an effective field theory for doubly heavy baryons that goes beyond the compact heavy diquark approximation. The heavy quark distance $r$ is only restricted to $m_Q\gg 1/r \gg E_{bin}$, where $m_Q$ is the mass of the heavy quark and $E_{bin}$ the typical binding energy. This means that the size of the heavy diquark can be as large as the typical size of a light hadron. We start from nonrelativistic QCD, and build the effective field theory at next-to-leading order in the $1/m_Q$ expansion. At leading order the effective field theory reduces to the Born-Oppenheimer approximation. The Born-Oppenheimer potentials are obtained from available lattice QCD data. The spectrum for double charm baryons below threshold is compatible with most of the lattice QCD results. We present for the first time the full spin averaged double bottom baryon spectrum below threshold based on QCD. We also present model-independent formulas for the spin splittings. 
\end{abstract}

\maketitle

\section{Introduction}

The recent discovery of the $\Xi^{++}_{cc}$ baryon by the LHCb Collaboration~\cite{Aaij:2017ueg,Aaij:2018gfl}, together with the expectation that other states can be confirmed or discovered in the near future, has revitalized the interest of the theoretical community on double heavy baryons. Earlier SELEX claims~\cite{Mattson:2002vu,Ocherashvili:2004hi} on the discovery of $\Xi^{+}_{cc}$ appear to clash with LHCb searches~\cite{Aaij:2013voa,Aaij:2019jfq}, as well as earlier ones by BABAR~\cite{Aubert:2006qw} and BELLE~\cite{Kato:2013ynr}.

From the theoretical side, a QCD based approach to double heavy baryons was already considered in the early days of Heavy Quark Effective Theory~\cite{Savage:1990di}. The key observation was that a $QQ$ state at short distances has an attractive channel in the $3^\ast$ representation. Then, if the heavy quark masses are large enough, the $QQ$ would form a compact hard core and the lowest-lying excitations would be given by the typical hadronic scale $\lQ$. The spectrum would then be analogous to the one of heavy-light mesons. Predictions for the hyperfine splitting were put forward within this approach~\cite{Savage:1990di}, sometimes referred to as heavy quark-diquark duality. This framework was put in a more solid theoretical basis in Ref.~\cite{Brambilla:2005yk} by working out a lower energy effective theory (EFT) for the $QQ$ system, similar to potential nonrelativistic QCD (pNRQCD)~\cite{Pineda:1997bj,Brambilla:1999xf}. The quark-diquark duality assumption could then be quantified: it would hold when the typical binding energies of the $QQ$ systems, $E_{bin}$, are much larger than the typical hadronic scale ($E_{bin}\gg \lQ$). In practice, however, the duality hypothesis does not hold for double charm nor for double bottom baryons. Indeed, it is well known that $E_{bin}\gg \lQ$ does not hold for charmonium and bottomonium ($Q\bar Q$), the attractive channel of which is twice stronger than the one of their $QQ$ counterpart. The EFT built in Ref.~\cite{Brambilla:2005yk} is valid whenever the typical size of the system is smaller than typical hadronic size, $r\ll 1/\lQ$, and in particular it is still correct for $E_{bin}\sim \lQ$. In this case, the excitations due to the internal $QQ$ dynamics compete with the excitations of the light degrees of freedom (which include a light quark) surrounding the $QQ$ compact core, and the spectrum of the lower-lying states will not mimic the one of the heavy-light mesons anymore. For charmonium and bottomonium the hypothesis that $E_{bin}\sim \lQ$ is only reasonable for the ground state and the gross description of excited states clearly requires the introduction of a confining potential, in addition to the Coulomb-like potential that arises from the hypothesis $r\ll 1/\lQ$. Therefore, for doubly heavy charm or bottom baryons, for which the attraction of the Coulomb-like potential is twice weaker than that for quarkonium, the assumption $E_{bin}\sim \lQ$ seems unlikely to hold.

It is the aim of this paper to build an EFT for $QQq$ systems in which the hypothesis $r\ll 1/\lQ$ is released, along the lines suggested in Ref.~\cite{Soto:2003ft}. In fact, this paper may be considered a concrete example of a more general formalism developed in an accompanying one~\cite{Soto:2020xpm}. This EFT is built upon the heavy quark mass expansion, $m_Q\gg\lQ$, and an adiabatic expansion between the dynamics of the heavy quarks, and the light degrees of freedom, the gluons and light quarks, $\lQ\gg E_{bin}$. Under these assumptions, the heavy quark mass and the typical hadronic scale can be integrated out producing an EFT that at leading order (LO) consist of a set of wave function fields for the $QQ$ system with the quantum numbers of the light degrees of freedom, in addition to the ones of the $QQ$, interacting through a number of Born-Oppenheimer (BO) potentials. Since the BO potentials cannot be calculated in perturbation theory, we shall use available lattice data for them. Four different BO potentials turn out to be relevant for describing the spectrum of double charm and double bottom baryons below the first heavy-meson-heavy-baryon threshold. At LO, the BO potentials are flavor independent. They neither depend on the heavy quark mass, nor on light quark ones $m_q$, if $\lQ\gg m_q$ is assumed. Hence, the LO Lagrangian enjoys heavy quark spin symmetry and chiral symmetry. We calculate the spectrum using the lattice data of Refs.~\cite{Najjar:2009da,Najjarthesis} as the input for the BO potential. We also work out the EFT at next-to-leading order (NLO) in the $1/m_Q$ expansion for the terms depending on the heavy quark spin and angular momentum. Heavy quark spin symmetry is violated at this order. We put forward model-independent formulas for the spin splittings. In particular, we make a prediction for the spin partner of $\Xi^{++}_{cc}$.

Another QCD-based approach to double heavy baryons is lattice QCD. The study of double heavy baryons on the lattice is quite challenging due to the wide spread of the characteristic scales. The light quark and gluon dynamics occurs at low energies and requires of large lattices for accurate simulations, while the heavy quarks necessitate small lattice spacings. The combination of both requirements results in computationally demanding simulations. To reduce the computational cost, early studies relied on the quenched approximation and were carried out in lattice nonrelativistic QCD (NRQCD)~\cite{Alexandrou:1994dm,Bowler:1996ws,AliKhan:1999yb,Mathur:2002ce}. For doubly charmed baryons, relativistic actions were latter used in Refs.~\cite{Lewis:2001iz,Flynn:2003vz} and full QCD simulations in Refs.~\cite{Chiu:2005zc,Na:2007pv,Liu:2009jc,Lin:2010wb,Briceno:2012wt,Alexandrou:2012xk,Basak:2012py,Bali:2012ua,Namekawa:2013vu,Brown:2014ena,Bali:2015lka,Alexandrou:2017xwd,Can:2019wts}. The latter, however, were limited to the lowest-lying spin $1/2$ and $3/2$. The spectrum of a wider array of $j^{\eta_P}$ states was obtained in Refs.~\cite{Padmanath:2015jea,Mathur:2018rwu}. In the bottom sector unquenched computations have been carried out in Refs.~\cite{Na:2007pv,Lewis:2008fu,Brown:2014ena,Mohanta:2019mxo} but still using nonrelativistic bottom quarks. Lattice NRQCD, expanded about the static limit, is also necessary to compute the matching coefficients of the EFT presented here, namely the BO potentials. This can be done by using the expressions of the matching coefficients as operator insertions in the Wilson loop that we present in an accompanying paper~\cite{Soto:2020xpm}. An example of this are the static energies computed in Refs.~\cite{Najjar:2009da,Najjarthesis} that we use as input.  

We have organized the paper as follows. In Sec.~\ref{nreft} we construct the EFT at NLO. In Sec.~\ref{lo} we focus on the LO Lagrangian. We find suitable parametrizations of the lattice data of Refs.~\cite{Najjar:2009da,Najjarthesis} that fulfill expected short and long distance constraints, and calculate the spectrum. In Sec.~\ref{hf}, we discuss the hyperfine splittings and produce a number of model-independent results. We compare our findings with lattice QCD results in Sec.~\ref{latcomp}. Sec.~\ref{concl} is devoted to the conclusions. In Appendix~\ref{csed} we derive the coupled Schr\"odinger equations for the $\kappa^p=(3/2)^-$ states, which are a mixture of the $(1/2)_u$ and $(3/2)_u$ static energies, and in Appendix~\ref{rwfp} we collect the plots of the double heavy baryon radial wave functions.

\section{Nonrelativistic EFT for double heavy baryons}
\label{nreft}

Double heavy baryons are composed of two distinct components: a heavy quark pair in a $3^\ast$ color state and a light quark. The heavy quark mass is larger than the typical hadronic scale, $m_Q\gg\Lambda_{\rm QCD}$, and therefore the natural starting point to study double heavy baryons is NRQCD~\cite{Caswell:1985ui,Bodwin:1994jh,Manohar:1997qy}. At leading order in the $1/m_Q$ expansion the heavy quarks are static and the spectrum of the theory is given by the so-called static energies. The static energies are the energy eigenvalues of the static eigenstates which are characterized by the following set of quantum numbers: the flavor of the light quark, the heavy quark pair relative distance $\bm{r}$, and the representations of $D_{\infty h}$. The latter is a cylindrical symmetry group, also encountered in diatomic molecules. The representations of $D_{\infty h}$ are customarily written as $\Lambda_{\eta}$, with $\Lambda$ the absolute value of the projection of the light quark state angular momentum on the axis joining the two heavy quarks, $\hat{\bm{r}}$, and $\eta=\pm 1$ is the parity eigenvalue, denoted by $g = + 1$ and $u = - 1$\footnote{Additionally for $\Lambda=0$ there is a symmetry under reflection in any plane passing through the axis $\hat{\bm{r}}$, the eigenvalues of the corresponding symmetry operator being $\sigma=\pm 1$ and indicated as superscript. However, it is not needed for half-integer light quark spin states and we will omit it.}.

\begin{figure}[ht!]
   \centerline{\includegraphics[width=.6\textwidth]{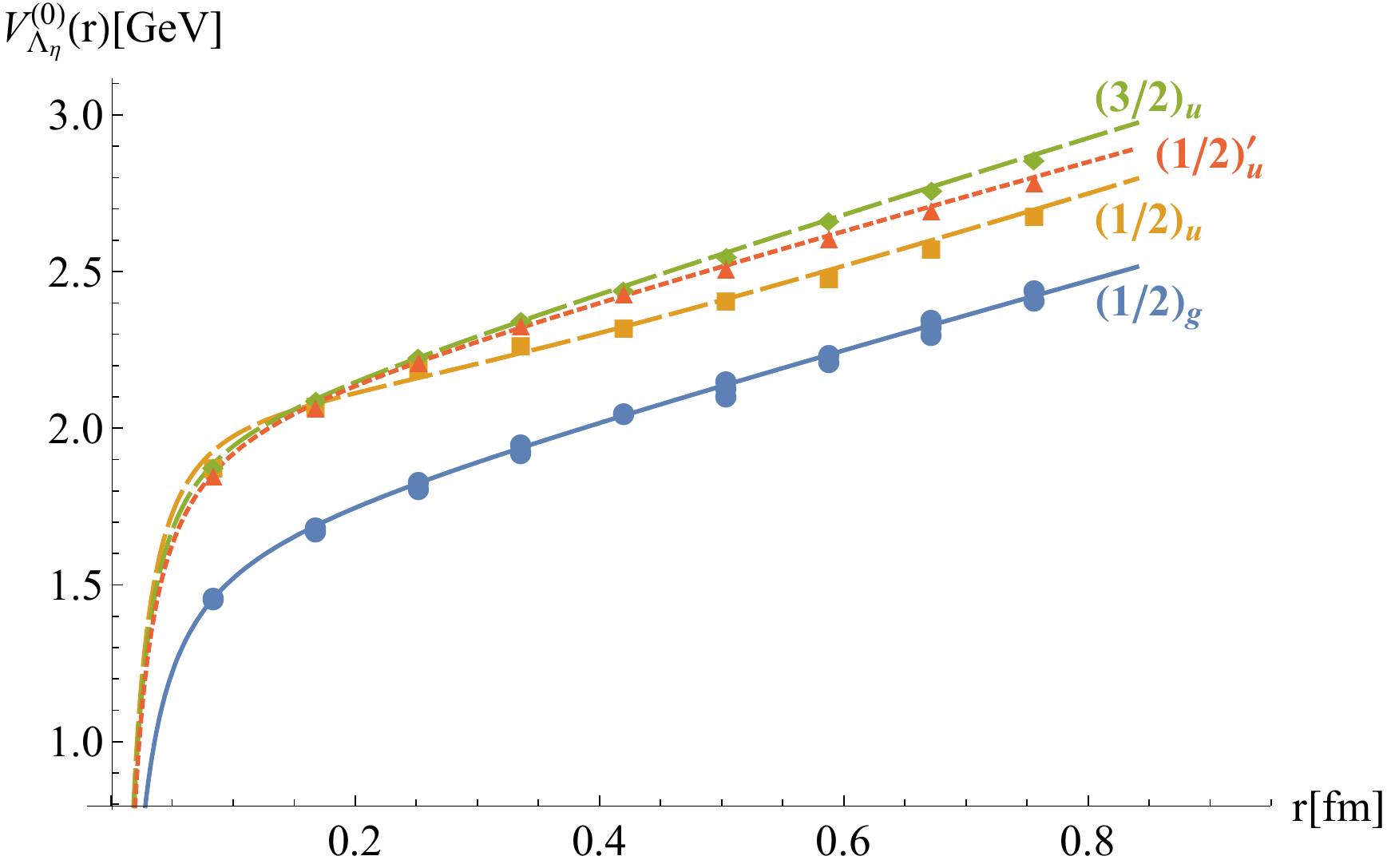}}
	\caption{Lattice data of Refs.~\cite{Najjar:2009da,Najjarthesis} for the four lowest lying double heavy baryon static energies together with the fitted potentials corresponding to the parametrizations of Eqs.~\eqref{v1o2g}-\eqref{v1o2up}.}
	\label{pfit}
 \end{figure}

The static energies are nonperturbative objects and have to be computed in lattice QCD. Preliminary calculations for the static energies of two heavy quarks and a light quark state in the isospin limit were presented in Refs.~\cite{Najjar:2009da,Najjarthesis}. The lattice data for the four lowest lying states is shown in Fig.~\ref{pfit}. The lowest lying static energy corresponds to the representation $(1/2)_g$ followed by three very close states corresponding to the representations $(1/2)_u$, $(3/2)_u$ and $(1/2)_u'$. The prime is used to indicate an excited state with the same representation as a lower lying one. In the short-distance limit the symmetry group is enlarged from $D_{\infty h}$ to $O(3)$ and the states can be labeled by their spin ($\kappa$) and parity ($p$). In this limit the $\kappa^p$ of the three lowest lying states found in Refs.~\cite{Najjar:2009da,Najjarthesis} correspond to $(1/2)^+$, $(3/2)^-$, and $(1/2)^-$. In the short-distance limit the two heavy quarks act as a single heavy antiquark; this is sometimes refereed to as quark-diquark symmetry~\cite{Savage:1990di,Hu:2005gf,Mehen:2017nrh,Mehen:2019cxn}. Therefore, the double heavy baryon system in the short-distance limit is equivalent to a heavy-light meson and in fact the spectrum found in Refs.~\cite{Najjar:2009da,Najjarthesis} is consistent with the D and B meson spectra within a few tenths of MeV in this limit. Projecting the $\kappa^p$ states into the heavy quark axis one can obtain states in representations of $D_{\infty h}$. We show the correspondence in Table~\ref{repcor}. The most significant feature in Table~\ref{repcor} is that $(3/2)^-$ projects both to $(1/2)_u$ and $(3/2)_u$, thus we expect these two static energies to be degenerate in the short-distance limit on symmetry grounds. This behavior can in fact be observed form the lattice data in Fig.~\ref{pfit}. The short-distance degeneracy between $(1/2)_u'$ and $(1/2)_u-(3/2)_u$ is a reflection of the degeneracy between the short distance states $(3/2)^-$ and $(1/2)^-$, which as far as we know is accidental.

\begin{table}[ht!]
\begin{tabular}{||c|c||} \hline\hline
$O(3)$ & $D_{\infty h}$ \\\hline
$(1/2)^+$ & $(1/2)_g$ \\
$(3/2)^-$ & $(1/2)_u,\,(3/2)_u$ \\
$(1/2)^-$ & $(1/2)_u'$ \\\hline\hline
\end{tabular}
\caption{Correspondence between short-distance $O(3)$ representations and $D_{\infty h}$ representations.}
\label{repcor}
\end{table}

The static energies can be computed in the lattice from the large time limit of correlators of appropriate operators that overlap dominantly with the static states. Such operators must have the same quantum numbers as the static states. The static energies are given by

\begin{align}
E_{\kappa^p\Lambda_\eta}(\bm{r})=\lim_{t\to \infty}\frac{i}{t}\ln\langle 0|{\rm Tr}\left[{\cal P}_{\kappa\Lambda} \mathcal{O}_{\kappa^p}(t/2,\,\bm{r},\,\bm{R})\mathcal{O}^{\dagger}_{\kappa^p}(-t/2,\,\bm{r},\,\bm{R})\right]|0 \rangle\label{stec}
\end{align}
where $\bm{r}$ and $\bm{R}$ are the relative and the center-of-mass coordinate of the heavy quark pair, respectively, and

\begin{align}
\mathcal{O}^{\alpha}_{(1/2)^+}(t,\,\bm{r},\,\bm{R})=&\psi^{\top}(t,\,\bm{x}_2)\phi^{\top}(t,\,\bm{R},\,\bm{x}_2) \underline{T}^l \left[P_+q^{l}(t,\bm{R})\right]^\alpha\phi(t,\,\bm{R},\,\bm{x}_1)\psi(t,\,\bm{x}_1)\,,\\
\mathcal{O}^{\alpha}_{(1/2)^-}(t,\,\bm{r},\,\bm{R})=&\psi^{\top}(t,\,\bm{x}_2)\phi^{\top}(t,\,\bm{R},\,\bm{x}_2) \underline{T}^l \left[P_+\gamma^5q^l(t,\bm{R})\right]^\alpha\phi(t,\,\bm{R},\,\bm{x}_1)\psi(t,\,\bm{x}_1)\,,\\
\mathcal{O}^{\alpha}_{(3/2)^-}(t,\,\bm{r},\,\bm{R})=&\psi^{\top}(t,\,\bm{x}_2)\phi^{\top}(t,\,\bm{R},\,\bm{x}_2){\cal C}^{3/2\,\alpha}_{1\,m\,1/2\,\alpha'}\underline{T}^l\left[\left(\bm{e}^{\dagger}_{m}\cdot\bm{D}\right) \left(P_+q(t,\bm{R})\right)^{\alpha'}\right]^l\phi(t,\,\bm{R},\,\bm{x}_1)\psi(t,\,\bm{x}_1)\,.
\end{align}
with $\psi$ the Pauli spinor fields that annihilate a heavy quark, $\underline{T}^l_{ij}=\epsilon_{lij}/\sqrt{2}$ is a $3^\ast$ irreducible tensor~\cite{Brambilla:2005yk}, ${\cal C}^{j_3\,m_3}_{j_1\,m_1\,j_2\,m_2}$ is a Clebsch-Gordan coefficient, $P_+=(1+\gamma^0)/2$ and the polarization vectors are $\bm{e}_{+1}=-(1,\,i,\,0)/\sqrt{2}$, $\bm{e}_{-1}=(1,\,-i,\,0)/\sqrt{2}$, $\bm{e}_{0}=(0,\,0,\,1)$. The light quark fields are standard Dirac fermions represented by $q_{\alpha}^l(t,\bm{R})$ where $l$ is the color index and $\alpha$ the spin index. $\phi$ is a Wilson line defined as
\begin{align}
\phi(t,\bm{y},\bm{x})=P\left\{e^{i\int_0^1 ds\left(\bm{x}-\bm{y}\right)\cdot g\bm{A}(t,\bm{x}-s(\bm{x}-\bm{y}))}\right\}\,,
\end{align}
where $P$ is the path-ordering operator. The projectors ${\cal P}_{\kappa \Lambda}$ in Eq.~\eqref{stec} act on the light quark spin indices and for spin-$1/2$ and spin-$3/2$ take the form
\begin{align}
{\cal P}_{\frac{1}{2}\frac{1}{2}}&=\mathbb{1}^{\rm lq}_{2}\,,\\
{\cal P}_{\frac{3}{2}\frac{1}{2}}&=\frac{9}{8}\mathbb{1}^{\rm lq}_{4}-\frac{1}{2}\left(\hat{\bm{r}}\cdot\bm{S}_{3/2}\right)^2\,,\\
{\cal P}_{\frac{3}{2}\frac{3}{2}}&=-\frac{1}{8}\mathbb{1}^{\rm lq}_{4}+\frac{1}{2}\left(\hat{\bm{r}}\cdot\bm{S}_{3/2}\right)^2\,,
\end{align}
with $\mathbb{1}^{\rm lq}_{n}$ an identity matrix in the light quark spin space of dimension $n=2\kappa+1$.

The spectrum of double heavy baryons corresponds the heavy-quark pair bound states on the static energies defined by Eq.~\eqref{stec}, thus the lowest lying states correspond to the static energies in Fig.~\ref{pfit}. The binding energies are expected to be smaller than $\Lambda_{\rm QCD}$ and therefore these bound states can be described in a BO-inspired approach~\cite{Brambilla:2017uyf}. That is, incorporating an adiabatic expansion in the energy scales of typical binding energies over the one of light quark and gluon degrees of freedom, $\Lambda_{\rm QCD}$.

The EFT describing heavy exotic hadrons and double heavy baryons up to $1/m_Q$ for any spin of the light-quark and gluonic degrees of freedom has been presented in Ref.~\cite{Soto:2020xpm}. In the case of double heavy baryons corresponding to the static states of Fig.~\ref{pfit} the Lagrangian is as follows:
\begin{align}
{\cal L}_{QQq}=&\Psi^{\dagger}_{(1/2)^+}\left[i\partial_t-h_{(1/2)^+}\right]\Psi_{(1/2)^+}+\Psi^{\dagger}_{(3/2)^-}\left[i\partial_t-h_{(3/2)^-}\right]\Psi_{(3/2)^-}+\Psi^{\dagger}_{(1/2)^-}\left[i\partial_t-h_{(1/2)^-}\right]\Psi_{(1/2)^-}\,,\label{boeft}
\end{align}
with the $\Psi$ fields understood as depending on $t,\,\bm{r},\,\bm{R}$, where $\bm{r}=\bm{x}_1-\bm{x}_2$ and $\bm{R}=(\bm{x}_1+\bm{x}_2)/2$ are the relative and center-of-mass coordinates of a heavy quark pair. The $\Psi$ fields live both in the light quark and heavy quark pair spin spaces. In the Lagrangian on Eq.~\eqref{boeft} we have chosen to leave the spin indices implicit.

The Hamiltonian densities $h_{\kappa^p}$ have the following expansion up to $1/m_Q$
\begin{align}
h_{\kappa^p}=\frac{\bm{p}^2}{m_Q}+\frac{\bm{P}^2}{4m_Q}+V_{\kappa^p}^{(0)}(\bm{r})+\frac{1}{m_Q}V_{\kappa^p}^{(1)}(\bm{r},\,\bm{p})\,,\label{hamden}
\end{align}
with $\bm{p}=-i\nabla_r$ and $\bm{P}=-i\nabla_R$. The kinetic terms in Eq.~\eqref{hamden} are diagonal in spin space while the potentials are not. The static potentials, $V^{(0)}$, are diagonal in the heavy quark spin space, due to heavy quark spin symmetry, while the light quark spin structure is determined by the representations of $D_{\infty h}$ that the $\kappa^p$ quantum numbers can be associated with:
\begin{align}
V_{(1/2)^{\pm}}^{(0)}(\bm{r})=&V_{(1/2)^{\pm}}^{(0)}(r)\,,\\
V_{(3/2)^-}^{(0)}(\bm{r})=&V_{(3/2)^-(3/2)}^{(0)}(r){\cal P}_{\frac{3}{2}\frac{3}{2}}+V_{(3/2)^-(1/2)}^{(0)}(r){\cal P}_{\frac{3}{2}\frac{1}{2}}\,,\label{edihsp}
\end{align}
with ${\cal P}_{\kappa\Lambda}$ the projectors into representations of $D_{\infty h}$ in the spin-$\kappa$ space. These fulfill the usual projector properties: they are idempotent ${\cal P}^2_{\kappa\Lambda}={\cal P}_{\kappa\Lambda}$, orthogonal to each other ${\cal P}_{\kappa\Lambda}{\cal P}_{\kappa\Lambda'}=\delta_{\Lambda\Lambda'}$, and add up to the identity in the spin-$\kappa$ space $\sum_{\Lambda}{\cal P}_{\kappa\Lambda}=\mathbb{1}_{2\kappa+1}$.

The subleading potentials $V^{(1)}$ can be split into terms that depend on $\bm{S}_{QQ}$ or $\bm{L}_{QQ}$ and terms that do not. The former read
\begin{align}
V_{(1/2)^{\pm}{\rm SD}}^{(1)}(\bm{r})=&V^{s1}_{(1/2)^{\pm}}(r)\bm{S}_{QQ}\cdot\bm{S}_{1/2}+V^{s2}_{(1/2)^{\pm}}(r)\bm{S}_{QQ}\cdot\left(\bm{{\cal T}}_{2}\cdot\bm{S}_{1/2}\right)+V^{l}_{(1/2)^{\pm}}(r)\left(\bm{L}_{QQ}\cdot\bm{S}_{1/2}\right)\,,\label{sdp12}\\
V_{(3/2)^-{\rm SD}}^{(1)}(\bm{r})=&\sum_{\Lambda\Lambda'=\frac{1}{2},\frac{3}{2}}{\cal P}_{\frac{3}{2}\Lambda}\left[V^{s1}_{(3/2)^-\Lambda\Lambda'}(r)\bm{S}_{QQ}\bm{S}_{3/2}+V^{s2}_{(3/2)^-\Lambda\Lambda'}(r)\bm{S}_{QQ}\cdot\left(\bm{{\cal T}}_{2}\cdot\bm{S}_{3/2}\right)\right.\nn\\
&\left.+V^{l}_{(3/2)^-\Lambda\Lambda'}(r)\left(\bm{L}_{QQ}\cdot\bm{S}_{3/2}\right)\right]{\cal P}_{\frac{3}{2}\Lambda'}\,.\label{sdp32}
\end{align}
with the total heavy quark spin defined as $2\bm{S}_{QQ}=\bm{\sigma}_{QQ}=\bm{\sigma}_{Q_1}\mathbb{1}_{2\,Q_2}+\mathbb{1}_{2\,Q_1}\bm{\sigma}_{Q_2}$ where the $\mathbb{1}_2$ are identity matrices in the heavy quark spin space for the heavy-quark labeled in the subindex, $\bm{L}_{QQ}=\bm{r}\times \bm{p}$ and the spin-$2$ irreducible tensor is defined as
\begin{align}
\left(\bm{{\cal T}}_{2}\right)^{ij}=\hat{\bm{r}}^i\hat{\bm{r}}^j-\frac{1}{3}\delta^{ij}\,.
\end{align}

The heavy quark spin component of the $\Psi$ fields is given by $\chi_{s}^{Q_1}\chi_{r}^{Q_2}$ with $\chi_{s}$ the usual spin-$1/2$ two-component spinors. The light quark spin component, $\chi^{lq}_{\alpha}$, is a $2$- or $4$-component spinor for the $\Psi_{(1/2)^{\pm}}$ and $\Psi_{(3/2)^-}$ fields, respectively.

The matching of the potentials in Eqs.~\eqref{edihsp}, \eqref{sdp12}, and \eqref{sdp32} in terms of static Wilson loops has been presented in Ref.~\cite{Soto:2020xpm}. These Wilson loops are nonperturbative quantities and should be computed on the lattice. The only ones available are the ones corresponding to the static energies, from Refs.~\cite{Najjar:2009da,Najjarthesis} shown in Fig.~\ref{pfit}, which match to the static potentials  
\begin{align}
V_{\kappa^p\Lambda}^{(0)}(\bm{r})=E_{\kappa^p\Lambda_\eta}(\bm{r})\,.\label{spmatching}
\end{align}

Furthermore the form of the static potentials can be constrained in the short- and long-distance limits from general grounds. In the short-distance regime, $r\lesssim 1/\Lambda_{\rm QCD}$, one can integrate out the relative momentum scale perturbatively and build weakly coupled pNRQCD~\cite{Pineda:1997bj,Brambilla:1999xf} for double heavy baryons as was done in Ref.~\cite{Brambilla:2005yk}. In this regime the potential at leading order in the multipole expansion is the sum of the Coulomb-like potential for two heavy quarks in a triplet state plus a nonperturbative constant
\begin{align}
V_{\kappa^p\Lambda}^{(0)}(\bm{r})=-\frac{2}{3}\frac{\alpha_s}{r}+\overline{\Lambda}_{\kappa^p}+{\cal O}(r^2)\label{shortdp}
\end{align}
with
\begin{align}
\overline{\Lambda}_{\kappa^p}=&\lim_{t \to \infty}\frac{i}{t}\log\langle \mathcal{Q}_{\kappa^p}(t/2,\,\bm{R})e^{ig\int^{t/2}_{-t/2}dt'A_0^l(t,\,\bm{R})\left(
{T}^{*}\right)^l} \mathcal{Q}^{\dag}_{\kappa^p}(-t/2,\,\bm{R})\rangle\,,
\end{align}
where $\mathcal{Q}_{\kappa^p}$ is the light-quark piece of the interpolating operator $\mathcal{O}_{\kappa^p}$.

Using quark-diquark symmetry the values of $\overline{\Lambda}_{\kappa^p}$ can be obtained from analysis of the D and B meson masses~\cite{Pineda:2001zq,Bazavov:2018omf}. On the other hand, in the long-distance regime, $r\gg 1/\Lambda_{\rm QCD}$, we expect the formation of a flux tube that behaves as a quantum string. The formation of such flux tubes has been observed from lattice QCD~\cite{Juge:2002br,Bali:2005fu,Mueller:2019mkh} in standard and hybrid quarkonium, but is not yet confirmed for double heavy baryons. Nevertheless, the data from Refs.~\cite{Yamamoto:2008jz,Najjar:2009da,Najjarthesis}, although it does not strictly reach the long-distance regime, does show a remarkable linear behavior for the larger heavy quark pair distances. Therefore in the long-distance regime we expect the potential to be linear in $r$ plus a possible additive constant depending on the light quark mass
\begin{align}
V_{\kappa^p\Lambda}^{(0)}(\bm{r})=\sigma r+c(m_q)+{\cal O}(1/r)\,.
\end{align}

\section{Double heavy baryon spectrum at leading order}
\label{lo}

The spectrum of double heavy baryons is obtained by solving the Schr\"odinger equations resulting from the LO Lagrangian
\begin{align}
{\cal L}^{\rm LO}_{QQq}=&\Psi^{\dag}_{(1/2)^+}\left(i\partial_t-\frac{\bm{p}^2}{m_Q}+V_{(1/2)^+}^{(0)}(r)\right)\Psi^{\dag}_{(1/2)^+}+\Psi^{\dag}_{(1/2)^-}\left(i\partial_t-\frac{\bm{p}^2}{m_Q}+V_{(1/2)^-}^{(0)}(r)\right)\Psi^{\dag}_{(1/2)^-}\nn\\
&+\Psi^{\dag}_{(3/2)^-}\left(i\partial_t-\frac{\bm{p}^2}{m_Q}+V_{(3/2)^-(1/2)}^{(0)}(r){\cal P}_{\frac{3}{2}\frac{1}{2}}+V_{(3/2)^-(3/2)}^{(0)}(r){\cal P}_{\frac{3}{2}\frac{3}{2}}\right)\Psi^{\dag}_{(3/2)^-}\label{boeftlo}
\end{align}

The first two terms in Eq.~\eqref{boeftlo} define standard Schr\"odinger equations, while the last term corresponds to two sets of coupled Schr\"odinger equations corresponding to the two possible parities of the double heavy baryon states. The coupled Schr\"odinger equations for the latter case are derived in Appendix~\ref{csed}. The main point to keep in mind is that while the states associated with $\kappa^p=(1/2)^\pm$ are eigenstates of $\bm{L}^2_{QQ}$ with eigenvalue $l(l+1)$, the states associated with $\kappa^p=(3/2)^-$ are eigenstates of $\bm{L}^2=(\bm{L}_{QQ}+\bm{S}_{3/2})^2$ with eigenvalue $\ell(\ell+1)$. Additionally, in the latter case there are states with positive and negative parity for each $\ell$ with different masses due to the mixing that leads into the coupled Schr\"odinger equations. This is the so-called $\Lambda$-doubling effect known from molecular physics.

Using Eq.~\eqref{spmatching}, the static potentials in Eq.~\eqref{boeftlo} can be obtained by fitting the lattice data from Refs.~\cite{Najjar:2009da,Najjarthesis} on the corresponding static energies. To fit them we use the following parametrizations of the potentials which interpolate between the expected short- and long-distance behavior:
\begin{align}
&V_{(1/2)^+}=E_{(1/2)_g}=-\frac{2}{3}\frac{\alpha_s(\nu_{\rm lat})}{r}+\frac{c_2 r+c_1}{c_3 r+1}+\sigma r\,,\label{v1o2g}\\
&V_{(3/2)^-(1/2)}=E_{(1/2)_u}=-\frac{2}{3}\frac{\alpha_s(\nu_{\rm lat})}{r}+\frac{b_3 r^2+b_2 r+b_1}{b_5 r^2+b_4 r+1}+\sigma r\,,\label{v1o2u}\\
&V_{(3/2)^-(3/2)}=E_{(3/2)_u}=-\frac{2}{3}\frac{\alpha_s(\nu_{\rm lat})}{r}+\frac{b_7 r^2+b_6 r+b_1}{b_9 r^2+b_8 r+1}+\sigma r\,,\label{v3o2u}\\
&V_{(1/2)^-}=E_{(1/2)_u'}-\frac{2}{3}\frac{\alpha_s(\nu_{\rm lat})}{r}+\frac{c_5 r+c_4}{c_6 r+1}+\sigma r\,,\label{v1o2up}
\end{align}
with the $\nu$ scale taken as the inverse of the lattice spacing $\nu_{\rm lat}=1/a=2.16$~GeV ($a=0.084$fm). We constrain the parameters $b_1=c_1+E^{\rm latt}(1a)_{(1/2)_u}-E^{\rm latt}(1a)_{(1/2)_g}$ and $c_4=c_1+E^{\rm latt}(1a)_{(1/2)^{\prime}_u}-E^{\rm latt}(1a)_{(1/2)_g}$. Notice that, numerically, $E^{\rm latt}(1a)_{(1/2)_u}\simeq E^{\rm latt}(1a)_{(3/2)_u}$ hence in practice $b_1\simeq c_4$. 

The energy offset $c_1$ is an additive constant that affects the masses of all the double heavy baryon states; therefore, it is important to determine it accurately. To do so, we fit the short-distance data ($r=a,\,2a,\,3a$) to the one-loop expression of the heavy-quark-heavy-quark $3^\ast$ potential added to $c_1$
\begin{align}
&V_{(1/2)^+}=-\frac{2}{3}\frac{\alpha_s(\nu_{\rm lat})}{r}\left(1+\frac{\alpha_s(\nu_{\rm lat})}{4\pi}\left(2\beta_0\log\left(\nu_{\rm lat}r e^{\gamma_E}\right)+a_1\right)\right)+c_1\,.
\end{align}
The value we obtain is
\begin{align}
c_1=1.948\,{\rm GeV}\,,
\end{align}
and therefore
\begin{align}
b_1=2.370\,{\rm GeV}\,,\quad c_4=2.343\,{\rm GeV}\,.
\end{align}

The long-distance behavior is given by the linear term whose constant is fixed at $\sigma=0.21$~GeV$^2$ \cite{Luscher:2002qv}. One could obtain similar values by fitting the longer distance lattice points to a straight line plus a constant term; however, there is some correlation between the two parameters that makes this procedure undesirable. The rest of the parameters are obtained by fitting the potentials in Eqs.~\eqref{v1o2g}-\eqref{v1o2up} to the lattice data. The fits yield the following values:
\begin{align}
&c_2=15.782~{\rm GeV^2},\,c_3=9.580~{\rm GeV},\,c_5=13.265~{\rm GeV^2},\,c_6=6.560~{\rm GeV}\,,\nn\\
& b_2=1.196~{\rm GeV^2},\,b_3=0.123~{\rm GeV^3},\,b_4=0.763~{\rm GeV},\,b_5=0.041~{\rm GeV^2},\,\nn\\
& b_6=5.560~{\rm GeV^2},\,b_7=1.066~{\rm GeV^3},\,b_8=2.879~{\rm GeV},\,b_9=0.452~{\rm GeV^2}\,. 
\end{align}
In Fig.~\ref{pfit} we show the fitted potentials together with the lattice data.

\begin{figure}[ht!]
   \centerline{\includegraphics[width=.6\textwidth]{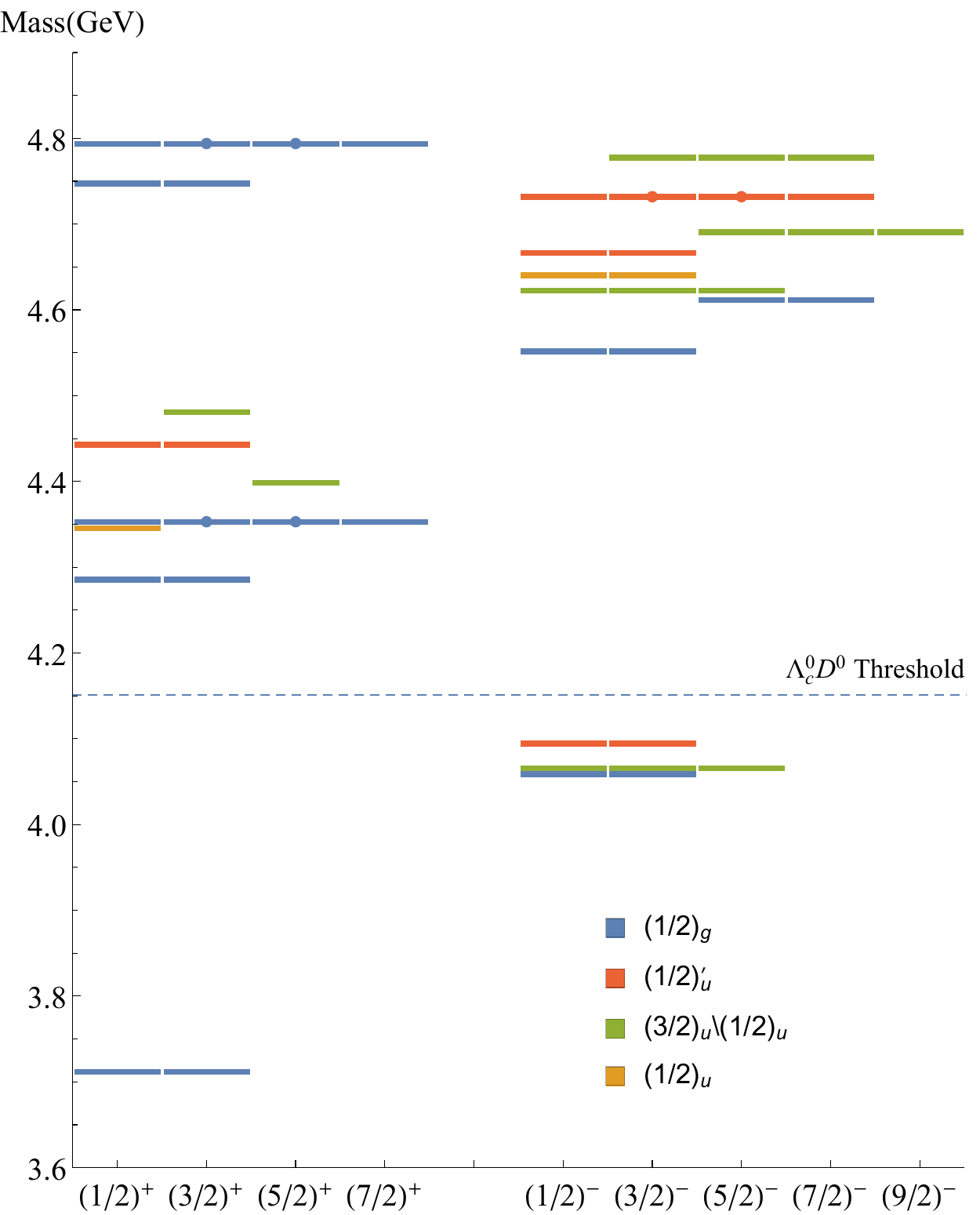}}
	\caption{Spectrum of $ccq$ double heavy baryons in terms of $j^{\eta_P}$ states. The spectrum corresponds to the results of Table~\ref{ccspc} and the corresponding $j^{\eta_P}$ multiplets from Table~\ref{quantumnumbers}. Each line represents a state; the lines with a dot indicate two degenerate states. The color indicates the static energies that generate the state.}
	\label{ccplot}
 \end{figure}

\begin{figure}[ht!]
   \centerline{\includegraphics[width=.6\textwidth]{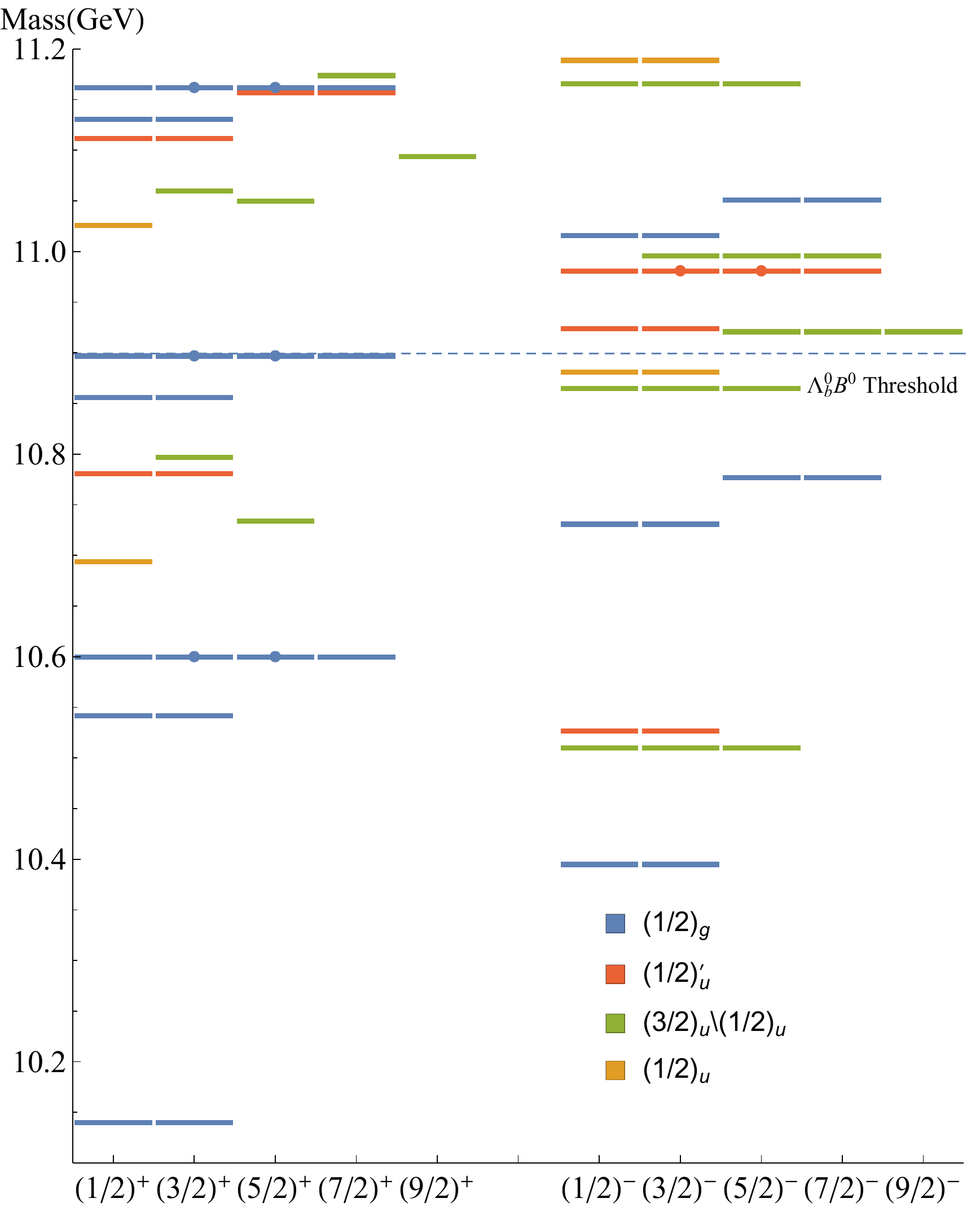}}
	\caption{Spectrum of $bbq$ double heavy baryons in terms of $j^{\eta_P}$ states. The spectrum corresponds to the results of Table~\ref{bbspc} and the corresponding $j^{\eta_P}$ multiplets from Table~\ref{quantumnumbers}. Each line represents a state; the lines with a dot indicate two degenerate states. The color indicates the static energies that generate the state.}
	\label{bbplot}
 \end{figure}

We solve numerically the radial Schr\"odinger equations in Eqs.~\eqref{v1o2g}-\eqref{v1o2up} with the short-distance constants of each potential ($c_1,\,c_4$ or $b_1$) subtracted. The effect of this is just to set a common origin of energies so we can compare the binding energies obtained from the Schr\"odinger equations. To obtain the total masses of the baryons we add to the binding energy two times the heavy quark mass and $\overline{\Lambda}_{(1/2)^+}$
\begin{align}
&M^{(0)}_{(1/2)_g}=2m_Q+E_b+\overline{\Lambda}_{(1/2)^+} \,,\\
&M^{(0)}_{(3/2)_u\backslash(1/2)_u}=2m_Q+E_b+\overline{\Lambda}_{(1/2)^+}+E^{\rm latt}(1a)_{(1/2)_u}-E^{\rm latt}(1a)_{(1/2)_g} \,,\\
&M^{(0)}_{(1/2)^{\prime}_u}=2m_Q+E_b+\overline{\Lambda}_{(1/2)^+}+E^{\rm latt}(1a)_{(1/2)^{\prime}_u}-E^{\rm latt}(1a)_{(1/2)_g}\,.
\end{align}
In practice what we are doing is removing the ambiguity on the origin of energies of the lattice data on the static energies by rescaling them by a factor $\overline{\Lambda}_{(1/2)^+}-c_1$ so that the short-distance behavior of $V_{(1/2)^+}$ matches exactly Eq.~\eqref{shortdp} since $\overline{\Lambda}_{(1/2)^+}$ can be obtained independently from lattice studies of D and B mesons masses. We take the values for the heavy quark masses and $\overline{\Lambda}_{(1/2)^+}$ from Ref.~\cite{Bazavov:2018omf}\footnote{Notice, that in Ref.~\cite{Bazavov:2018omf} $\overline{\Lambda}_{(1/2)^+}$ is simply called $\overline{\Lambda}$.}
\begin{align}
&m_c=1.392(11)~{\rm GeV}\,, \label{mc}\\
&m_b=4.749(18)~{\rm GeV}\,, \label{mb}\\
&\overline{\Lambda}_{(1/2)^+}=0.555(31)~{\rm GeV} \,.\label{lbar12}
\end{align}
The same masses have been used in the kinetic terms of the Schr\"odinger equations. In the Tables~\ref{ccspc} and \ref{bbspc} we present the results for the spectrum of $cc$ and $bb$ double heavy baryons. In Table~\ref{quantumnumbers} we show the full quantum numbers of the double baryon states including mixings and degenerate spin multiplets. We also represent the spectra in terms of $j^{\eta_P}$ states graphically in Figs.~\ref{ccplot} and \ref{bbplot} for double charm and double bottom baryons, respectively.

The BO approach, corresponding to our LO Lagrangian in Eq.~\eqref{boeftlo}, may only be completely consistent for states below the first heavy-meson-heavy-baryon threshold. Above threshold, the states predicted from the Schr\"odinger equations are expected to acquire widths, corresponding to decays into the particles that form the threshold, and the double heavy baryon masses can also vary due to coupled channel effects. Therefore, above heavy-baryon-meson thresholds our results should be taken with care. In the charm sector the first heavy-baryon-meson threshold is formed by the $\Lambda^0_c$ and $D_0$ at $\sim\,4.151$~GeV. There are only four spin multiplets below this threshold with quantum numbers $\Lambda_\eta=(1/2)_g$ with $l=0,1$, $n=1$, $\Lambda_\eta=(1/2)'_u$ with $l=0$ $n=1$ , and $\Lambda_\eta=(3/2)_{u}\backslash(1/2)_{u}$ with $\ell^{\eta_P}=(3/2)^+$, $n=1$. In the bottom sector the first heavy-baryon-meson threshold is formed by $\Lambda^0_b$ and $B_0$ at $\sim 10.899$~GeV. In this case sixteen spin multiplets are found below threshold: eight multiplets for $\Lambda_\eta=(1/2)_g$, two for $\Lambda_\eta=(1/2)_u'$, four for $\Lambda_\eta=(3/2)_{u}\backslash(1/2)_{u}$ and two for $(1/2)_{u}$.

\begin{table}[ht!]
\begin{tabular}{||c|c|c|c|c|c|c||}
 \hline\hline
$\kappa^p$                     & $\Lambda_\eta$                                  & $l$             & $\ell$        & $s_{QQ}$ & $j$                                       & $\eta_P$   \\ \hline
\multirow{4}{*}{$(1/2)^{\pm}$} & \multirow{4}{*}{$(1/2)_{g/u'}$}                 & $0$             & $1/2$         & $1$      & $(1/2,\,3/2)$                             & $\pm$ \\ 
                               &                                                 & $1$             & $(1/2,\,3/2)$ & $0$      & $(1/2,\,3/2)$                             & $\mp$ \\ 
                               &                                                 & $2$             & $(3/2,\,5/2)$ & $1$      & $((1/2,\,3/2,\,5/2)\,,(3/2,\,5/2,\,7/2))$ & $\pm$ \\ 
															 &                                                 & $3$             & $(5/2,\,7/2)$ & $0$      & $(5/2,\,7/2)$                             & $\mp$ \\ \hline
\multirow{10}{*}{$(3/2)^{-}$}  & \multirow{8}{*}{$(3/2)_{u}\backslash(1/2)_{u}$} & $0\backslash 2$ & $3/2$         & $1$      & $(1/2,\,3/2,\,5/2)$                       & $-$   \\ 
                               &                                                 & $1\backslash 3$ & $5/2$         & $0$      & $5/2$                                     & $+$   \\ 
                               &                                                 & $2\backslash 4$ & $7/2$         & $1$      & $(5/2,\,7/2,\,9/2)$                       & $-$   \\ 
															 &                                                 & $3\backslash 5$ & $9/2$         & $0$      & $9/2$                                     & $+$   \\ \cline{3-7}
															 &                                                 & $1\backslash 3$ & $3/2$         & $0$      & $3/2$                                     & $+$   \\ 
                               &                                                 & $2\backslash 4$ & $5/2$         & $1$      & $(3/2,\,5/2,\,7/2)$                       & $-$   \\  
                               &                                                 & $3\backslash 5$ & $7/2$         & $0$      & $7/2$                                     & $+$   \\ 
	  												   &                                                 & $4\backslash 6$ & $9/2$         & $1$      & $(7/2,\,9/2,\,11/2)$                      & $-$   \\ \cline{2-7}
                               & \multirow{2}{*}{$(1/2)_{u}$}                    & $1$             & $1/2$         & $0$      & $1/2$                                     & $+$   \\ \cline{3-7}
                               &                                                 & $2$             & $1/2$         & $1$      & $(1/2,\,3/2)$                             & $-$   \\ \hline\hline
\end{tabular}
\caption{Quantum numbers of double heavy baryons associated with the four lowest static energies displayed in Fig.~\ref{pfit}. The quantum numbers are as follows: $l(l+1)$ is the eigenvalue of $\bm{L}^2_{QQ}$, $\ell(\ell+1)$ is the eigenvalue of $\bm{L}^2=(\bm{L}_{QQ}+\bm{S}_\kappa)^2$, $s_{QQ}(s_{QQ}+1)$ is the eigenvalue of $\bm{S}^2_{QQ}$. Note that the Pauli exclusion principle constrains $s_{QQ}=0$ for odd $l$ and $s_{QQ}=1$ for even $l$. The total angular momentum $\bm{J}^2=(\bm{L}+\bm{S}_{QQ})^2$ has eigenvalue $j(j+1)$. Finally, $\eta_P$ stands for the parity eigenvalue. Numbers in parentheses correspond to degenerate multiplets at LO, numbers separated by backslashes indicate mixing of the physical state on those quantum numbers. Notice that $\pm$ in the parity column does not indicate degeneracy in that quantum number but correlates to the $\pm$ parity of the light quark operator in the first column.}
\label{quantumnumbers}
\end{table}

\begin{table}[ht!]
 \begin{tabular}{||c|c|c|c|c|c|c|c||}
 \hline\hline
   \multicolumn{8}{||c||}{$ccq$} \\ \hline
 $\Lambda_\eta$ & $n$ & $l(\ell^{\eta_P})$ & $E_{bin}$ & $M^{(0)}$   & $\langle1/r\rangle$ & $E_{kin}$ & $P_{1/2}$ \\ \hline
 $(1/2)_g$      & $1$ & $0$       & $0.373$   & $3.712$ & $0.586$             & $0.270$   & $1$       \\ 
 $(1/2)_g$      & $2$ & $0$       & $0.947$   & $4.286$ & $0.396$             & $0.447$   & $1$       \\ 
 $(1/2)_g$      & $3$ & $0$       & $1.409$   & $4.748$ & $0.319$             & $0.595$   & $1$       \\ 
% $(1/2)_g$      & $4$ & $0$       & $1.815$   & $5.154$ & $0.273$             & $0.727$   & $1$       \\ 
 $(1/2)_g$      & $1$ & $1$       & $0.723$   & $4.062$ & $0.347$             & $0.368$   & $1$       \\ 
 $(1/2)_g$      & $2$ & $1$       & $1.213$   & $4.552$ & $0.276$             & $0.526$   & $1$       \\  
% $(1/2)_g$      & $3$ & $1$       & $1.636$   & $4.975$ & $0.236$             & $0.6638$   & $1$       \\
 $(1/2)_g$      & $1$ & $2$       & $1.014$   & $4.353$ & $0.260$             & $0.458$   & $1$       \\ 
 $(1/2)_g$      & $2$ & $2$       & $1.455$   & $4.794$ & $0.221$             & $0.602$   & $1$       \\ 
 $(1/2)_g$      & $1$ & $3$       & $1.273$   & $4.612$ & $0.213$             & $0.541$   & $1$       \\ \hline 
% $(1/2)_g$      & $1$ & $4$       & $1.512$   & $4.851$ & $0.183$             & $0.618$   & $1$       \\ \hline 
 $(1/2)^{\prime}_u$ & $1$ & $0$   & $0.362$   & $4.095$ & $0.583$             & $0.268$   & $1$       \\ 
 $(1/2)^{\prime}_u$ & $2$ & $0$   & $0.933$   & $4.667$ & $0.396$             & $0.445$   & $1$       \\ 
 $(1/2)^{\prime}_u$ & $1$ & $1$   & $0.709$   & $4.443$ & $0.346$             & $0.366$   & $1$       \\
% $(1/2)^{\prime}_u$ & $2$ & $1$   & $1.198$   & $4.931$ & $0.276$             & $0.524$   & $1$       \\ 
 $(1/2)^{\prime}_u$ & $1$ & $2$   & $0.998$   & $4.732$ & $0.260$             & $0.457$   & $1$       \\ \hline 
% $(1/2)^{\prime}_u$ & $1$ & $3$   & $1.257$   & $4.990$ & $0.213$             & $0.540$   & $1$       \\ \hline 
 $(3/2)_u-(1/2)_u$ & $1$ & $(3/2)^-$ & $0.305$  & $4.066$ & $0.579$           & $0.267$   & $0.571$   \\ 
 $(3/2)_u-(1/2)_u$ & $2$ & $(3/2)^-$ & $0.863$  & $4.623$ & $0.381$           & $0.473$   & $0.828$   \\ 
 $(3/2)_u-(1/2)_u$ & $1$ & $(5/2)^+$ & $0.639$  & $4.399$ & $0.348$           & $0.380$   & $0.697$   \\ 
% $(3/2)_u-(1/2)_u$ & $2$ & $(5/2)^+$ & $1.139$  & $4.899$ & $0.277$           & $0.562$   & $0.821$   \\ 
 $(3/2)_u-(1/2)_u$ & $1$ & $(7/2)^-$ & $0.931$  & $4.691$ & $0.265$           & $0.484$   & $0.750$   \\ \hline
% $(3/2)_u-(1/2)_u$ & $2$ & $(7/2)^-$ & $1.392$  & $5.153$ & $0.225$           & $0.647$   & $0.825$   \\
% $(3/2)_u-(1/2)_u$ & $1$ & $(9/2)^+$ & $1.199$  & $4.959$ & $0.218$           & $0.578$   & $0.778$   \\ \hline 
 $(3/2)_u-(1/2)_u$ & $1$ & $(3/2)^+$ & $0.720$  & $4.481$ & $0.355$           & $0.390$   & $0.155$   \\ 
% $(3/2)_u-(1/2)_u$ & $2$ & $(3/2)^+$ & $1.159$  & $4.919$ & $0.233$           & $0.574$   & $0.883$   \\ 
 $(3/2)_u-(1/2)_u$ & $1$ & $(5/2)^-$ & $1.017$  & $4.778$ & $0.266$           & $0.488$   & $0.235$   \\ \hline
% $(3/2)_u-(1/2)_u$ & $1$ & $(7/2)^-$ & $1.286$  & $5.046$ & $0.218$           & $0.577$   & $0.283$   \\  \hline
 $(1/2)_u$     & $1$ & $(1/2)^+$  & $0.585$   & $4.346$ & $0.347$             & $0.371$   & $1$       \\  \hline
% $(1/2)_u$     & $2$ & $(1/2)^+$  & $1.105$   & $4.865$ & $0.282$             & $0.554$   & $1$       \\ \hline
 $(1/2)_u$     & $1$ & $(1/2)^-$  & $0.880$   & $4.641$ & $0.265$             & $0.478$   & $1$       \\ \hline\hline
\end{tabular}
 \caption{Double charm baryon spectrum. All dimensionful entries are in GeV. $E_{bin}$ is the binding energy of the heavy quarks, $\langle1/r\rangle$ is the expected value of $1/r$ and $E_{kin}$ is the expected value of the kinetic energy of the heavy quarks. $P_{1/2}$ is the probability of finding the state in a $\Lambda=1/2$ state, and therefore a measure of the mixing between $\Lambda=3/2$ and $1/2$. The first charmed-meson-charmed-baryon threshold is $M(\Lambda^0_c)+M(D^0)=4.151$~GeV.}
\label{ccspc}
\end{table}

\begin{table}[ht!]
\begin{tabular}{||c|c|c|c|c|c|c|c||}
 \hline\hline
   \multicolumn{8}{||c||}{$bbq$} \\ \hline
 $\Lambda_\eta$ & $n$ & $l(\ell^{\eta_P})$ & $E_{bin}$ & $M^{(0)}$    & $\langle1/r\rangle$ & $E_{kin}$ & $P_{1/2}$ \\ \hline
 $(1/2)_g$      & $1$ & $0$       & $0.087$   & $10.140$ & $0.952$             & $0.204$   & $1$       \\ 
 $(1/2)_g$      & $2$ & $0$       & $0.489$   & $10.542$ & $0.617$             & $0.312$   & $1$       \\ 
 $(1/2)_g$      & $3$ & $0$       & $0.803$   & $10.856$ & $0.490$             & $0.407$   & $1$       \\ 
 $(1/2)_g$      & $4$ & $0$       & $1.078$   & $11.131$ & $0.417$             & $0.492$   & $1$       \\ 
 $(1/2)_g$      & $1$ & $1$       & $0.345$   & $10.398$ & $0.537$             & $0.257$   & $1$       \\ 
 $(1/2)_g$      & $2$ & $1$       & $0.678$   & $10.731$ & $0.423$             & $0.359$   & $1$       \\  
 $(1/2)_g$      & $3$ & $1$       & $0.963$   & $11.016$ & $0.360$             & $0.449$   & $1$       \\  
 $(1/2)_g$      & $1$ & $2$       & $0.547$   & $10.600$ & $0.398$             & $0.313$   & $1$       \\ 
 $(1/2)_g$      & $2$ & $2$       & $0.844$   & $10.897$ & $0.337$             & $0.407$   & $1$       \\ 
 $(1/2)_g$      & $3$ & $2$       & $1.109$   & $11.162$ & $0.297$             & $0.493$   & $1$       \\ 
 $(1/2)_g$      & $1$ & $3$       & $0.724$   & $10.777$ & $0.324$             & $0.366$   & $1$       \\
 $(1/2)_g$      & $2$ & $3$       & $0.998$   & $11.051$ & $0.285$             & $0.455$   & $1$       \\ \hline 
% $(1/2)_g$      & $3$ & $3$       & $1.247$   & $11.300$ & $0.257$             & $0.536$   & $1$       \\ 
% $(1/2)_g$      & $1$ & $4$       & $0.886$   & $10.939$ & $0.277$             & $0.417$   & $1$       \\ 
 $(1/2)^{\prime}_u$ & $1$ & $0$   & $0.079$   & $10.527$ & $0.942$             & $0.193$   & $1$       \\ 
 $(1/2)^{\prime}_u$ & $2$ & $0$   & $0.477$   & $10.924$ & $0.615$             & $0.298$   & $1$       \\ 
% $(1/2)^{\prime}_u$ & $3$ & $0$   & $0.790$   & $11.238$ & $0.489$             & $0.404$   & $1$       \\ 
 $(1/2)^{\prime}_u$ & $1$ & $1$   & $0.334$   & $10.781$ & $0.534$             & $0.242$   & $1$       \\ 
 $(1/2)^{\prime}_u$ & $2$ & $1$   & $0.664$   & $11.112$ & $0.422$             & $0.357$   & $1$       \\ 
 $(1/2)^{\prime}_u$ & $1$ & $2$   & $0.534$   & $10.981$ & $0.397$             & $0.311$   & $1$       \\ 
% $(1/2)^{\prime}_u$ & $2$ & $2$   & $0.830$   & $11.277$ & $0.336$             & $0.406$   & $1$       \\ 
 $(1/2)^{\prime}_u$ & $1$ & $3$   & $0.709$   & $11.157$ & $0.323$             & $0.365$   & $1$       \\ \hline
 $(3/2)_u-(1/2)_u$ & $1$ & $(3/2)^-$ & $0.036$  & $10.510$ & $0.890$           & $0.184$   & $0.574$   \\ 
 $(3/2)_u-(1/2)_u$ & $2$ & $(3/2)^-$ & $0.390$  & $10.865$ & $0.568$           & $0.310$   & $0.842$   \\ 
 $(3/2)_u-(1/2)_u$ & $3$ & $(3/2)^-$ & $0.692$  & $11.166$ & $0.495$           & $0.423$   & $0.906$   \\ 
 $(3/2)_u-(1/2)_u$ & $1$ & $(5/2)^+$ & $0.259$  & $10.734$ & $0.517$           & $0.248$   & $0.718$   \\ 
 $(3/2)_u-(1/2)_u$ & $2$ & $(5/2)^+$ & $0.576$  & $11.050$ & $0.414$           & $0.369$   & $0.833$   \\ 
 $(3/2)_u-(1/2)_u$ & $1$ & $(7/2)^-$ & $0.447$  & $10.921$ & $0.392$           & $0.315$   & $0.781$   \\ 
% $(3/2)_u-(1/2)_u$ & $2$ & $(7/2)^-$ & $0.743$  & $11.217$ & $0.337$           & $0.426$   & $0.844$   \\ 
 $(3/2)_u-(1/2)_u$ & $1$ & $(9/2)^+$ & $0.620$  & $11.094$ & $0.324$           & $0.378$   & $0.816$   \\ \hline 
 $(3/2)_u-(1/2)_u$ & $1$ & $(3/2)^+$ & $0.323$  & $10.797$ & $0.540$           & $0.266$   & $0.175$   \\ 
 $(3/2)_u-(1/2)_u$ & $2$ & $(3/2)^+$ & $0.586$  & $11.060$ & $0.350$           & $0.373$   & $0.876$   \\ 
 $(3/2)_u-(1/2)_u$ & $1$ & $(5/2)^-$ & $0.522$  & $10.996$ & $0.400$           & $0.330$   & $0.286$   \\ 
% $(3/2)_u-(1/2)_u$ & $2$ & $(5/2)^-$ & $0.759$  & $11.233$ & $0.303$           & $0.432$   & $0.794$   \\ \hline
 $(3/2)_u-(1/2)_u$ & $1$ & $(7/2)^+$ & $0.700$  & $11.174$ & $0.327$           & $0.390$   & $0.363$   \\ \hline
 $(1/2)_u$     & $1$ & $(1/2)^+$  & $0.219$   & $10.694$ & $0.510$             & $0.235$   & $1$       \\ 
 $(1/2)_u$     & $2$ & $(1/2)^+$  & $0.552$   & $11.026$ & $0.421$             & $0.360$   & $1$       \\ \hline
% $(1/2)_u$     & $3$ & $(1/2)^+$  & $0.848$   & $11.322$ & $0.366$             & $0.4660$   & $1$       \\ \hline
 $(1/2)_u$     & $1$ & $(1/2)^-$  & $0.406$   & $10.881$ & $0.391$             & $0.305$   & $1$       \\
 $(1/2)_u$     & $2$ & $(1/2)^-$  & $0.714$   & $11.189$ & $0.340$             & $0.418$   & $1$       \\ \hline\hline
\end{tabular}
 \caption{Double bottom baryon spectrum. All dimensionful entries are in GeV. $E_{bin}$ is the binding energy of the heavy quarks, $\langle1/r\rangle$ is the expected value of $1/r$ and $E_{kin}$ is the expected value of the kinetic energy of the heavy quarks. $P_{1/2}$ is the probability of finding the state in a $\Lambda=1/2$ state, and therefore a measure of the mixing between $\Lambda=3/2$ and $1/2$. The first bottomed-meson-bottomed-baryon threshold is $M(\Lambda^0_b)+M(B^0)=10.899$~GeV.}
\label{bbspc}
\end{table}

There are two possible different choices for the origin of energies of the potentials that would affect the number of multiplets below heavy-baryon-meson thresholds. First, instead of using $\overline{\Lambda}_{(1/2)^+}$ one could adjust the short-distance constant $c_1$ to reproduce the physical mass of $\Xi^{++}_{cc}$ from Ref.~\cite{Aaij:2017ueg}, which in practice is equivalent to shift down the masses of all the multiplets by $91$~MeV. Second, the energy gaps $E^{\rm latt}(1a)_{(1/2)_u}-E^{\rm latt}(1a)_{(1/2)_g}=421$~MeV and $E^{\rm latt}(1a)_{(1/2)^{\prime}_u}-E^{\rm latt}(1a)_{(1/2)_g}=394$~MeV correspond to the mass gaps of the light quark states $(3/2)^-$ and $(1/2)^-$ with respect to $(1/2)^+$ which corresponds the mass difference of the ground and first excited heavy-light mesons. From the PDG~\cite{Tanabashi:2018oca}, these mass gaps read
\begin{align}
&m_{B^0_1}-m_{B^0}=446{\rm MeV}\,,\quad m_{B^{*0}_2}-m_{B^{*}}=415{\rm MeV}\,, \\
&m_{D^0_1}-m_{D^0}=556{\rm MeV}\,,\quad m_{D^{*0}_2}-m_{D^{*}}=454{\rm MeV}\,. 
 \end{align}
The B meson mass gap, which should give the most accurate value, is compatible with the lattice data and any shifting of the states would be small. 

To close this section let us comment on the uncertainties of our results. The main source of uncertainty on the values of $M^{(0)}$ are the uncertainties on the value of the parameters in Eqs.~\eqref{mc}-\eqref{lbar12}. These account for an uncertainty of $53$~MeV and $67$~MeV for $M^{(0)}_{ccq}$ and $M^{(0)}_{bbq}$, respectively. However, this uncertainty drops in the mass differences. Additionally, there is the uncertainty on the lattice data for the static energies of Refs.~\cite{Najjar:2009da,Najjarthesis} and the model dependence on the long-distance parametrization of the static potentials. We estimate the effect of these two sources of uncertainty to be no larger than $10$~MeV. Finally, one should keep in mind that $M^{(0)}$ is only the LO contribution to the full double heavy baryon masses. The NLO corrections can be estimated to be of parametrical size $\lQ^2/m_Q$, which amounts to $\sim 100$~MeV and $\sim 30$~MeV for $M_{ccq}$ and $M_{bbq}$, respectively. In the next section we study in more detail the heavy quark spin and angular momentum NLO contributions to the double heavy baryon masses. However, one should keep in mind that NLO contributions independent of the spin and angular momentum do also exist.

\section{Spin splittings for \texorpdfstring{$\kappa=1/2$}{k=1/2} states}
\label{hf}

At LO the potential is independent of the spin of the heavy quarks, hence the final $j^{\eta_P}$ states appear in degenerate multiplets. This degeneracy is broken by the spin- and angular-momentum-dependent operators in the Lagrangian in Eqs.~\eqref{sdp12} and \eqref{sdp32}. At the moment the potentials corresponding to these spin-dependent operators are unknown and therefore the full computation of this spin-splitting corrections is not possible. However, in the case of the $\kappa=1/2$ states, due to the relatively small number of operators in the Lagrangian in Eq.~\eqref{sdp12} it is possible to obtain relations between the masses of states belonging to multiplets of a given $l$ independent of the shape of the potentials. We summarize the states in the multiplets for $l=0,\,1,\,2$ and the corresponding angular expected values of the spin-dependent operators in Table~\ref{aexval}.

\begin{table}[ht!]
\begin{tabular}{||c|c|c|c|c|c|c||} \hline \hline
 $l$ & $s_{QQ}$ & $\ell$ & $j$ & $\langle\bm{S}_{1/2}\cdot\bm{S}_{QQ}\rangle$ & $\langle \bm{S}_{1/2}\cdot\left(\bm{{\cal T}}_2\cdot \bm{S}_{QQ}\right)\rangle $ & $\langle\bm{S}_{1/2}\cdot\bm{L}_{QQ}\rangle$ \\ \hline 
\multirow{2}{*}{$0$} & \multirow{2}{*}{$1$} & \multirow{2}{*}{$1/2$} & $1/2$ & $-1$                 & \multirow{2}{*}{$0$} & \multirow{2}{*}{$0$}     \\ \cline{4-5}
                     &                      &                        & $3/2$ & $1/2$                &                      &                          \\ \hline
\multirow{2}{*}{$1$} & \multirow{2}{*}{$0$} & $1/2$                  & $1/2$ & \multirow{2}{*}{$0$} & \multirow{2}{*}{$0$} & $-1$                     \\ \cline{3-4}\cline{7-7}
                     &                      & $3/2$                  & $3/2$ &                      &                      & $1/2$                    \\ \hline
\multirow{6}{*}{$2$} & \multirow{6}{*}{$1$} & \multirow{3}{*}{$3/2$} & $1/2$ & $1/2$                & $-1/3$               & $\multirow{3}{*}{$-3/2$}$ \\ \cline{4-6}
                     &                      &                        & $3/2$ & $1/5$                & $-2/15$              &                          \\ \cline{4-6}
                     &                      &                        & $5/2$ & $-3/10$              & $1/5$                &                          \\ \cline{3-7}
                     &                      & \multirow{3}{*}{$5/2$} & $3/2$ & $-7/10$               & $2/15$               & $\multirow{3}{*}{$1$}$   \\ \cline{4-6}
                     &                      &                        & $5/2$ & $-1/5$               & $4/105$              &                          \\ \cline{4-6}
                     &                      &                        & $7/2$ & $1/2$                & $-2/21$              &                          \\ \hline\hline
\end{tabular}
\caption{Angular matrix elements for $\kappa=1/2$ double heavy baryon states. The heavy quark spin state is fixed by the Pauli exclusion principle.}
\label{aexval}
\end{table}

Let us label the mass of the states as $M_{njl\ell}=M^{(0)}_{nl}+M^{(1)}_{njl\ell}+\dots$ with $M^{(0)}_{nl}$ the mass solution of the Schr\"odinger equation with the static potential and $M^{(1)}_{njl\ell}$ the $1/m_Q$ spin-dependent contributions. Notice that the Pauli exclusion principle constrains $s_{QQ}=0$ for odd $l$ and $s_{QQ}=1$ for even $l$; therefore, $s_{QQ}$ is not needed to label the states.

First, let us look at the states with $l=0$. In this case the expected values of both $\bm{S}_{1/2}\cdot\bm{L}_{QQ}$ and $\bm{S}_{1/2}\cdot\left(\bm{{\cal T}}_2\cdot \bm{S}_{QQ}\right)$ are zero. Therefore the spin splittings are produced only by $\bm{S}_{1/2}\cdot\bm{S}_{QQ}$ and take the form
\begin{align}
M^{(1)}_{nj0\frac{1}{2}}=\frac{1}{2}\left(j(j+1)-\frac{11}{4}\right)\frac{\langle V^{s1}_{(1/2)^{\pm}}\rangle_{n0}}{m_Q}\,,\label{l0sp}
\end{align}
where we use the bracket notation to denote the expected values of the potentials between the radial wave functions
\begin{align}
\langle V^{i}_{(1/2)^{\pm}}\rangle_{nl}=\int_0^{\infty}dr\,r^2\,\psi^{nl\,\dagger}(r)V^{i}_{(1/2)^{\pm}}(r)\psi^{nl}(r)\,,\quad i=s1,\,s2,\,l.
\end{align}

Although the potentials are unknown we can use Eq.~\eqref{l0sp} to write the following relation between the masses of the two $l=0$ states and their LO mass:
\begin{align}
2M_{n\frac{3}{2}0\frac{1}{2}}+M_{n\frac{1}{2}0\frac{1}{2}}=3M^{(0)}_{n0}\,.\label{l0hf}
\end{align}

In the case with $l=1$ the Pauli principle fixes the heavy quark spin in a singlet state, and both the expected values of $\bm{S}_{1/2}\cdot\bm{S}_{QQ}$ and $\bm{S}_{1/2}\cdot\left(\bm{{\cal T}}_2\cdot \bm{S}_{QQ}\right)$ vanish. Only the contribution of $\bm{S}_{1/2}\cdot\bm{L}_{QQ}$ remains which is found to be
\begin{align}
M^{(1)}_{nj1j}=\frac{1}{2}\left(j(j+1)-\frac{11}{4}\right)\frac{\langle V^{l}_{(1/2)^{\pm}}\rangle_{n1}}{m_Q}\,,
\end{align}
that is, the same pattern as the $l=0$ states. We can also express the splitting as a mass relation between the two states that form the $l=1$ multiplet
\begin{align}
2M_{n\frac{3}{2}1\frac{3}{2}}+M_{n\frac{1}{2}1\frac{1}{2}}=3M^{(0)}_{n1}\,.\label{l1hf}
\end{align}

Notice, that both Eqs.~\eqref{l0hf} and \eqref{l1hf} are equivalent to the statement that the spin average of the $l=0$ and $l=1$ multiplets is equal to our LO masses $M^{(0)}_{nl}$.

For the case $l=2$ all three operators in the Lagrangian in Eq.~\eqref{sdp12} contribute to the spin splittings. In this case the computations of the angular expected values are slightly more involved. Let us write the general structure of the spin-dependent contributions for each state. For $j=1/2$ and $7/2$ the contributions are
\begin{align}
&M^{(1)}_{n\frac{1}{2}2\frac{3}{2}}=\frac{1}{2}\frac{\langle V^{s1}_{(1/2)^{\pm}}\rangle_{n2}}{m_Q}-\frac{1}{3}\frac{\langle V^{s2}_{(1/2)^{\pm}}\rangle_{n2}}{m_Q}-\frac{3}{2}\frac{\langle V^{l}_{(1/2)^{\pm}}\rangle_{n2}}{m_Q}\,,\label{ml21}\\
&M^{(1)}_{n\frac{7}{2}2\frac{5}{2}}=\frac{1}{2}\frac{\langle V^{s1}_{(1/2)^{\pm}}\rangle_{n2}}{m_Q}-\frac{2}{21}\frac{\langle V^{s2}_{(1/2)^{\pm}}\rangle_{n2}}{m_Q}+\frac{\langle V^{l}_{(1/2)^{\pm}}\rangle_{n2}}{m_Q}\,.\label{ml26}
\end{align}
For $j=3/2,5/2$ we have the mixing matrices for $\ell=3/2$ and $\ell=5/2$ states
\begin{align}
&M^{(1)}_{n\frac{3}{2}2}=\frac{1}{m_Q}\left(\begin{array}{cc} \frac{1}{5}\langle V^{s1}_{(1/2)^{\pm}}\rangle_{n2}-\frac{2}{15}\langle V^{s2}_{(1/2)^{\pm}}\rangle_{n2}-\frac{3}{2}\langle V^{l}_{(1/2)^{\pm}}\rangle_{n2} & \frac{3}{5}\langle V^{s1}_{(1/2)^{\pm}}\rangle_{n2}+\frac{1}{10}\langle V^{s2}_{(1/2)^{\pm}}\rangle_{n2} \\ \frac{3}{5}\langle V^{s1}_{(1/2)^{\pm}}\rangle_{n2}+\frac{1}{10}\langle V^{s2}_{(1/2)^{\pm}}\rangle_{n2} & -\frac{7}{10}\langle V^{s1}_{(1/2)^{\pm}}\rangle_{n2}+\frac{2}{15}\langle V^{s2}_{(1/2)^{\pm}}\rangle_{n2}+\langle V^{l}_{(1/2)^{\pm}}\rangle_{n2}\end{array}\right)\,,\\
&M^{(1)}_{n\frac{5}{2}2}=\frac{1}{m_Q}\left(\begin{array}{cc} -\frac{3}{10}\langle V^{s1}_{(1/2)^{\pm}}\rangle_{n2}+\frac{1}{5}\langle V^{s2}_{(1/2)^{\pm}}\rangle_{n2}-\frac{3}{2}\langle V^{l}_{(1/2)^{\pm}}\rangle_{n2} & \frac{\sqrt{14}}{5}\langle V^{s1}_{(1/2)^{\pm}}\rangle_{n2}+\frac{1}{15}\sqrt{\frac{7}{2}}\langle V^{s2}_{(1/2)^{\pm}}\rangle_{n2} \\ \frac{\sqrt{14}}{5}\langle V^{s1}_{(1/2)^{\pm}}\rangle_{n2}+\frac{1}{15}\sqrt{\frac{7}{2}}\langle V^{s2}_{(1/2)^{\pm}}\rangle_{n2} & -\frac{1}{5}\langle V^{s1}_{(1/2)^{\pm}}\rangle_{n2}+\frac{4}{105}\langle V^{s2}_{(1/2)^{\pm}}\rangle_{n2}+\langle V^{l}_{(1/2)^{\pm}}\rangle_{n2}\end{array}\right)\,.
\end{align}
We diagonalize to obtain the physical states
\begin{align}
M^{(1)}_{n\frac{3}{2}2\pm}=&-\frac{1}{4m_Q}\left\{\langle V^{s1}_{(1/2)^{\pm}}\rangle_{n2}+\langle V^{l}_{(1/2)^{\pm}}\rangle_{n2}\pm\frac{1}{3}\left[81\left(\langle V^{s1}_{(1/2)^{\pm}}\rangle_{n2}\right)^2+4\left(\langle V^{s2}_{(1/2)^{\pm}}\rangle_{n2}\right)^2+225\left(\langle V^{l}_{(1/2)^{\pm}}\rangle_{n2}\right)^2 \right.\right.\nn\\
&\left.\left.-6\langle V^{l}_{(1/2)^{\pm}}\rangle_{n2}\left(27\langle V^{s1}_{(1/2)^{\pm}}\rangle_{n2}-8\langle V^{s2}_{(1/2)^{\pm}}\rangle_{n2}\right)\right]^{1/2}\right\}\,,\\
M^{(1)}_{n\frac{5}{2}2\pm}=&-\frac{1}{84m_Q}\left\{21\langle V^{s1}_{(1/2)^{\pm}}\rangle_{n2}-10\langle V^{s2}_{(1/2)^{\pm}}\rangle_{n2}+21\langle V^{l}_{(1/2)^{\pm}}\rangle_{n2}\pm\left[3969\left(\langle V^{s1}_{(1/2)^{\pm}}\rangle_{n2}\right)^2+156\left(\langle V^{s2}_{(1/2)^{\pm}}\rangle_{n2}\right)^2 \right.\right.\nn\\
&\left.\left.+11025\left(\langle V^{l}_{(1/2)^{\pm}}\rangle_{n2}\right)^2+126\langle V^{s1}_{(1/2)^{\pm}}\rangle_{n2}\left(10\langle V^{s2}_{(1/2)^{\pm}}\rangle_{n2}+7\langle V^{l}_{(1/2)^{\pm}}\rangle_{n2}\right)\right.\right.\nn\\
&\left.\left.-1428\langle V^{s2}_{(1/2)^{\pm}}\rangle_{n2}\langle V^{l}_{(1/2)^{\pm}}\rangle_{n2}\right]^{1/2}\right\}\,.
\end{align}
Let us consider the following hyperfine splittings among $l=2$ which are linear in the expectation values of the potentials
\begin{align}
M_{n\frac{5}{2}2+}+M_{n\frac{5}{2}2-}-M_{n\frac{3}{2}2+}-M_{n\frac{3}{2}2-}=&\frac{5}{21m_Q}\langle V^{s2}_{(1/2)^{\pm}}\rangle_{n2}\,,\\
M_{n\frac{1}{2}2\frac{3}{2}}-\frac{1}{2}\left(M_{n\frac{3}{2}2+}+M_{n\frac{3}{2}2-}\right)=&\frac{1}{12m_Q}\left(9\langle V^{s1}_{(1/2)^{\pm}}\rangle_{n2}-4\langle V^{s2}_{(1/2)^{\pm}}\rangle_{n2}-15\langle V^{l}_{(1/2)^{\pm}}\rangle_{n2}\right)\,,\\
M_{n\frac{7}{2}2\frac{5}{2}}-\frac{1}{2}\left(M_{n\frac{3}{2}2+}+M_{n\frac{3}{2}2-}\right)=&\frac{1}{m_Q}\left(\frac{3}{4}\langle V^{s1}_{(1/2)^{\pm}}\rangle_{n2}-\frac{2}{21}\langle V^{s2}_{(1/2)^{\pm}}\rangle_{n2}+\frac{5}{4}\langle V^{l}_{(1/2)^{\pm}}\rangle_{n2}\right)\,.
\end{align}
These formulas fix $\langle V^{s1}_{(1/2)^{\pm}}\rangle_{n2}$, $\langle V^{s2}_{(1/2)^{\pm}}\rangle_{n2}$ and $\langle V^{l}_{(1/2)^{\pm}}\rangle_{n2}$ in terms of physical masses. Then, we have the following model-independent predictions
\begin{align}
M^{(1)}_{n\frac{3}{2}2+}-M^{(1)}_{n\frac{3}{2}2-}=&-\frac{1}{6m_Q}\left[81\left(\langle V^{s1}_{(1/2)^{\pm}}\rangle_{n2}\right)^2+4\left(\langle V^{s2}_{(1/2)^{\pm}}\rangle_{n2}\right)^2+225\left(\langle V^{l}_{(1/2)^{\pm}}\rangle_{n2}\right)^2-6\langle V^{l}_{(1/2)^{\pm}}\rangle_{n2}\left(27\langle V^{s1}_{(1/2)^{\pm}}\rangle_{n2}\right.\right.\nn\\
&\left.\left.-8\langle V^{s2}_{(1/2)^{\pm}}\rangle_{n2}\right)\right]^{1/2}\,,\label{l2hf1}\\ 
M^{(1)}_{n\frac{5}{2}2+}-M^{(1)}_{n\frac{5}{2}2-}=&-\frac{1}{42m_Q}\left[3969\left(\langle V^{s1}_{(1/2)^{\pm}}\rangle_{n2}\right)^2+156\left(\langle V^{s2}_{(1/2)^{\pm}}\rangle_{n2}\right)^2+11025\left(\langle V^{l}_{(1/2)^{\pm}}\rangle_{n2}\right)^2+126\langle V^{s1}_{(1/2)^{\pm}}\rangle_{n2}\left(\right.\right.\nn \\
&\left.\left.10\langle V^{s2}_{(1/2)^{\pm}}\rangle_{n2}+7\langle V^{l}_{(1/2)^{\pm}}\rangle_{n2}\right)-1428\langle V^{s2}_{(1/2)^{\pm}}\rangle_{n2}\langle V^{l}_{(1/2)^{\pm}}\rangle_{n2}\right]^{1/2}\,.\label{l2hf2}
\end{align}

\section{Comparison with lattice QCD}\label{latcomp}

Let us first discuss double charm baryons. For the ground state spin multiplet $(1/2,\,3/2)^+$ we get $M_{\Xi_{cc}}=3712(63)$~MeV. If we assign to the $(1/2)^+$ state the mass of the $\Xi_{cc}^{++}$ found by LHCb~\cite{Aaij:2017ueg}, $M_{\Xi_{cc}}=3621.2\pm 0.7$~MeV, and use it in Eq.~\eqref{l0hf} we obtain a prediction for the mass of the $(3/2)^+$ state, $M_{\Xi_{cc}^\ast}=3757(68)$~MeV. This implies a hyperfine splitting $\delta_{hf}=136(44)$~MeV. We estimate the uncertainty as corrections of ${\cal O}(\lQ^3/m^2_c)$ added quadratically to the uncertainties in the masses, for which we take into account that the uncertainties in the $\overline{\Lambda}_{(1/2)^+}$ and $m_c$ vanish for mass differences. This is about $30$~MeV larger than what is obtained by assuming heavy quark-diquark symmetry, namely a very compact diquark~\cite{Brambilla:2005yk,Savage:1990di,Fleming:2005pd,Mehen:2019cxn}, using the most up to date PDG values. In Table~\ref{splitlat} we compare this mass splitting to the ones obtained from lattice QCD calculations with fully relativistic charm quarks\footnote{We thank Stefan Meinel for pointing out a misquotation of the hyperfine splitting error of Ref.~\cite{Brown:2014ena} in an earlier version of this paper.}. We observe that our value is about $50$~MeV higher than that of Refs.~\cite{Namekawa:2013vu,Brown:2014ena,Alexandrou:2014sha,Bali:2015lka,Padmanath:2015jea} except for Refs.~\cite{Briceno:2012wt,Alexandrou:2017xwd} for which the discrepancy is higher. In most cases, including the uncertainties, the results are compatible. According to Eq.~\eqref{l0hf} the spin average of the ground state multiplet should coincide with our LO mass. We can check this by comparing the spin averages of the ground state multiplets from the lattice, summarized in Table~\ref{splitlat}, with our result $M^{(0)}_{10}=3712(63)$~MeV. We find that our results are compatible with Refs.~\cite{Briceno:2012wt,Namekawa:2013vu,Brown:2014ena,Bali:2015lka,Padmanath:2015jea,Alexandrou:2017xwd} and very close to compatible with Ref.~\cite{Alexandrou:2014sha}.

\begin{table}[ht!]
\begin{tabular}{||ccc||}\hline\hline
Ref. & $\delta_{hf}~[{\rm MeV}]$ & spin avg.\\ \hline

\cite{Briceno:2012wt} & $53(94)$ & $3630(50)$ \\

\cite{Namekawa:2013vu} & $101(36)$ & $3672(20)$\\

\cite{Brown:2014ena} & $82.8 (9.2)$ & $3665(36)$ \\

\cite{Alexandrou:2014sha} & $84(58)$ & $3624(33)$ \\

\cite{Bali:2015lka} & $85(9)$ & $3666(13)$ \\

\cite{Padmanath:2015jea} & $94(12)$ & $3700(6)$ \\

\cite{Alexandrou:2017xwd} & $76(41)$ & $3657(25)$ \\ \hline
Our values                 & $136(44)$ & $3712(63)$ \\\hline\hline
\end{tabular}
\caption{Lattice results with fully relativistic charm quarks for the hyperfine splitting $\delta_{hf}=M_{\Xi^*_{cc}}-M_{\Xi_{cc}}$ and spin average compared to our results.}
\label{splitlat}
\end{table}

For the first excitation, we have three spin multiplets very close in mass: two $(1/2,\,3/2)^-$ doublets, the first one corresponding to the  P-wave excitation of the heavy quarks ($4062$~MeV) and the second one to the S-wave ground state of the $(1/2)'_u$ BO potential ($4095$~MeV ), and one $(1/2,\,3/2,\,5/2)^-$ triplet corresponding to the S-wave ground state of the $(3/2)_u\backslash(1/2)_u$ BO potentials ($4066$~MeV). This structure qualitatively agrees with the results of Ref.~\cite{Padmanath:2015jea} (see Fig.~11 in that reference). It also agrees quantitatively for the first doublet, but the second doublet and the triplet are about $100$~MeV and $150$~MeV lower, respectively, than in Ref.~\cite{Padmanath:2015jea}. In Refs.~\cite{Can:2019wts,Bali:2015lka}, only a doublet is reported. The first reference is compatible with our results whereas the ones of the second reference lie about $100$ MeV below ours.

The higher excitations are all above the $\Lambda_c-D$ threshold. The only lattice calculation reporting on them is Ref.~\cite{Padmanath:2015jea}. For the positive parity states, our lowest $l=2$ spin multiplet $(1/2,\,3/2,\,3/2,\,5/2,\,5/2,\,7/2)^+$ lies very close to the $(1/2)_u$ singlet $(1/2)^+$ and both are above the $n=2,\,l=0$ spin doublet $(1/2,\,3/2)^+$ (see Fig.~\ref{ccplot}) whereas in Ref.~\cite{Padmanath:2015jea} all these states lay in the same energy range. The spin average of the $l=2$ multiplet is about $40$~MeV below ours, the spin singlet $(1/2)^+$ lays $13$~MeV above ours and $n=2,\,l=0$ spin doublet has a spin average $47$~MeV above ours. In any case, we can check our model-independent formulas in Eqs.~\eqref{l2hf1} and \eqref{l2hf2} for the $l=2$ spin multiplet against this data. Both formulas are compatible with the lattice masses of Ref.~\cite{Padmanath:2015jea}. Recall however that the uncertainties in the masses of the states involved are large. For positive parity, we can qualitatively accommodate the rest of our states below $4450$~MeV to those of Ref.~\cite{Padmanath:2015jea}, though precise identifications are difficult due to the large errors in that reference. Beyond that point we have much less positive parity states than Ref.~\cite{Padmanath:2015jea}. This is easy to understand because the first BO potential that we left out starts contributing about this energy. The negative parity states below $4600$~MeV (two spin doublets and a spin triplet) agree reasonably well with the spin averages of those of Ref.~\cite{Padmanath:2015jea}, within $72$~MeV. Beyond that point we have more states of higher spin. This is so even if we still miss some states that would arise from the first BO potential left out, which are expected to contribute in this region.

Let us next discuss double bottom baryons, for which no experimental evidence exists yet. The fact that the bottom quark mass is more than three times larger than the charm mass poses further difficulties to direct lattice QCD calculations. A way out is to factor out the bottom quark mass from the problem and do the calculations in lattice NRQCD. So far only results from the ground state spin doublet $(1/2,3/2)^+$ are available. Following Eq.~\eqref{l1hf} we compare the lattice spin average (s.a.) for the ground state to our value $M^{(0)}_{10}=10.140(77)$~MeV. This is in agreement with the results of Ref.~\cite{Lewis:2008fu}, $M_{\rm s.a.}=10.166(40)$~MeV, and Ref.~\cite{Brown:2014ena}, $M_{\rm s.a.}=10.143(29)$~MeV, but about $\sim 40$ MeV higher than the spin average of those reported in Ref.~\cite{Mohanta:2019mxo}, $M_{\rm s.a.}=10.099(17)$~MeV, but still compatible with our result.

\section{Conclusions}
\label{concl}

We have put forward an EFT for double heavy baryons that goes beyond the compact diquark approximation. This EFT is built upon two expansions on the small ratios between the characteristic energy scales of double heavy baryon systems. These are: the heavy quark mass expansion, $m_Q\gg \lQ\,, 1/r$, and an adiabatic expansion between the heavy quark and light degree of freedom dynamics, $\lQ\,, 1/r \gg E_{bin}$. At LO the EFT reproduces the Born-Oppenheimer approximation with a set of static energies. In that sense, it bridges smoothly between QCD and potential model calculations. The spectrum of static energies is obtained from lattice QCD data from Refs.~\cite{Najjar:2009da,Najjarthesis} and general constraints of the shape of these in the short- and long-distance regimes. We take into account the four lowest-lying static energies below the first heavy baryon-meson threshold, which correspond to the representations $(1/2)_g$, $(1/2)_u$, $(3/2)_u$ and $(1/2)_u'$ of $D_{\infty h}$. These are plotted together with the lattice data in Fig.~\ref{pfit}. In the short distance limit, $r\to 0$, the compact diquark approximation becomes exact and the spectrum of static energies matches the heavy meson spectrum (see Table~\ref{repcor} for the short-distance quantum numbers).

In Sec.~\ref{lo}, we obtained the spectrum of $ccq$ and $bbq$ baryons by numerically solving the Schr\"odinger equations with the aforementioned static energies including the mixing between $(1/2)_u\backslash(3/2)_u$ produced by the leading nonadiabatic corrections. A clear advantage of our method in comparison to lattice QCD is that obtaining higher excitations is very easy, even though the systematic errors are less under control. Indeed, the condition $E_{bin}\ll 1/r$ is not fulfilled for some multiplets close to or beyond threshold. For $ccq$ ($bbq$) only three (ten) of the four (sixteen) multiplets below threshold fulfill it. In the bottom case, there is also a multiplet slightly above threshold that fulfills it. However, $E_{bin}\sim 500$ MeV for this multiplet, and for two more below threshold, and hence they fail to meet the $E_{bin}\ll \lQ$ condition. In any case, we provide here for the first time the full spectrum below threshold for double bottom baryons based on lattice QCD data. 

Our results do not support the heavy quark-diquark approximation, neither for double charm baryons nor for double bottom ones. For $ccq$, the first angular excitation of the diquark core (negative parity lower blue bands in Fig.~\ref{ccplot}) is almost degenerate with the two lower light quark excitations (green and red lines in Fig.~\ref{ccplot}). For bottomonium, the situation is even worse as the first angular excitation of the diquark core (negative parity lower blue bands in Fig.~\ref{bbplot}) is clearly below the two lower light quark excitations (green and red lines in Fig.~\ref{bbplot}) and the first principal quantum number excitation has about the same energy as the latter. This is due to the fact that both for $ccq$ and $bbq$ the ground state already lies on the non-Coulombic part of the potential, which is clearly reflected in the positive binding energies displayed in Tables~\ref{ccspc} and \ref{bbspc}. Nevertheless, it is still remarkable that the gap between the ground state and the first light quark excitation ($383$~MeV for $ccq$ and $387$~MeV for $bbq$) is reasonably close to the gap for the spin average of the heavy-light meson systems ($427$~MeV for $\bar c q$ and $406$~MeV for $\bar b q$).

In Sec.~\ref{hf}, we discussed the hyperfine splittings generated by the NLO heavy quark spin- and angular-momentum-dependent operators in the Lagrangian of Eq.~\eqref{sdp12}. Unlike quarkonium, where these corrections start at $1/m^2_Q$ suppression, in double heavy baryons these start at only $1/m_Q$ suppression. This is a feature shared with exotic heavy hadrons~\cite{Soto:2020xpm}, such as heavy hybrids~\cite{Soto:2017one,Brambilla:2018pyn,Brambilla:2019jfi}. These operators are accompanied with, so far, unknown potentials. Despite this, it is possible to obtain relations between the hyperfine splittings of different states that are independent of the specific shape of the potentials. These are presented in Eqs.~\eqref{l0hf},\, \eqref{l1hf}\,, \eqref{l2hf1}, and \eqref{l2hf2}.

We have compared our results for the LO masses with the available lattice calculations for the ground state spin multiplet making use of Eq.~\eqref{l0hf}. Both for double charm and bottom baryons we find results compatible with most of the lattice references when uncertainties are taken into account. In the case of double bottom baryons we find a remarkably close agreement. For double charm baryons, the large array of states computed on the lattice in Ref.~\cite{Padmanath:2015jea} allows us to compare the general features of the spectrum finding a good qualitative agreement. 

The light flavor dependence of the double heavy baryon spectrum enters through the lattice data from which the potentials are obtained. The data of Refs.~\cite{Najjar:2009da,Najjarthesis} that we use is for degenerate $u$ and $d$ quarks corresponding to a pion mass of $783$~MeV. Data at more realistic pion masses would be highly desirable. The EFT used in this paper may be taken in the chiral limit and therefore our results would be independent of the flavor of the light quark forming the double heavy baryon. The implementation of chiral symmetry breaking to this EFT is straightforward and can be done using the standard techniques from Heavy Baryon Chiral Perturbation Theory~\cite{Burdman:1992gh}. However, the couplings with the pseudo-Goldstone bosons would introduce a number of unknown potentials that limit the predictability of the EFT, unless data for different low enough values of the pion mass become available for the potentials. Interestingly, upon implementing chiral symmetry to this EFT and the expansion of the fields in the eigenstates of the Schr\"ordinger equations, our EFT will match the standard chiral hadronic theories. These EFTs have been used abundantly to discuss threshold effects and decays~\cite{Hu:2005gf,Mehen:2017nrh,Shi:2020qde}. 

The main obstacle in the development of the EFT presented in this paper is the dependence of the potentials in nonperturbative dynamics. In an accompanying paper~\cite{Soto:2020xpm} we have presented the matching of the potentials of the EFT of Sec.~\ref{nreft} in terms of insertions of NRQCD operators in the Wilson loop. These Wilson loops are suitable to be computed on the lattice with light quarks and gluons only, sidestepping the issue of having to deal with widely separated scales on the lattice. Therefore, combining lattice QCD and EFT the major obstacles of both approaches can be avoided.

\section*{Acknowledgements}

We thank M.~Padmanath and the rest of the authors of Ref.~\cite{Padmanath:2015jea} for providing their numerical data of the double charmed baryon spectrum. J.S. also thanks Johannes Najjar and Gunnar Bali for providing the data of Refs.~\cite{Najjar:2009da,Najjarthesis}, and for discussions on those references when this paper was not even a project. He acknowledges financial support from  the 2017-SGR-929 grant from the Generalitat de Catalunya and the FPA2016-76005-C2-1-P and FPA2016-81114-P projects from Ministerio de Ciencia, Innovaci\'on y Universidades.  J.T.C acknowledges the financial support from the European Union's Horizon 2020 research and innovation program under the Marie Sk\l{}odowska--Curie Grant Agreement No. 665919. He has also been supported in part by the Spanish Grants No. FPA2017-86989-P and SEV-2016-0588 from the Ministerio de Ciencia, Innovaci\'on y Universidades, and the Grants No. 2017-SGR-1069 from the Generalitat de Catalunya. This research was supported by the Munich Institute for Astro- and Particle Physics (MIAPP) which is funded by the Deutsche Forschungsgemeinschaft (DFG, German Research Foundation) under Germany's Excellence Strategy – EXC-2094 – 390783311.

\appendix

\section{Coupled Schrödinger equations for spin-\texorpdfstring{$3/2$}{3/2}}\label{csed}

Let us introduce the projection vectors $P_{\frac{3}{2}\la}$ as the eigenvectors of
\begin{align}
\left(\hat{\bm{r}}\cdot\bm{S}_{3/2}\right)\bm{P}_{\frac{3}{2}\la}=\la\, \bm{P}_{\frac{3}{2}\la}\,,\quad \la=\frac{3}{2},\frac{1}{2},-\frac{1}{2},-\frac{3}{2}\label{eigenvp}
\end{align}
The projectors used in the Lagrangians in Eqs.~\eqref{edihsp} and \eqref{sdp32} can be constructed from the projection vectors defined in Eq.~\eqref{eigenvp}.
\begin{align}
{\cal P}_{\frac{3}{2}\Lambda}=\sum_{\lambda=\pm\Lambda}\bm{P}_{\frac{3}{2}\la}\bm{P}^{\dagger}_{\frac{3}{2}\la}\,,\quad \Lambda=\frac{1}{2},\,\frac{3}{2}\,.
\end{align}
We can diagonalize the potential matrix for the spin-$3/2$ field in Eq.~\eqref{boeftlo} by projecting the field into a basis of eigenstates of $\hat{\bm{r}}\cdot\bm{S}_{3/2}$
\begin{align}
\Psi_{(3/2)^-\la}&=\bm{P}^{\dagger}_{\frac{3}{2}\la}\cdot\Psi_{(3/2)^-}\,,\\
\Psi_{(3/2)^-}&=\sum_{\la}\bm{P}_{\frac{3}{2}\la}\Psi_{(3/2)^-\la}\,.
\end{align}
Note that the projection vectors act onto the light quark spin inside of the $\Psi_{(3/2)^-}$ field. In this basis the LO Lagrangian for the $\Psi_{(3/2)^-}$ field can be written as
\begin{align}
{\cal L}^{\rm LO}_{(3/2)^-}=&\sum_{\la\lap}\Psi^{\dag}_{(3/2)^-\la}\left(i\partial_t-\frac{1}{m_Q r^2}\,\partial_rr^2\partial_r-\frac{1}{m_Q}\bm{P}^{\alpha\,\dagger}_{\frac{3}{2}\la}\bm{L}^2_{QQ}\bm{P}^{\alpha}_{\frac{3}{2}\lap}+V_{(3/2)^-\lambda}^{(0)}(\bm{r})\delta_{\la\lap}\right)\Psi^{\dag}_{(3/2)^-\lap}\label{boeftlo32}
\end{align}

Now we want to obtain the matrix $\bm{P}^{\alpha\,\dagger}_{\frac{3}{2}\la}\bm{L}^2_{QQ}\bm{P}^{\alpha}_{\frac{3}{2}\lap}$ using the techniques of Refs.~\cite{Berwein:2015vca,Brambilla:2019jfi}. Let us start with the following auxiliary formulas. The angular momentum operator in spherical coordinates is
\begin{align}
\bm{L}_{QQ}=-i\hat{\bm{\phi}}\pa_{\theta}+\frac{i}{\sin\theta}\hat{\bm{\theta}}\pa_{\phi}\,.
\end{align}

The unit vectors in spherical coordinates read
\begin{align}
\hat{\bm{r}}&=(\sin(\theta)\cos(\phi),\,\sin(\theta)\sin(\phi)\,,\cos(\theta)) \,, \nn \\
\hat{\bm\theta}&=(\cos(\theta)\cos(\phi),\,\cos(\theta)\sin(\phi)\,,-\sin(\theta)) \,, \nn \\
\hat{\bm\phi}&=(-\sin(\phi),\,\cos(\phi)\,,0)\,.
\end{align}
from which we define
\begin{align}
\hat{\bm{r}}_0^i&= \hat{\bm{r}}^i\,,\label{pr10}\\
\hat{\bm{r}}^i_{\pm}&=\mp\left(\hat{\bm{\theta}}^i\pm i\hat{\bm{\phi}}^i\right)/\sqrt{2}\,,\label{pr11}
\end{align}

Let us compute the commutators of the angular momentum operator and the unit vectors in spherical coordinates
\begin{align}
\left[\bm{L}_{QQ}^i,\hat{\bm{r}}_{0}^j\right]&=\hat{\bm{r}}^i_+\hat{\bm{r}}^j_--\hat{\bm{r}}^i_-\hat{\bm{r}}^j_+\,, \\
\left[\bm{L}_{QQ}^i,\hat{\bm{\theta}}^j\right]&=i \hat{\bm{\phi}}^i\hat{\bm{r}}_0^j+i\cot(\theta)\hat{\bm{\theta}}^i\hat{\bm{\phi}}^j\,, \\
\left[\bm{L}_{QQ}^i,\hat{\bm{\phi}}^j\right]&=-i\hat{\bm{\theta}}^i\left(\hat{\bm{r}}_0^j+\cot(\theta)\hat{\bm{\theta}}^j\right)\,,
\end{align}
from which one can obtain 
\begin{align}
\left[\bm{L}_{QQ}^i,\hat{\bm{r}}_{\pm}^j\right]=\pm\left(\hat{\bm{r}}^j_0\hat{\bm{r}}^i_{\pm}+\cot(\theta)\hat{\bm{\theta}}^i\hat{\bm{r}}^j_{\pm}\right)\,.
\end{align}
Therefore, 
\begin{align}
&\left[\bm{L}_{QQ}^i,\hat{\bm{r}}_{\pm}^i\right]=-\frac{\cot(\theta)}{\sqrt{2}}\,,\label{low}\\
&\left[\bm{L}_{QQ}^i,\hat{\bm{r}}^{\dagger\,i}_{\pm}\right]=\frac{\cot(\theta)}{\sqrt{2}}\label{up}\,.
\end{align}

From an explicit computation we obtain
\begin{align}
\bm{P}^{\alpha\,\dagger}_{\frac{3}{2}\la}\bm{L}_{QQ}\bm{P}^\alpha_{\frac{3}{2}\lap}=\left(\begin{array}{cccc}
\frac{3}{2}\cot\theta\hat{\bm{\theta}} & -\sqrt{\frac{3}{2}}\hat{\bm{r}}^{\dagger}_- & 0 & 0 \\
-\sqrt{\frac{3}{2}}\hat{\bm{r}}_- & \frac{1}{2}\cot\theta\hat{\bm{\theta}} & -\sqrt{2}\hat{\bm{r}}^{\dagger}_- & 0 \\
0 & -\sqrt{2}\hat{\bm{r}}_- & -\frac{1}{2}\cot\theta\hat{\bm{\theta}} & -\sqrt{\frac{3}{2}}\hat{\bm{r}}^{\dagger}_- \\
0 & 0 & -\sqrt{\frac{3}{2}}\hat{\bm{r}}_- & -\frac{3}{2}\cot\theta\hat{\bm{\theta}} \\
\end{array}\right)\,.\label{vLQQv}
\end{align}
and
\begin{align}
\bm{P}^{\alpha\,\dagger}_{\frac{3}{2}\la}\bm{S}_{3/2}\bm{P}^\alpha_{\frac{3}{2}\lap}=\left(\begin{array}{cccc}
\frac{3}{2}\hat{\bm{r}}_0 & \sqrt{\frac{3}{2}}\hat{\bm{r}}^{\dagger}_- & 0 & 0 \\
\sqrt{\frac{3}{2}}\hat{\bm{r}}_- & \frac{1}{2}\hat{\bm{r}}_0 & \sqrt{2}\hat{\bm{r}}^{\dagger}_- & 0 \\
0 & \sqrt{2}\hat{\bm{r}}_- & -\frac{1}{2}\hat{\bm{r}}_0 & \sqrt{\frac{3}{2}}\hat{\bm{r}}^{\dagger}_- \\
0 & 0 & \sqrt{\frac{3}{2}}\hat{\bm{r}}_- & -\frac{3}{2}\hat{\bm{r}}_0 \\
\end{array}\right)\,,\label{vkv}
\end{align}
and we note that
\begin{align}
\bm{P}^{\alpha\,\dagger}_{\frac{3}{2}\la}\bm{S}^2_{3/2}\bm{P}^\alpha_{\frac{3}{2}\lap}=\sum_{\lapp}\bm{P}^{\alpha\,\dagger}_{\frac{3}{2}\la}\bm{S}^2_{3/2}\bm{P}^\alpha_{\frac{3}{2}\lapp}\bm{P}^{\beta\,\dagger}_{\frac{3}{2}\lapp}\bm{S}^2_{3/2}\bm{P}^\beta_{\frac{3}{2}\lap}=\frac{15}{4}\delta_{\la\lap}=\frac{3}{2}\left(\frac{3}{2}+1\right)\delta_{\la\lap}\,.
\end{align}
If we define $\bm{L}=\bm{L}_{QQ}+\bm{S}_{3/2}$, from Eqs.~\eqref{vkv} and \eqref{vLQQv} we can write
\begin{align}
\bm{P}^{\alpha\,\dagger}_{\frac{3}{2}\la}\bm{L}\bm{P}^\alpha_{\frac{3}{2}\lap}=\left[\bm{L}_{QQ}+\la\left(\cot\theta\hat{\bm{\theta}}+\hat{\bm{r}}_0\right)\right]\delta_{\la\lap}\,.\label{melqqpk}
\end{align}
We are in position now to compute the matrix elements of the square of the heavy-quark angular momentum in between the projector vectors 
\begin{align}
&\bm{P}^{\alpha\,\dagger}_{\frac{3}{2}\la}\bm{L}^2_{QQ}\bm{P}^\alpha_{\frac{3}{2}\lap}=\bm{L}^2\delta_{\la\lap}+\frac{3}{2}\left(\frac{3}{2}+1\right)\delta_{\la\lap}-\bm{L}\bm{P}^{\alpha\,\dagger}_{\frac{3}{2}\la}\bm{S}_{3/2}\bm{P}^\alpha_{\frac{3}{2}\lap}-\bm{P}^{\alpha\,\dagger}_{\frac{3}{2}\la}\bm{S}_{3/2}\bm{P}^\alpha_{\frac{3}{2}\lap}\bm{L}\,.\label{amodec}
\end{align}
The square of the right-hand side of Eq.~\eqref{melqqpk} is
\begin{align}
&\bm{L}^2=\left[\bm{L}_{QQ}+\la\left(\cot\theta \hat{\bm{\theta}}+\hat{\bm{r}}_0\right)\right]^2=\bm{L}^2_{QQ}+\frac{\la^2}{\sin^2\theta}+2i\la\frac{\cos\theta}{\sin^2\theta}\partial_{\theta}\,,
\end{align}
which is the operator whose eigenfunctions are our angular wave functions,
\begin{align}
\left(\bm{L}^2_{QQ}+\frac{\la^2}{\sin^2\theta}+2i\la\frac{\cos\theta}{\sin^2\theta}\partial_{\theta}\right)v^{\la}_{\ell m_\ell}(\theta,\phi)=\ell(\ell+1)v^{\la}_{\ell m_\ell}(\theta,\phi)\,,
\end{align}
with
\begin{align}
v^{\lambda}_{\ell m_\ell}(\theta,\phi)&=\frac{(-1)^{m_\ell+\la}}{2^\ell}\sqrt{\frac{2\ell+1}{4\pi}\frac{(\ell-m_\ell)!}{(\ell+m_\ell)!(\ell-\la)!(\ell+\la)!}}\,P^\la_{\ell m_\ell}(\cos\theta)e^{im_\ell\phi}\,,\\
P^\la_{\ell m_\ell}(x)&=(1-x)^{\frac{m_\ell-\la}{2}}(1+x)^{\frac{m_\ell+\la}{2}}
\partial_x^{\ell+m_\ell}(x-1)^{\ell+\la}(x+1)^{\ell-\la}\,,
\end{align}
with $|m|<\ell$ and $|\lambda|<\ell$. The operators $\mathcal{K}_{\pm}$ act as the $\la$-raising and -lowering operators for the angular wave functions $v^{\la}_{\ell m_\ell}$,
\begin{align}
&{\cal K}_{\pm}=\left(\mp\pa_{\theta}+\frac{i}{\sin\theta}\pa_{\phi}\mp\cot\theta\right)\,,\\
&{\cal K}_{\pm}v^{\la}_{\ell m_\ell}(\theta,\phi)=\sqrt{\ell(\ell+1)-\la(\la\pm 1)}v^{\la \pm 1}_{\ell m_\ell}(\theta,\phi)\,.
\end{align}

The mixing terms in Eq.~\eqref{amodec} can be written as a matrix in the $\la$-$\lap$ indices as
\begin{align}
&\bm{L}\bm{P}^{\alpha\,\dagger}_{\frac{3}{2}\la}\bm{S}_{3/2}\bm{P}^\alpha_{\frac{3}{2}\lap}+\bm{P}^{\alpha\,\dagger}_{\frac{3}{2}\la}\bm{S}_{3/2}\bm{P}^\alpha_{\frac{3}{2}\lap}\bm{L}=\left(\begin{array}{cccc}
2\left(\frac{3}{2}\right)^2 & \sqrt{3}{\cal K}_-  & 0 & 0 \\
\sqrt{3}{\cal K}_+ & 2\left(\frac{1}{2}\right)^2 & 2{\cal K}_- & 0 \\
0 & {\cal K}_+ & 2\left(-\frac{1}{2}\right)^2 & \sqrt{3}{\cal K}_- \\
0 & 0 & \sqrt{3}{\cal K}_+ & 2\left(-\frac{3}{2}\right)^2\end{array}\right)\nn\\
&=2\la^2\delta_{\la\lap}+{\cal K}_-\sqrt{\frac{3}{2}\left(\frac{3}{2}+1\right)-\la(\la+1)}\delta_{(\la+1)\lap}+{\cal K}_+\sqrt{\frac{3}{2}\left(\frac{3}{2}+1\right)-\la(\la-1)}\delta_{(\la-1)\lap}\,.\label{ammt}
\end{align}

Introducing Eq.~\eqref{ammt} into Eq.~\eqref{amodec} we arrive at
\begin{align}
&\bm{P}^{\alpha\,\dagger}_{\frac{3}{2}\la}\bm{L}^2_{QQ}\bm{P}^\alpha_{\frac{3}{2}\lap}\nn\\
&=\left[\bm{L}^2+\frac{3}{2}\left(\frac{3}{2}+1\right)-2\la^2\right]\delta_{\la\lap}-{\cal K}_-\sqrt{\frac{3}{2}\left(\frac{3}{2}+1\right)-\la(\la+1)}\delta_{(\la+1)\lap}-{\cal K}_+\sqrt{\frac{3}{2}\left(\frac{3}{2}+1\right)-\la(\la-1)}\delta_{(\la-1)\lap}\,.
\end{align}

Now, let us look at the expected value of the square angular momentum operator between the angular wave functions
\begin{align}
&\int d\Omega\, v^{\la *}_{\ell m_\ell}\bm{P}^{\alpha\,\dagger}_{\frac{3}{2}\la}\bm{L}^2_{QQ}\bm{P}^\alpha_{\frac{3}{2}\lap}v^{\lap}_{\ell m_\ell}\nn\\
&=
\left(\begin{array}{cccc}
\ell(\ell+1)-\frac{3}{4} & -\sqrt{3}\sqrt{\ell(\ell+1)-\frac{3}{4}} & 0 & 0\\
-\sqrt{3}\sqrt{\ell(\ell+1)-\frac{3}{4}} & \ell(\ell+1)+\frac{13}{4} & -2\sqrt{\ell(\ell+1)+\frac{1}{4}} & 0 \\
0 & -2\sqrt{\ell(\ell+1)+\frac{1}{4}}& \ell(\ell+1)+\frac{13}{4} & -\sqrt{3}\sqrt{\ell(\ell+1)-\frac{3}{4}}\\
0 & 0 & -\sqrt{3}\sqrt{\ell(\ell+1)-\frac{3}{4}} & \ell(\ell+1)-\frac{3}{4} \\
\end{array}\right)\,.\label{evamwf}
\end{align}

The system decouples into two two-state coupled equations in the basis given by the following transformation matrix:
\begin{align}
R&=\frac{1}{\sqrt{2}}\left(\begin{array}{cccc}
1 & 0 & 0 & 1 \\
0 & 1 & 1 & 0 \\
0 & 1 & -1 & 0 \\
1 & 0 & 0 & -1\\
\end{array}\right)\,,\\
R\eqref{evamwf}R^{-1}&=
\left(\begin{array}{cccc}
\ell(\ell+1)-\frac{3}{4} & -\sqrt{3}\sqrt{\ell(\ell+1)-\frac{3}{4}} & 0 & 0\\
-\sqrt{3}\sqrt{\ell(\ell+1)-\frac{3}{4}} & \ell(\ell-1)+\frac{9}{4} & 0 & 0 \\
0 & 0 & \ell(\ell+3)+\frac{17}{4} & -\sqrt{3}\sqrt{\ell(\ell+1)-\frac{3}{4}}\\
0 & 0 & -\sqrt{3}\sqrt{\ell(\ell+1)-\frac{3}{4}} & \ell(\ell+1)-\frac{3}{4} \\
\end{array}\right)\,.
\end{align}

Therefore, we arrive at two sets of coupled Schr\"odinger equations
\begin{align}
& \hspace{-4pt}\left[-\frac{1}{m_Q r^2}\,\partial_rr^2\partial_r+\frac{1}{m_Qr^2}\begin{pmatrix} \ell(\ell+1)-\frac{3}{4} & -\sqrt{3}\sqrt{\ell(\ell+1)-\frac{3}{4}} \\ -\sqrt{3}\sqrt{\ell(\ell+1)-\frac{3}{4}} & \ell(\ell-1)+\frac{9}{4} \end{pmatrix}+\begin{pmatrix} V^{(0)}_{(3/2)^-(3/2)}(r) & 0 \\ 0 & V^{(0)}_{(3/2)^-(1/2)}(r) \end{pmatrix}\right]\hspace{-2pt}\begin{pmatrix} \psi_{3/2+}^{(n)}(r) \\ \psi_{1/2+}^{(n)}(r)\end{pmatrix}\nn\\
&=\mathcal{E}_n\begin{pmatrix} \psi_{3/2+}^{(n)}(r) \\ \psi_{1/2+}^{(n)}(r)\end{pmatrix}\label{csequ}\,,
\end{align}
\begin{align}
& \hspace{-4pt}\left[-\frac{1}{m_Q r^2}\,\partial_rr^2\partial_r+\frac{1}{m_Qr^2}\begin{pmatrix} \ell(\ell+1)-\frac{3}{4} & -\sqrt{3}\sqrt{\ell(\ell+1)-\frac{3}{4}} \\ -\sqrt{3}\sqrt{\ell(\ell+1)-\frac{3}{4}} & \ell(\ell+3)+\frac{17}{4} \end{pmatrix}+\begin{pmatrix} V^{(0)}_{(3/2)^-(3/2)}(r) & 0 \\ 0 & V^{(0)}_{(3/2)^-(1/2)}(r) \end{pmatrix}\right]\hspace{-2pt}\begin{pmatrix} \psi_{3/2-}^{(n)}(r) \\ \psi_{1/2-}^{(n)}(r)\end{pmatrix}\nn\\
&=\mathcal{E}_n\begin{pmatrix} \psi_{3/2-}^{(n)}(r) \\ \psi_{1/2-}^{(n)}(r)\end{pmatrix}\label{cseqd}\,,
\end{align}
where $\psi_{3/2+}^{(n)}(r)$, $\psi_{1/2+}^{(n)}(r)$, $\psi_{3/2-}^{(n)}(r)$ and $\psi_{1/2-}^{(n)}(r)$ are radial wave functions with $(n)$ the quantum number labeling the radial eigenstates. The $\mathcal{E}_n$ are the energy eigenvalues. Notice that for the lowest value of $\ell$ allowed, $\ell=1/2$ the upper entries of the coupled system vanish and only the component corresponding to $V^{(0)}_{(3/2)^-(1/2)}$ survives decoupled.

The full wave function solutions of the Schr\"odinger equation for the $\Psi_{(3/2)^-}$ field are then
\begin{align}
&\Psi^{n j m_j \ell s_{QQ}}_{+}(\bm{r})=\frac{1}{\sqrt{2}}\sum_{m_\ell m_{s_{QQ}}} \mathcal{C}^{j m_j}_{\ell\,m_\ell\,s_{QQ}\,m_{s_{QQ}}}\left(
\begin{array}{c}
\psi_{3/2+}^{(n)}(r)v_{\ell\,m_\ell}^{+3/2}(\theta,\phi) \\
\psi_{1/2+}^{(n)}(r)v^{+1/2}_{\ell\,m_\ell}(\theta,\phi) \\
\psi_{1/2+}^{(n)}(r)v^{-1/2}_{\ell\,m_\ell}(\theta,\phi) \\
\psi_{3/2+}^{(n)}(r)v_{\ell\,m_\ell}^{-3/2}(\theta,\phi) \\
\end{array}\right)\chi_{s_{QQ}\,m_{s_{QQ}}}\,,\label{psip}\\
&\Psi^{n j m_j \ell s_{QQ}}_{-}(\bm{r})=\frac{1}{\sqrt{2}}\sum_{m_\ell m_{s_{QQ}}} \mathcal{C}^{j m_j}_{\ell\,m_\ell\,s_{QQ}\,m_{s_{QQ}}}\left(
\begin{array}{c}
\psi_{3/2-}^{(n)}(r)v_{\ell\,m_\ell}^{+3/2}(\theta,\phi) \\
\psi_{1/2-}^{(n)}(r)v^{+1/2}_{\ell\,m_\ell}(\theta,\phi) \\
-\psi_{1/2-}^{(n)}(r)v^{-1/2}_{\ell\,m_\ell}(\theta,\phi) \\
-\psi_{3/2-}^{(n)}(r)v_{\ell\,m_\ell}^{-3/2}(\theta,\phi) \\
\end{array}\right)\chi_{s_{QQ}\,m_{s_{QQ}}}\,,\label{psim}
\end{align}
and the states are related to the wave functions in Eqs.~\eqref{psip} and \eqref{psim} as
\begin{align}
|n\,j\,m_j\,\ell\,s_{QQ}\,\pm\rangle=\sum_{\la}\left(\psi^{n j m_j \ell s_{QQ}}_\pm\right)_\la\Psi^{\dagger}_{(3/2)^-\la}|0\rangle\,.\label{states}
\end{align}

\subsection{Parity of the states}

Under a parity transformation we have
\begin{align}
&\Psi_{(3/2)^-}\left(\bm{r},\,\bm{R}\right)\stackrel{P}{\rightarrow} -\Psi_{(3/2)^-}\left(-\bm{r},\,-\bm{R}\right)\,,\\
&P^i_{\frac{3}{2}\la}\stackrel{P}{\rightarrow}(-1)^{3/2}P^i_{\frac{3}{2}-\la}\,.\label{ppf}
\end{align}
Notice, that the phase in Eq.~\eqref{ppf} is a complex number. Thus, the projected fields transform as 
\begin{align}
&\Psi_{\left(3/2\right)^-\la}\stackrel{P}{\rightarrow}(-1)^{1/2} \Psi_{\left(3/2\right)^--\la}\,,\\
&\Psi^{\dagger}_{\left(3/2\right)^-\la}\stackrel{P}{\rightarrow}(-1)^{-1/2} \Psi^{\dagger}_{\left(3/2\right)^--\la}\,.
\end{align}
Since the angular wave function transforms under parity as 
\begin{align}
v_{\ell\,m_\ell}^{\la}(\theta,\phi) \stackrel{P}{\rightarrow}(-1)^\ell v_{\ell\,m_\ell}^{-\la}(\pi-\theta,\phi+\pi)\,,
\end{align} 
the whole wave function components transform as
\begin{align}
\left(\psi^{n j m_j \ell s_{QQ}}_\pm\right)_\la \stackrel{P}{\rightarrow}\pm(-1)^\ell \left(\psi^{n j m_j \ell s_{QQ}}_\pm\right)_{-\la}\,.
\end{align} 
Therefore the states in Eq.~\eqref{states} transform under parity as 
\begin{align}
&|n\,j\,m_j\,\ell\,s_{QQ},\pm\rangle \stackrel{P}{\rightarrow} \eta_P|n\,j\,m_j\,\ell\,s_{QQ},\pm\rangle\,,\\
&\eta_P=\pm (-1)^{\ell-1/2}
\end{align} 

\subsection{Boundary conditions}

A system of two linearly coupled differential equations of second order has in general four linearly independent solutions of which two will be singular at the origin. For the case of only one second order differential equation ($\ell=1/2$) there exists two linearly independent solutions, one of which will be singular at the origin. These solutions can be distinguished by their behavior at the origin which will also be important as initial conditions for the numerical solution of the Sch\"odinger equations. To obtain the behavior at the origin of the solutions of the differential equations we note that in the short distance the kinetic term dominates over the potential. In this limit the system can be diagonalized and the two resulting decoupled equations solved by a guess function $r^b$. Two solutions will be obtained for each equation, one of which diverges at the origin and should be discarded. The remaining one for each equation gives us one of the solutions for the system with the weights given by the components of the eigenvectors that diagonalized the system. These are the behaviors at the origin of the good solutions for $\ell\geq 3/2$
\begin{align}
&\left(\begin{array}{c}
\psi_{3/2+} \\
\psi_{1/2+} \\
\end{array}\right)\propto
\left(\begin{array}{c}
-\sqrt{12\ell(\ell+1)-9}r^{\ell+1/2} \\
(2l+3)r^{\ell+1/2} \\
\end{array}\right)\,,\\
&\left(\begin{array}{c}
\psi_{3/2+} \\
\psi_{1/2+} \\
\end{array}\right)\propto
\left(\begin{array}{c}
(2l+3)r^{\ell-3/2} \\
\sqrt{12\ell(\ell+1)-9}r^{\ell-3/2} \\
\end{array}\right)\,,\\
&\left(\begin{array}{c}
\psi_{3/2-} \\
\psi_{1/2-} \\
\end{array}\right)\propto
\left(\begin{array}{c}
\sqrt{12\ell(\ell+1)-9}r^{\ell-1/2} \\
(2\ell-1)r^{\ell-1/2} \\
\end{array}\right)\,,\\
&\left(\begin{array}{c}
\psi_{3/2-} \\
\psi_{1/2-} \\
\end{array}\right)\propto
\left(\begin{array}{c}
-\sqrt{12\ell(\ell+1)-9}r^{\ell+3/2} \\
3(2\ell+3)r^{\ell+3/2} \\
\end{array}\right)\,.
\end{align}
For the case $\ell=1/2$ the equations decouple and only $\psi_{1/2}$ remains
\begin{align}
&\psi_{1/2+}\propto r\,,\\
&\psi_{1/2-}\propto r^2\,.
\end{align}

\section{Radial wave function plots}\label{rwfp}

\begin{figure}[ht!]
\begin{tabular}{cc}
\includegraphics[width=.45\textwidth]{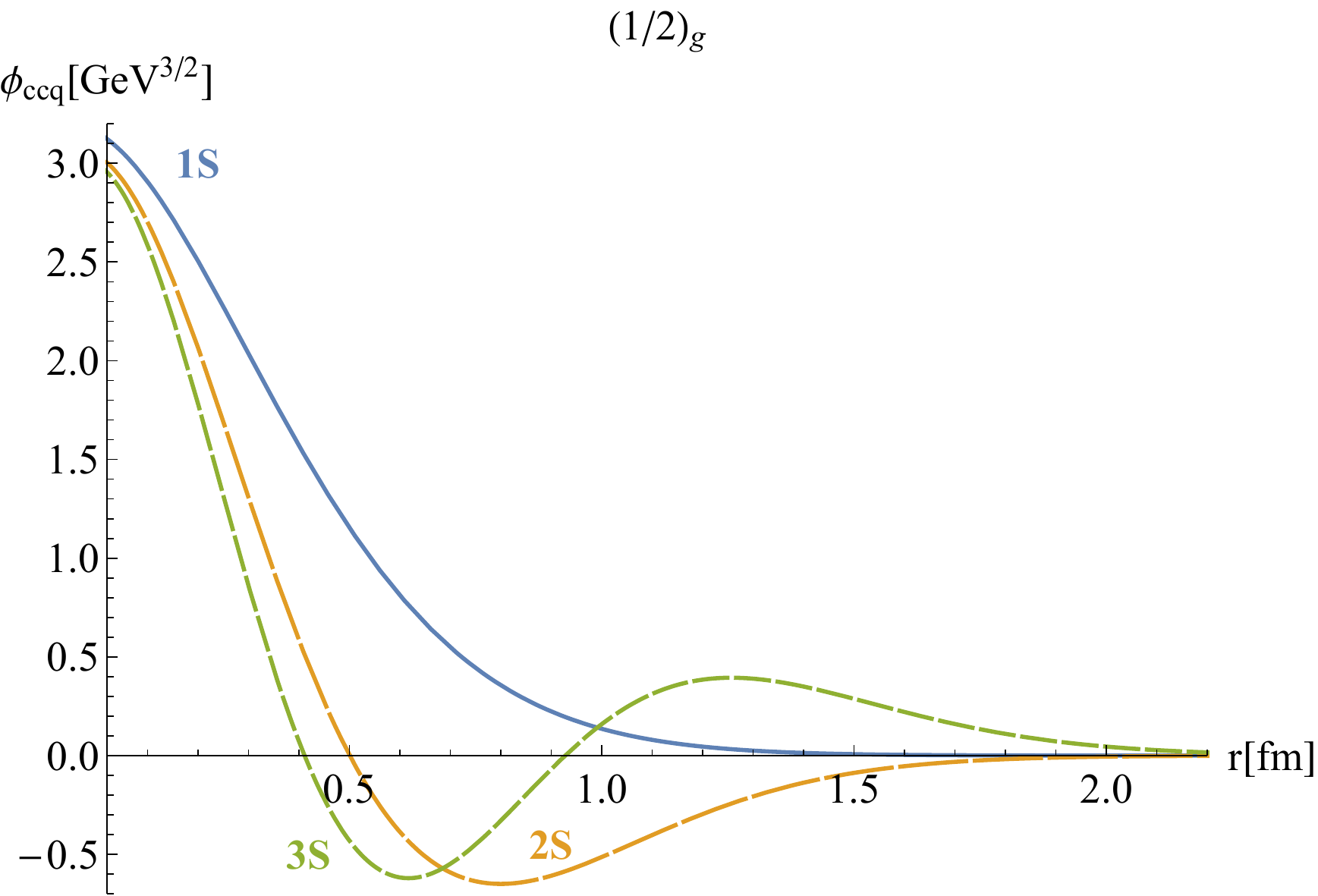} & \includegraphics[width=.45\textwidth]{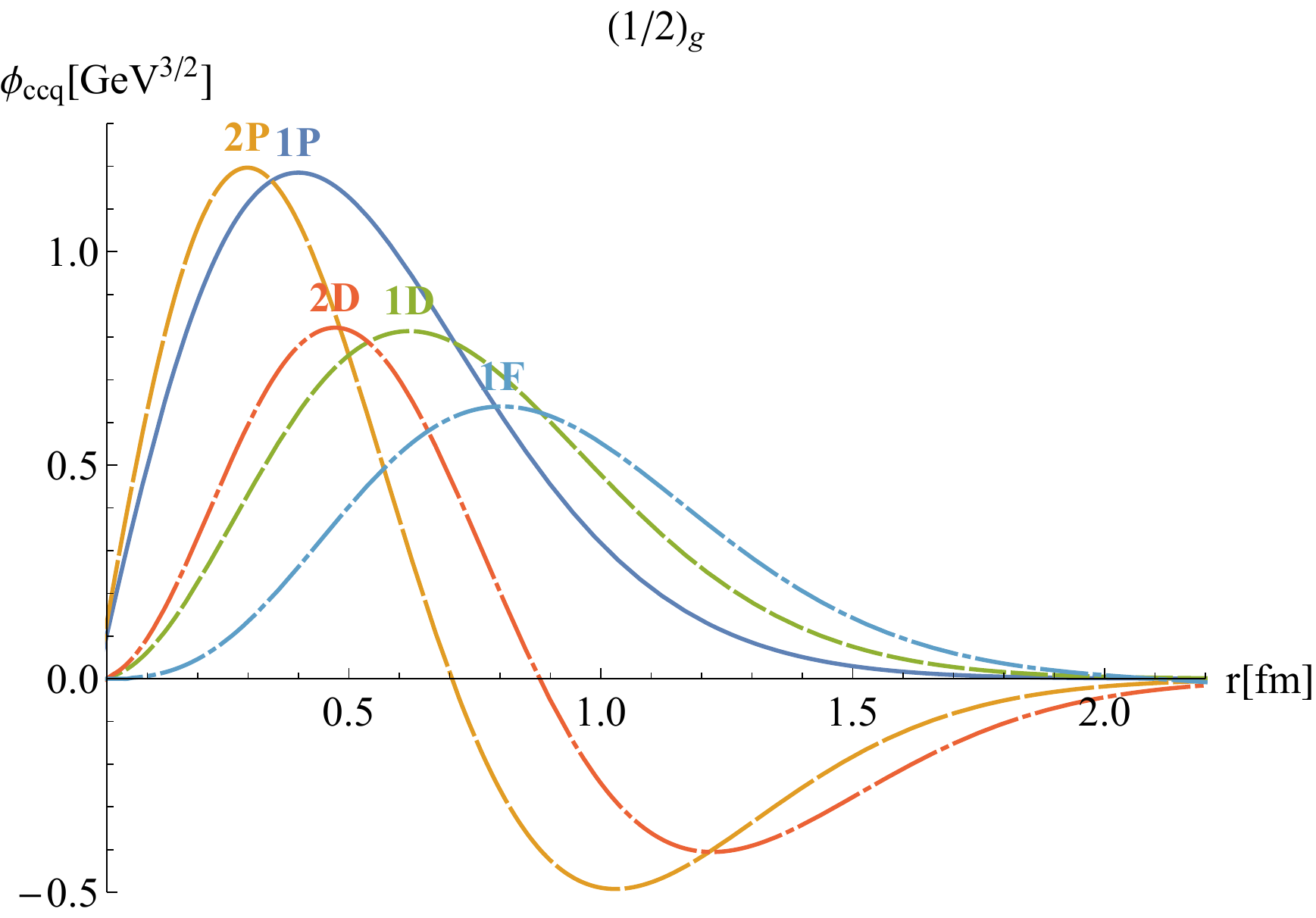} \\
\includegraphics[width=.45\textwidth]{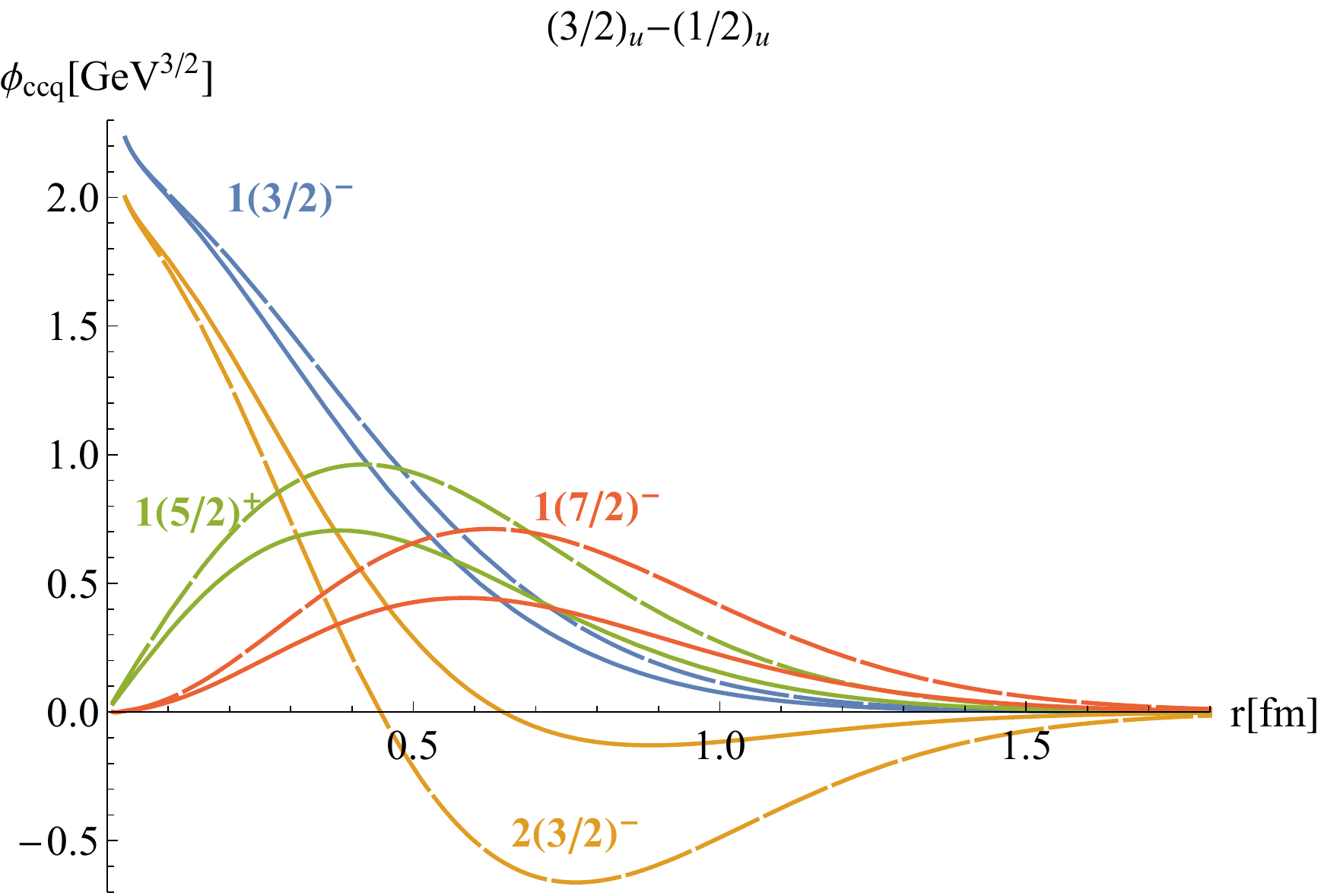} & \includegraphics[width=.45\textwidth]{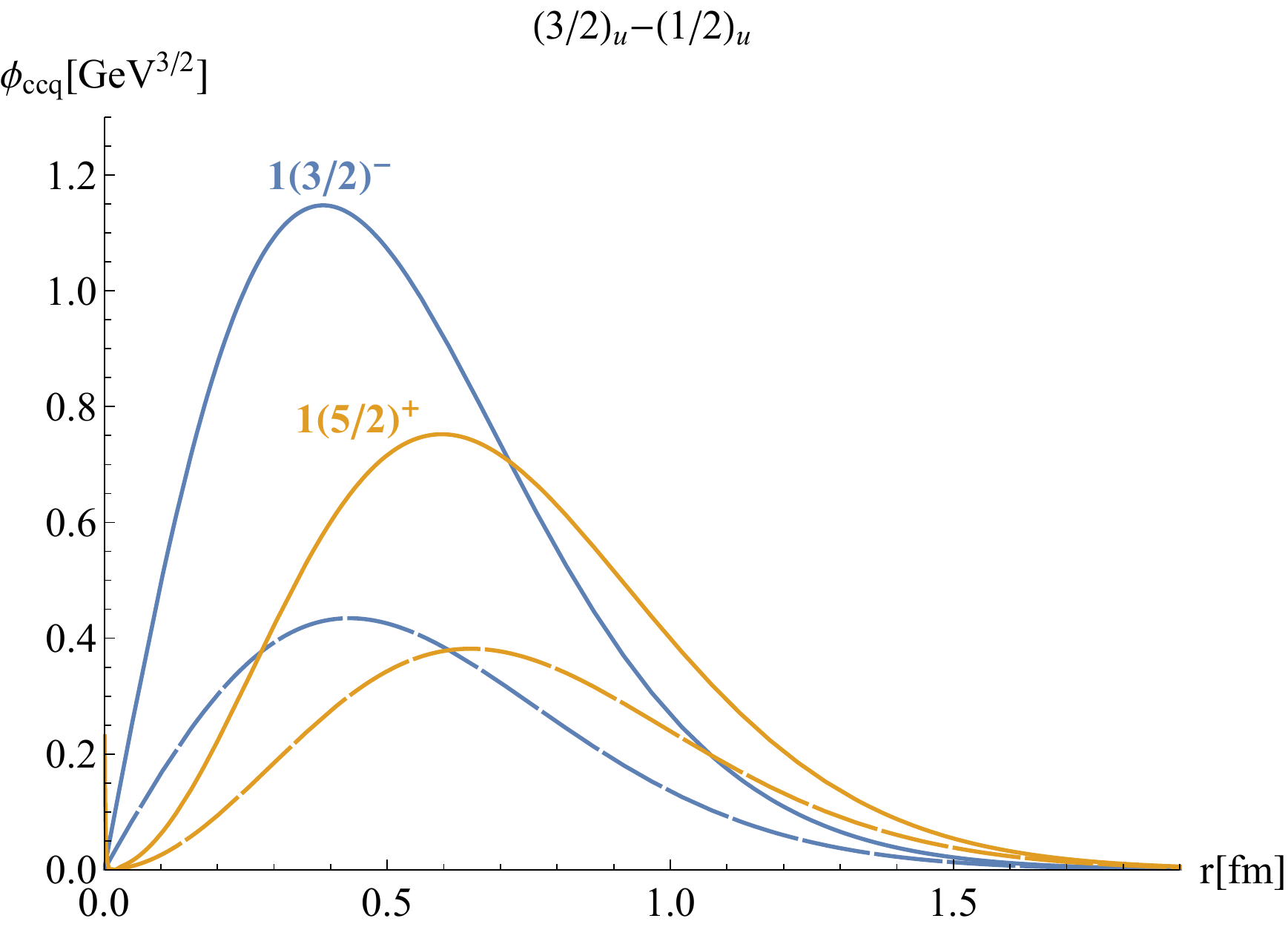} \\
\includegraphics[width=.45\textwidth]{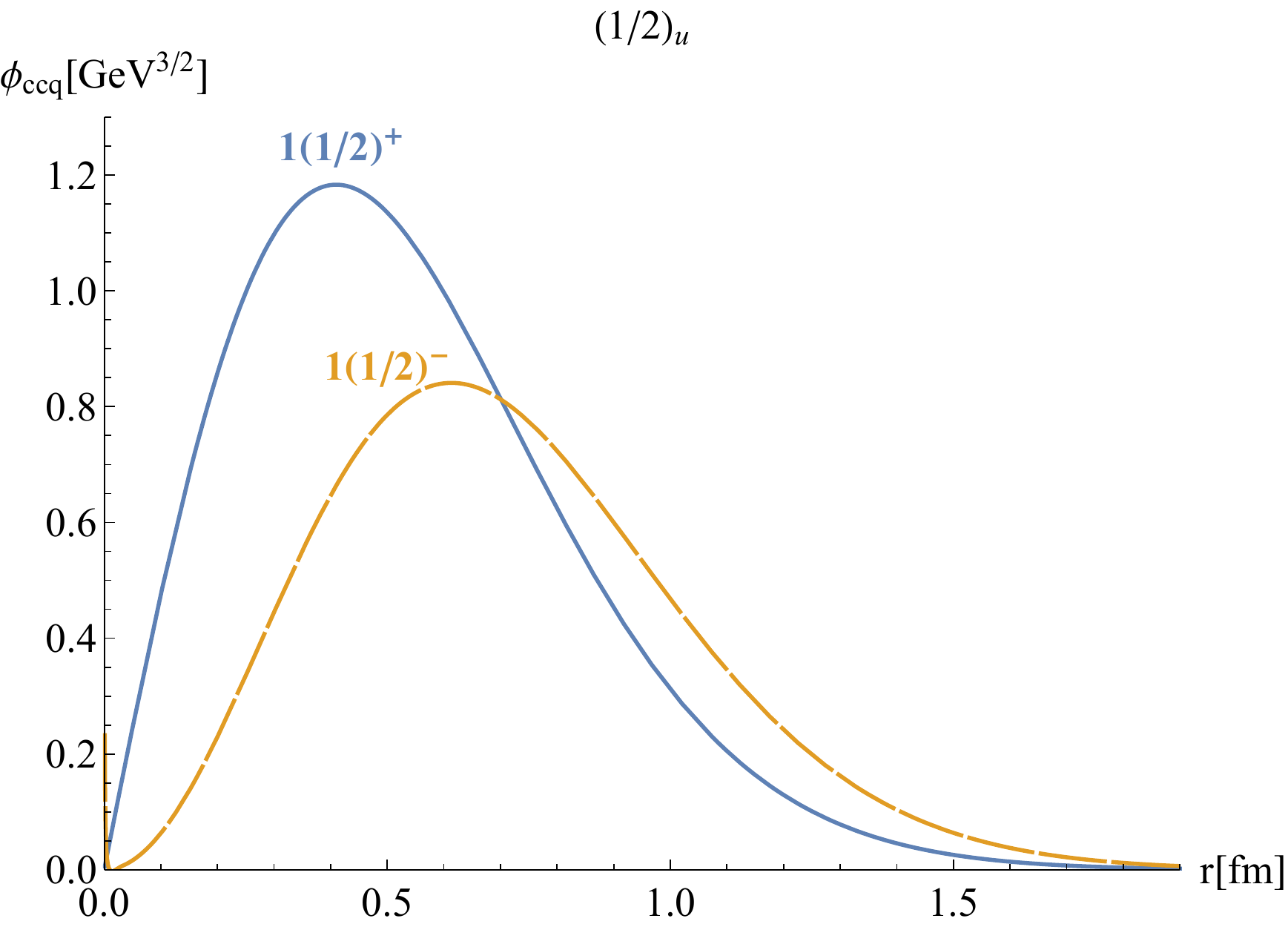} & \includegraphics[width=.45\textwidth]{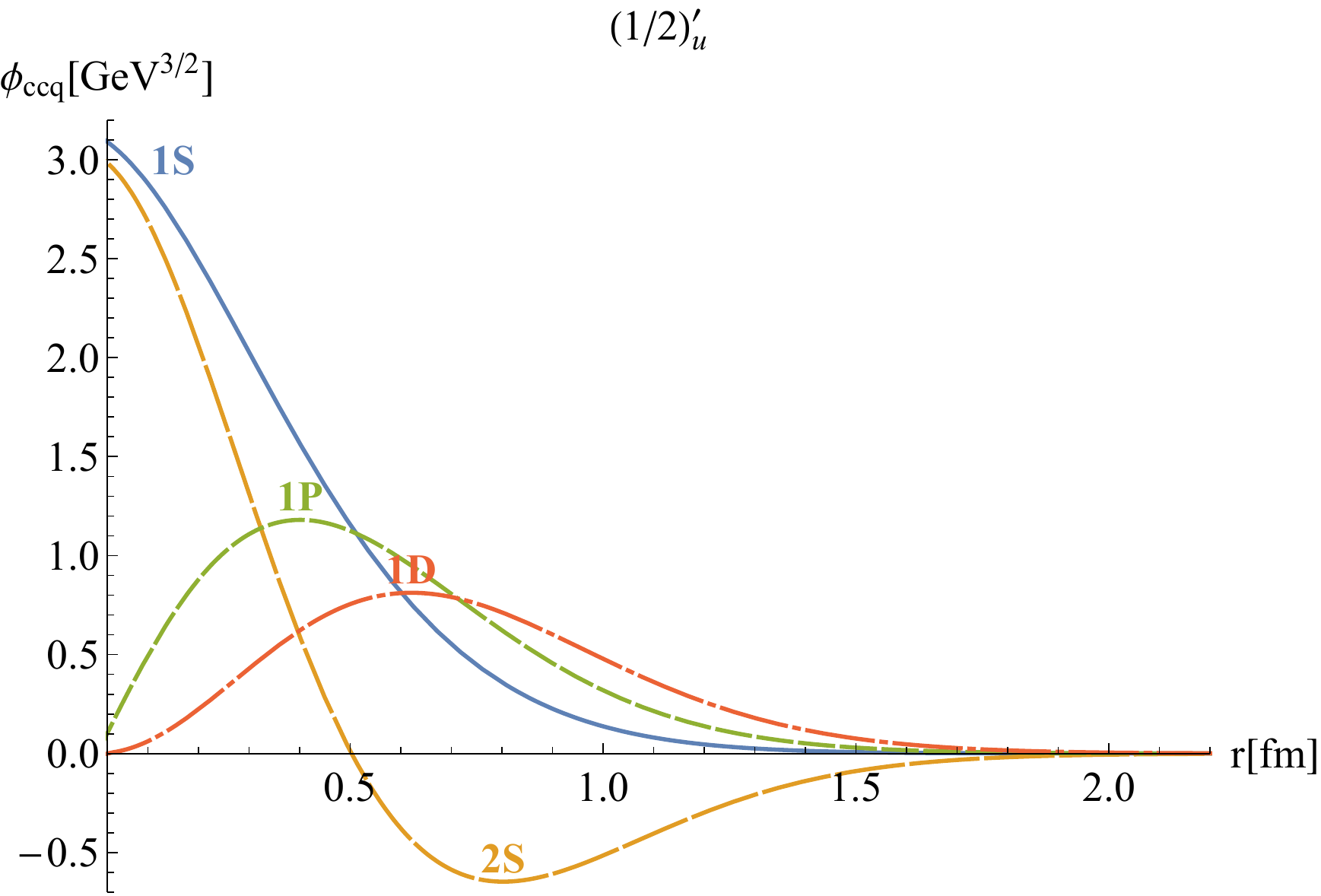}  \\
\end{tabular}
\caption{Radial wave functions $ccq$ double heavy baryons. The states are labeled as $nl$ for $(1/2)_g$ and $(1/2)^{\prime}_u$ and as $n\ell^{\eta_p}$ for $(3/2)_u-(1/2)_u$. In the cases of states with mixed contributions the solid and dashed lines correspond to the $(3/2)_u$ and $(1/2)_u$ contributions, respectively.}
\label{wfplotc}
\end{figure}

\begin{figure}[ht!]
\begin{tabular}{cc}
\includegraphics[width=.45\textwidth]{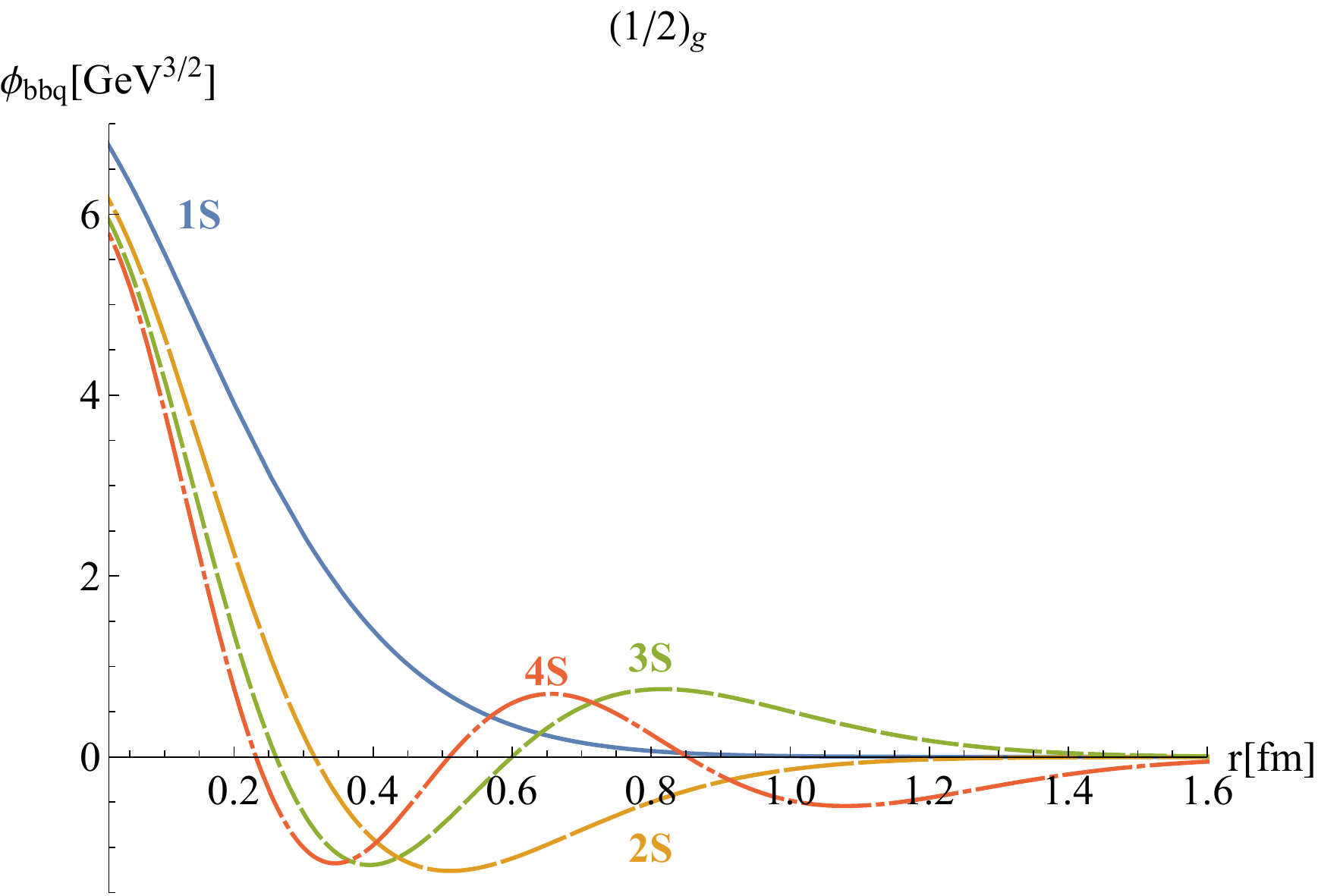} & \includegraphics[width=.45\textwidth]{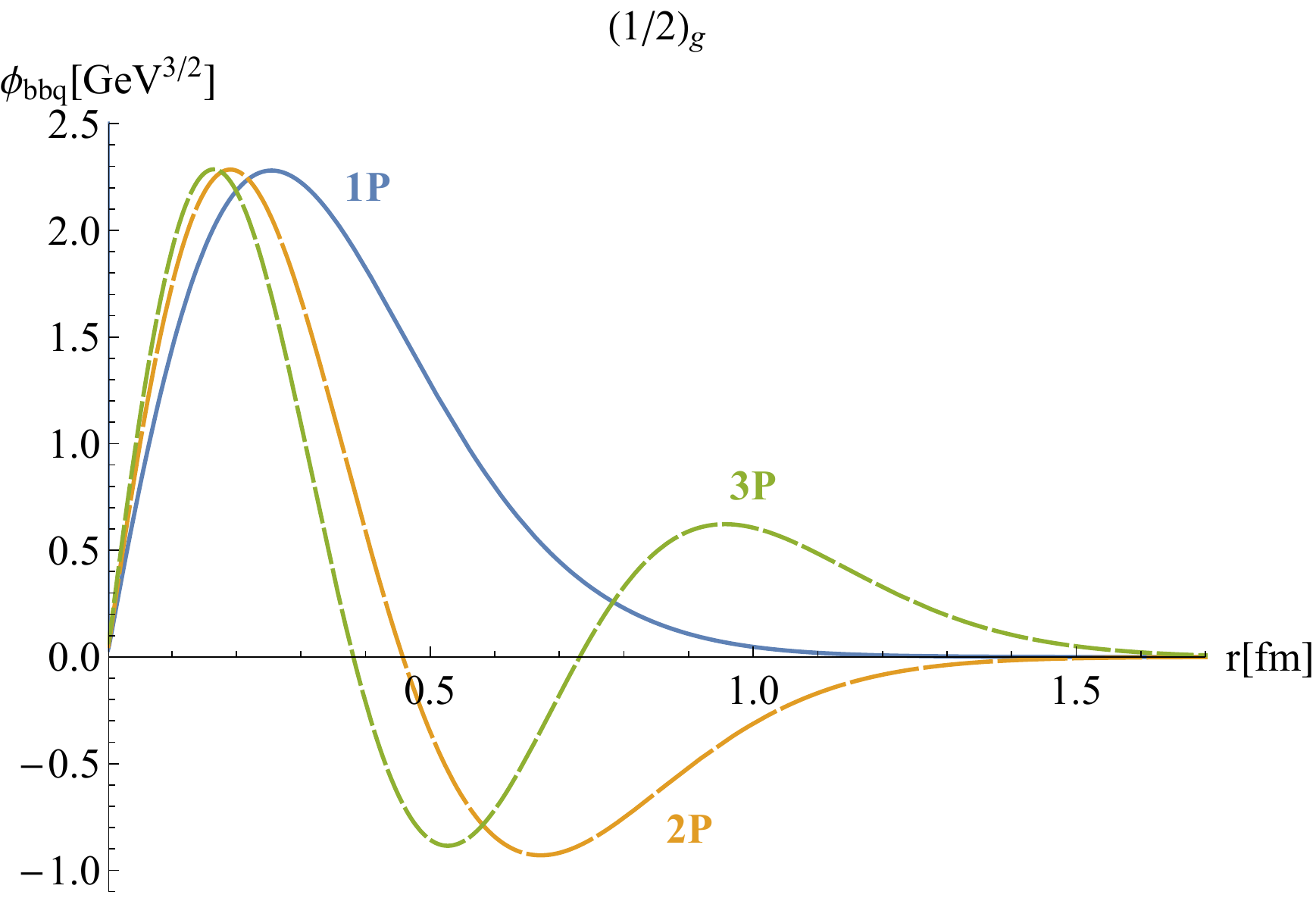} \\
\includegraphics[width=.45\textwidth]{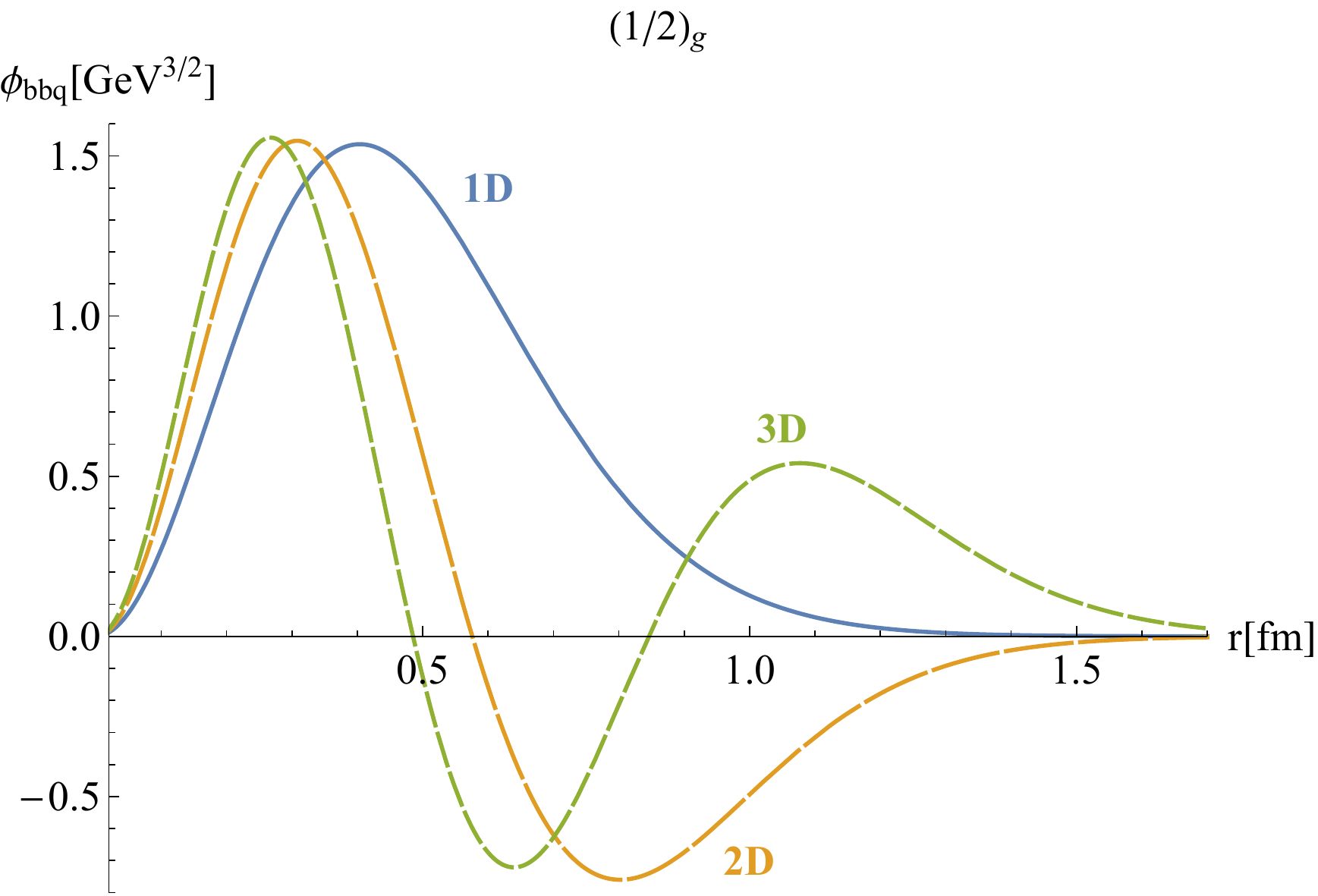} & \includegraphics[width=.45\textwidth]{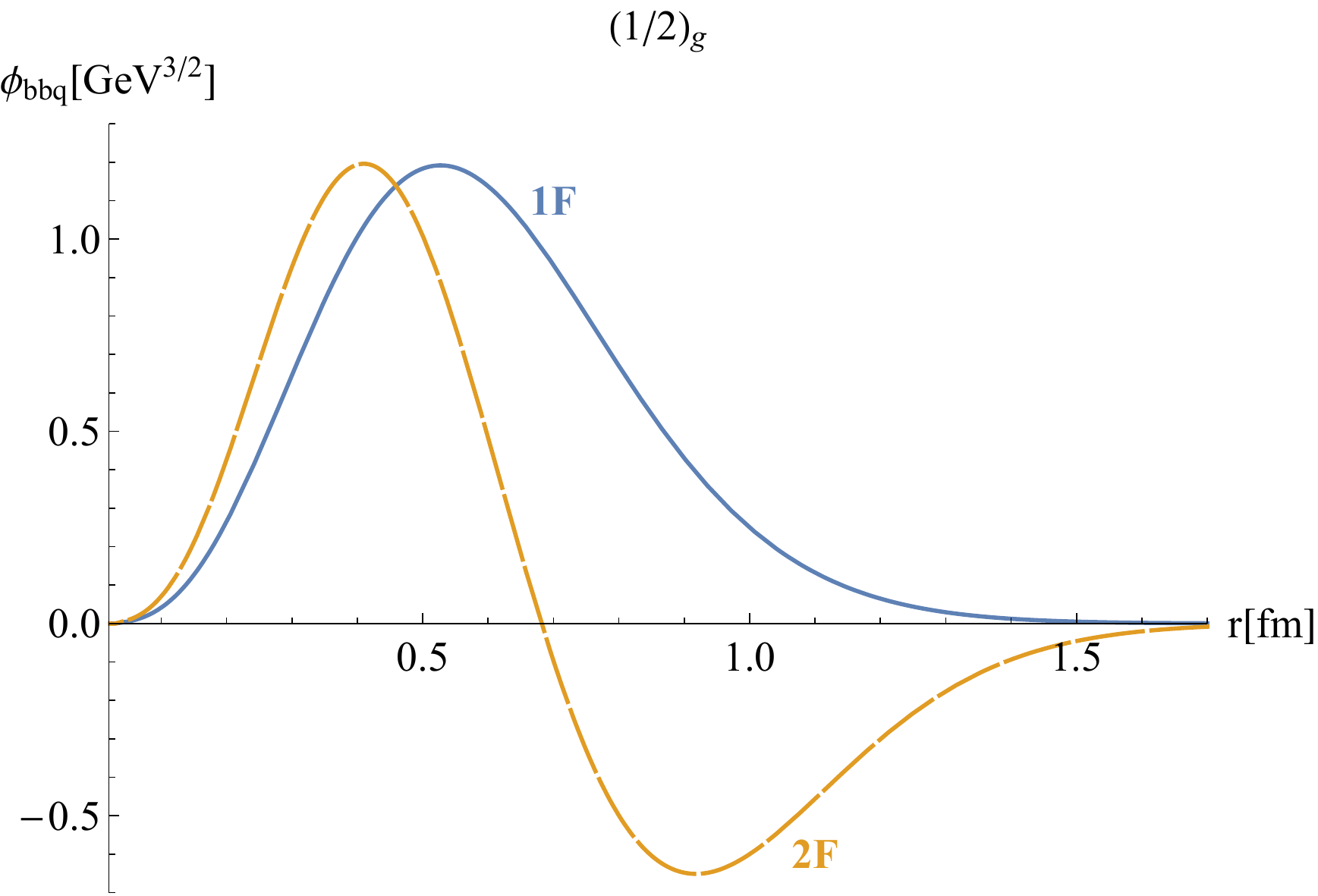} \\
\includegraphics[width=.45\textwidth]{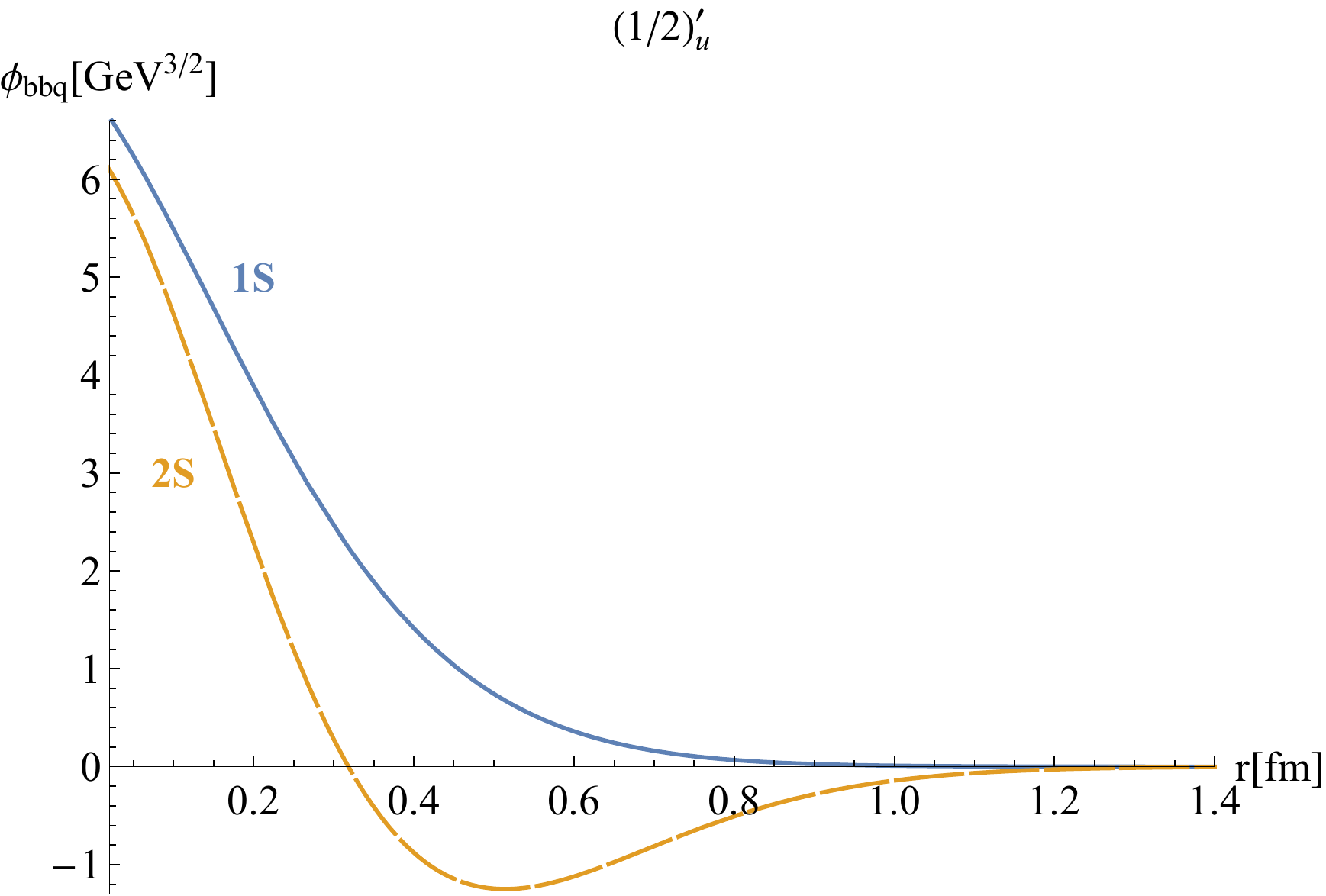} & \includegraphics[width=.45\textwidth]{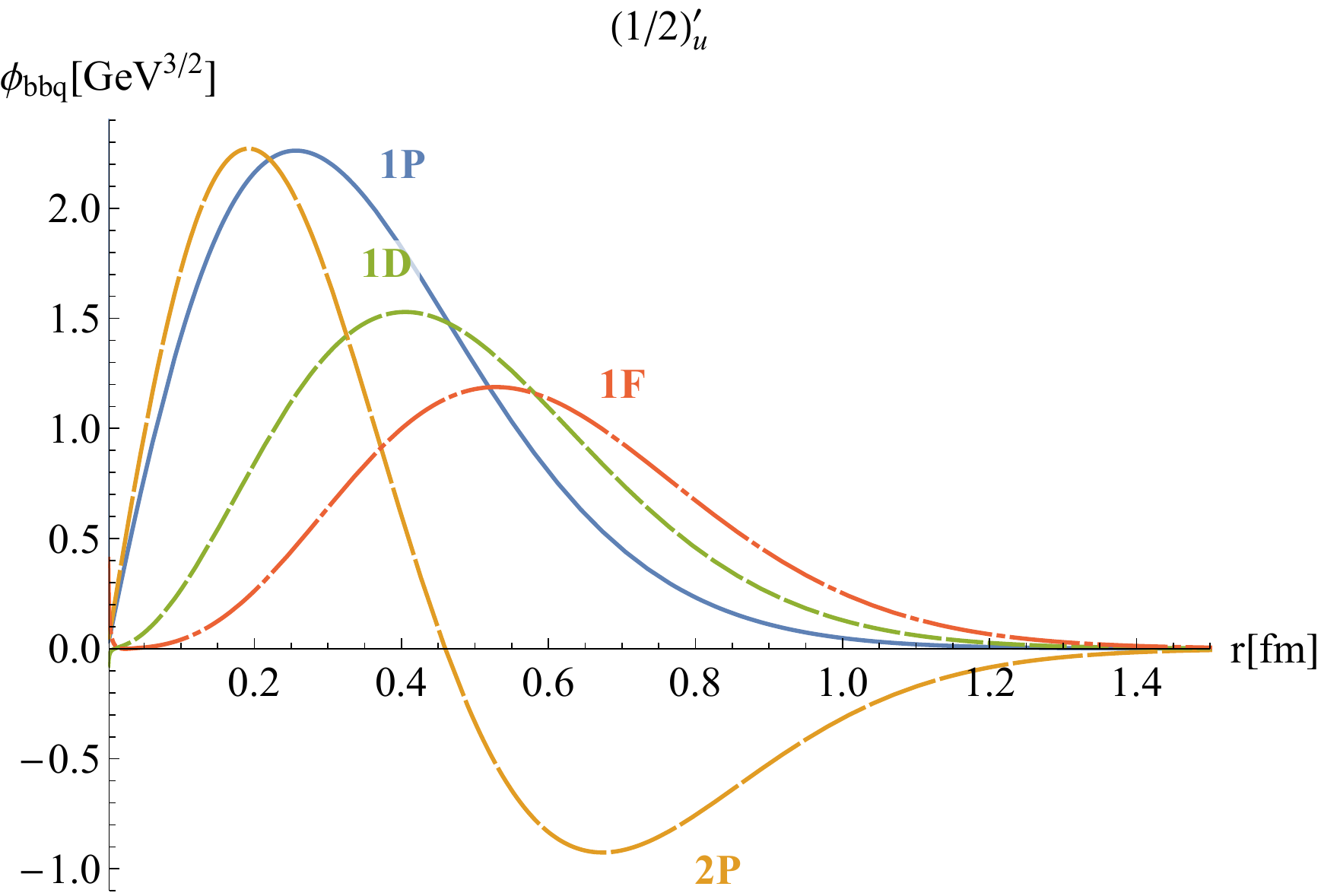}  \\
\end{tabular}
\caption{Radial wave functions $bbq$ double heavy baryons. The states are labeled as $nl$ for $(1/2)_g$ and $(1/2)^{\prime}_u$.}
\label{wfplotb}
\end{figure}

\begin{figure}[ht!]
\begin{tabular}{cc}
\includegraphics[width=.45\textwidth]{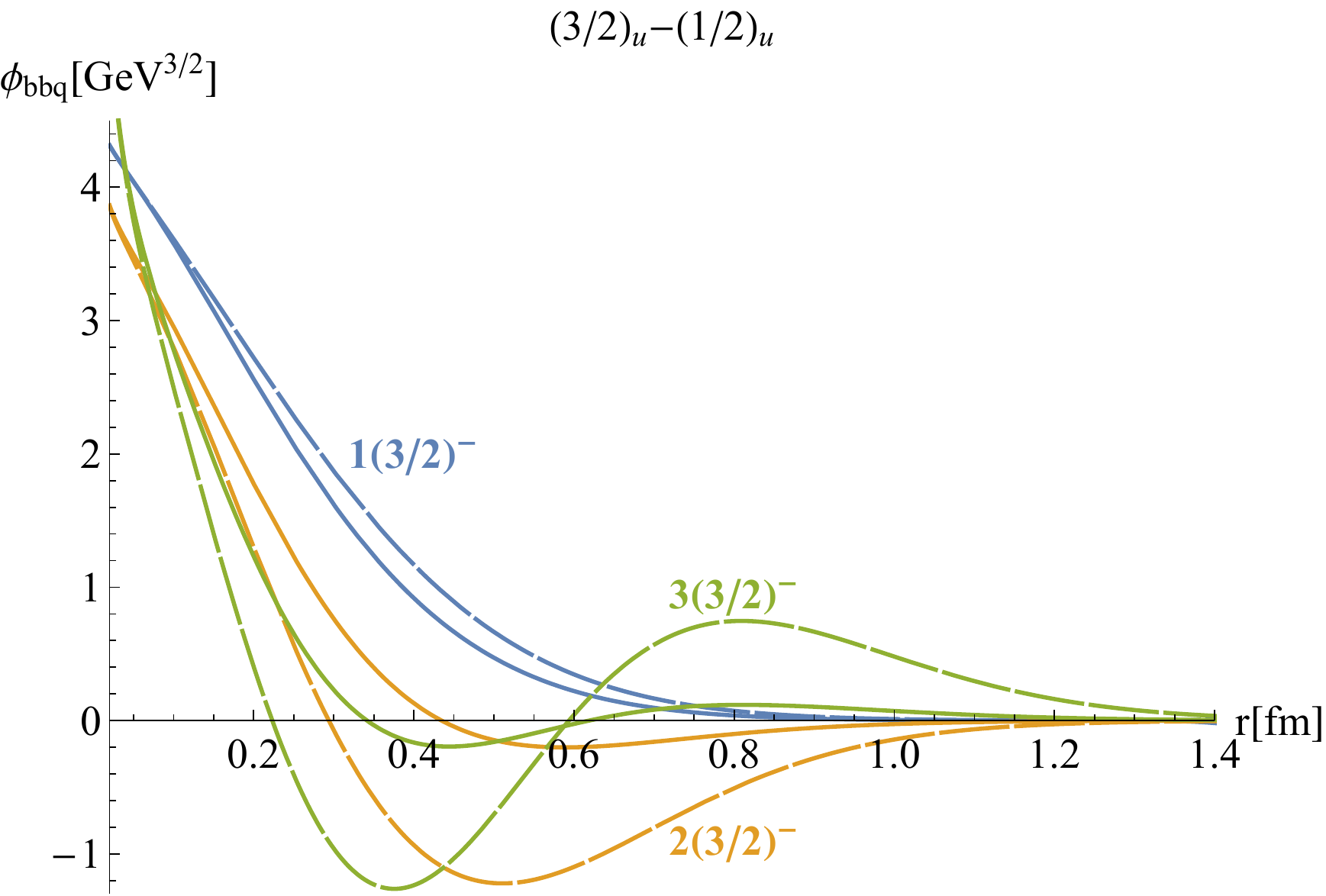} & \includegraphics[width=.45\textwidth]{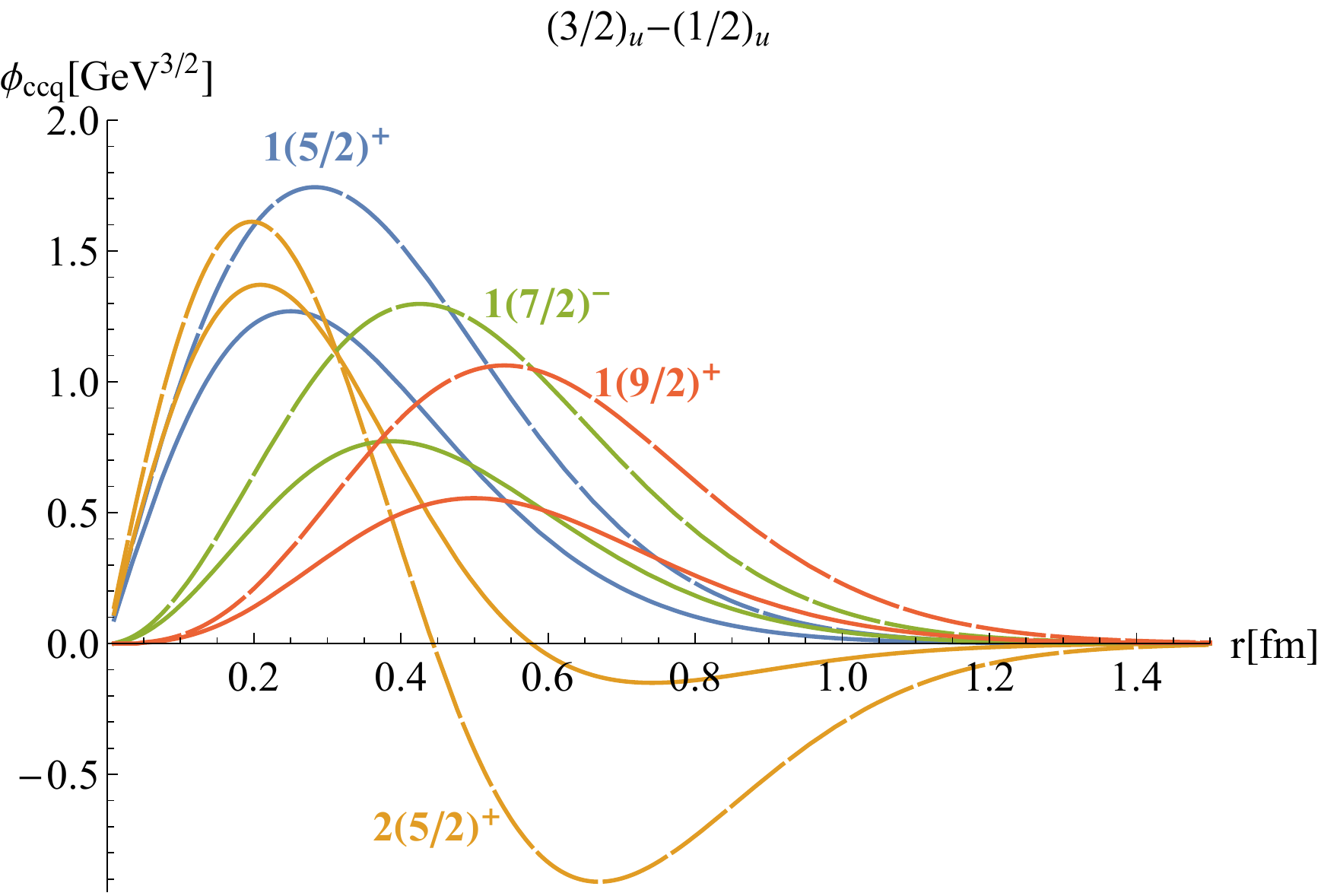} \\
\includegraphics[width=.45\textwidth]{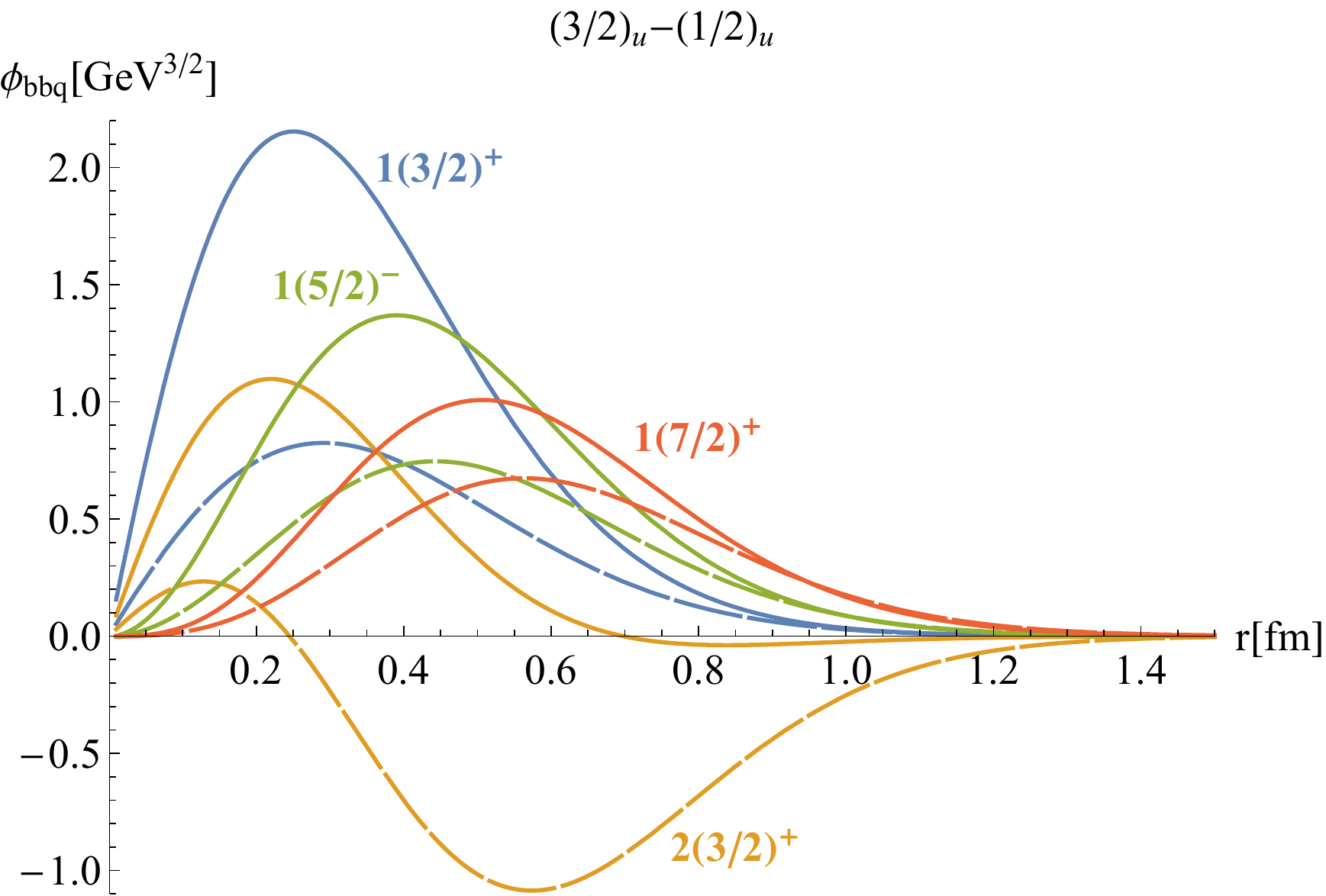} & \includegraphics[width=.45\textwidth]{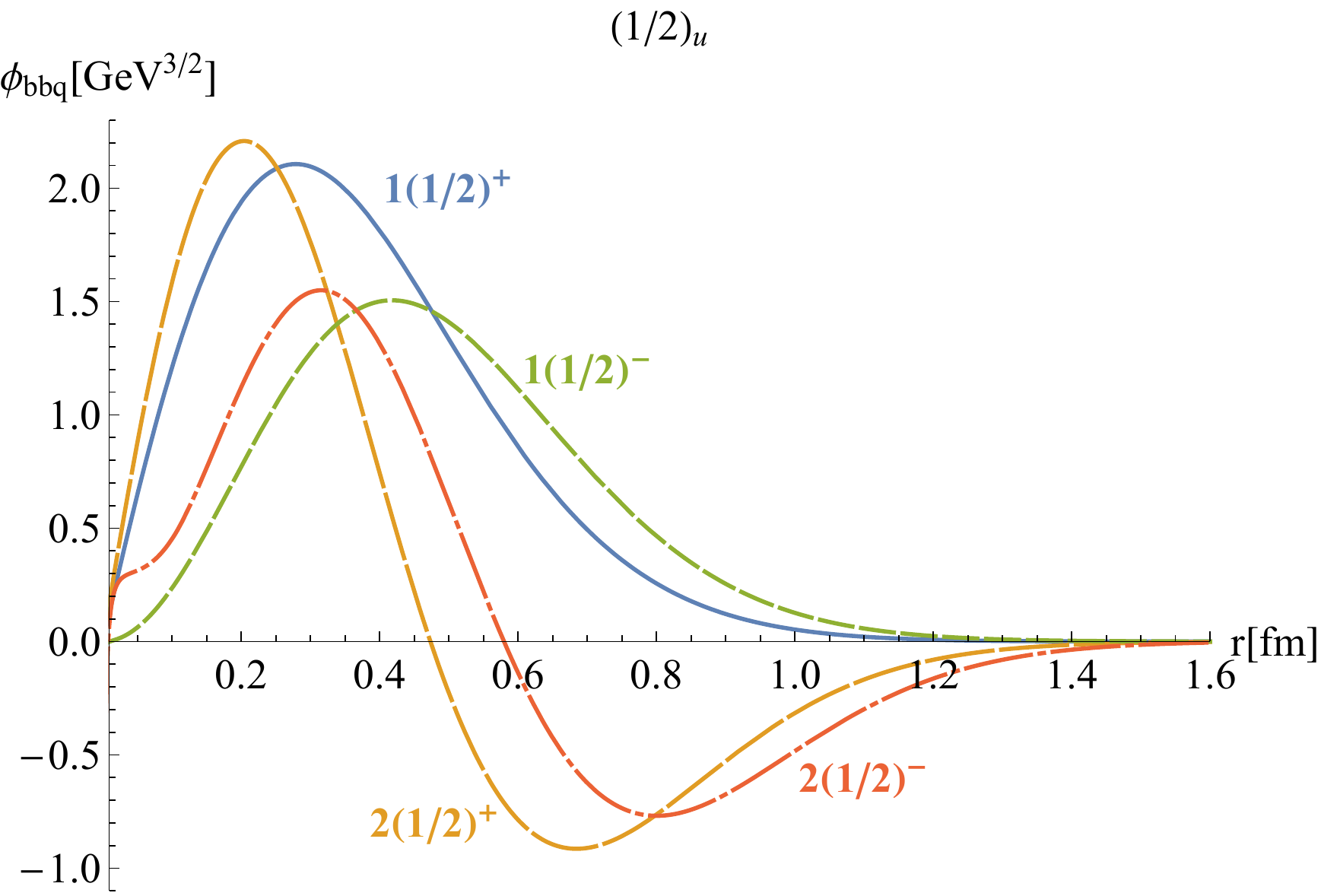} \\
\end{tabular}
\caption{Radial wave functions $bbq$ double heavy baryons. The states are labeled as $n\ell^{\eta_p}$. In the cases of states with mixed contributions the solid and dashed lines correspond to the $(3/2)_u$ and $(1/2)_u$ contributions, respectively.}
\label{wfplotb2}
\end{figure}

In this appendix we collect, in Figs.~\ref{wfplotc}, \ref{wfplotb} and, \ref{wfplotb2}, the plots of the radial wave functions of the states in Tables~\ref{ccspc} and \ref{bbspc}.

\newpage

\newpage

\bibliographystyle{apsrev4-1}
\bibliography{dhbbiblio}

\begin{thebibliography}{62}%
\makeatletter
\providecommand \@ifxundefined [1]{%
 \@ifx{#1\undefined}
}%
\providecommand \@ifnum [1]{%
 \ifnum #1\expandafter \@firstoftwo
 \else \expandafter \@secondoftwo
 \fi
}%
\providecommand \@ifx [1]{%
 \ifx #1\expandafter \@firstoftwo
 \else \expandafter \@secondoftwo
 \fi
}%
\providecommand \natexlab [1]{#1}%
\providecommand \enquote  [1]{``#1''}%
\providecommand \bibnamefont  [1]{#1}%
\providecommand \bibfnamefont [1]{#1}%
\providecommand \citenamefont [1]{#1}%
\providecommand \href@noop [0]{\@secondoftwo}%
\providecommand \href [0]{\begingroup \@sanitize@url \@href}%
\providecommand \@href[1]{\@@startlink{#1}\@@href}%
\providecommand \@@href[1]{\endgroup#1\@@endlink}%
\providecommand \@sanitize@url [0]{\catcode `\\12\catcode `\$12\catcode
  `\&12\catcode `\#12\catcode `\^12\catcode `\_12\catcode `\%12\relax}%
\providecommand \@@startlink[1]{}%
\providecommand \@@endlink[0]{}%
\providecommand \url  [0]{\begingroup\@sanitize@url \@url }%
\providecommand \@url [1]{\endgroup\@href {#1}{\urlprefix }}%
\providecommand \urlprefix  [0]{URL }%
\providecommand \Eprint [0]{\href }%
\providecommand \doibase [0]{http://dx.doi.org/}%
\providecommand \selectlanguage [0]{\@gobble}%
\providecommand \bibinfo  [0]{\@secondoftwo}%
\providecommand \bibfield  [0]{\@secondoftwo}%
\providecommand \translation [1]{[#1]}%
\providecommand \BibitemOpen [0]{}%
\providecommand \bibitemStop [0]{}%
\providecommand \bibitemNoStop [0]{.\EOS\space}%
\providecommand \EOS [0]{\spacefactor3000\relax}%
\providecommand \BibitemShut  [1]{\csname bibitem#1\endcsname}%
\let\auto@bib@innerbib\@empty
%</preamble>
\bibitem [{\citenamefont {Aaij}\ \emph {et~al.}(2017)\citenamefont {Aaij} \emph
  {et~al.}}]{Aaij:2017ueg}%
  \BibitemOpen
  \bibfield  {author} {\bibinfo {author} {\bibfnamefont {R.}~\bibnamefont
  {Aaij}} \emph {et~al.} (\bibinfo {collaboration} {LHCb}),\ }\href {\doibase
  10.1103/PhysRevLett.119.112001} {\bibfield  {journal} {\bibinfo  {journal}
  {Phys. Rev. Lett.}\ }\textbf {\bibinfo {volume} {119}},\ \bibinfo {pages}
  {112001} (\bibinfo {year} {2017})},\ \Eprint
  {http://arxiv.org/abs/1707.01621} {arXiv:1707.01621 [hep-ex]} \BibitemShut
  {NoStop}%
%%CITATION = ARXIV:1707.01621;%%
\bibitem [{\citenamefont {Aaij}\ \emph {et~al.}(2018)\citenamefont {Aaij} \emph
  {et~al.}}]{Aaij:2018gfl}%
  \BibitemOpen
  \bibfield  {author} {\bibinfo {author} {\bibfnamefont {R.}~\bibnamefont
  {Aaij}} \emph {et~al.} (\bibinfo {collaboration} {LHCb}),\ }\href {\doibase
  10.1103/PhysRevLett.121.162002} {\bibfield  {journal} {\bibinfo  {journal}
  {Phys. Rev. Lett.}\ }\textbf {\bibinfo {volume} {121}},\ \bibinfo {pages}
  {162002} (\bibinfo {year} {2018})},\ \Eprint
  {http://arxiv.org/abs/1807.01919} {arXiv:1807.01919 [hep-ex]} \BibitemShut
  {NoStop}%
%%CITATION = ARXIV:1807.01919;%%
\bibitem [{\citenamefont {Mattson}\ \emph {et~al.}(2002)\citenamefont {Mattson}
  \emph {et~al.}}]{Mattson:2002vu}%
  \BibitemOpen
  \bibfield  {author} {\bibinfo {author} {\bibfnamefont {M.}~\bibnamefont
  {Mattson}} \emph {et~al.} (\bibinfo {collaboration} {SELEX}),\ }\href
  {\doibase 10.1103/PhysRevLett.89.112001} {\bibfield  {journal} {\bibinfo
  {journal} {Phys. Rev. Lett.}\ }\textbf {\bibinfo {volume} {89}},\ \bibinfo
  {pages} {112001} (\bibinfo {year} {2002})},\ \Eprint
  {http://arxiv.org/abs/hep-ex/0208014} {arXiv:hep-ex/0208014 [hep-ex]}
  \BibitemShut {NoStop}%
%%CITATION = HEP-EX/0208014;%%
\bibitem [{\citenamefont {Ocherashvili}\ \emph {et~al.}(2005)\citenamefont
  {Ocherashvili} \emph {et~al.}}]{Ocherashvili:2004hi}%
  \BibitemOpen
  \bibfield  {author} {\bibinfo {author} {\bibfnamefont {A.}~\bibnamefont
  {Ocherashvili}} \emph {et~al.} (\bibinfo {collaboration} {SELEX}),\ }\href
  {\doibase 10.1016/j.physletb.2005.09.043} {\bibfield  {journal} {\bibinfo
  {journal} {Phys. Lett.}\ }\textbf {\bibinfo {volume} {B628}},\ \bibinfo
  {pages} {18} (\bibinfo {year} {2005})},\ \Eprint
  {http://arxiv.org/abs/hep-ex/0406033} {arXiv:hep-ex/0406033 [hep-ex]}
  \BibitemShut {NoStop}%
%%CITATION = HEP-EX/0406033;%%
\bibitem [{\citenamefont {Aaij}\ \emph {et~al.}(2013)\citenamefont {Aaij} \emph
  {et~al.}}]{Aaij:2013voa}%
  \BibitemOpen
  \bibfield  {author} {\bibinfo {author} {\bibfnamefont {R.}~\bibnamefont
  {Aaij}} \emph {et~al.} (\bibinfo {collaboration} {LHCb}),\ }\href {\doibase
  10.1007/JHEP12(2013)090} {\bibfield  {journal} {\bibinfo  {journal} {JHEP}\
  }\textbf {\bibinfo {volume} {12}},\ \bibinfo {pages} {090} (\bibinfo {year}
  {2013})},\ \Eprint {http://arxiv.org/abs/1310.2538} {arXiv:1310.2538
  [hep-ex]} \BibitemShut {NoStop}%
%%CITATION = ARXIV:1310.2538;%%
\bibitem [{\citenamefont {Aaij}\ \emph {et~al.}(2020)\citenamefont {Aaij} \emph
  {et~al.}}]{Aaij:2019jfq}%
  \BibitemOpen
  \bibfield  {author} {\bibinfo {author} {\bibfnamefont {R.}~\bibnamefont
  {Aaij}} \emph {et~al.} (\bibinfo {collaboration} {LHCb}),\ }\href {\doibase
  10.1007/s11433-019-1471-8} {\bibfield  {journal} {\bibinfo  {journal} {Sci.
  China Phys. Mech. Astron.}\ }\textbf {\bibinfo {volume} {63}},\ \bibinfo
  {pages} {221062} (\bibinfo {year} {2020})},\ \Eprint
  {http://arxiv.org/abs/1909.12273} {arXiv:1909.12273 [hep-ex]} \BibitemShut
  {NoStop}%
%%CITATION = ARXIV:1909.12273;%%
\bibitem [{\citenamefont {Aubert}\ \emph {et~al.}(2006)\citenamefont {Aubert}
  \emph {et~al.}}]{Aubert:2006qw}%
  \BibitemOpen
  \bibfield  {author} {\bibinfo {author} {\bibfnamefont {B.}~\bibnamefont
  {Aubert}} \emph {et~al.} (\bibinfo {collaboration} {BaBar}),\ }\href
  {\doibase 10.1103/PhysRevD.74.011103} {\bibfield  {journal} {\bibinfo
  {journal} {Phys. Rev.}\ }\textbf {\bibinfo {volume} {D74}},\ \bibinfo {pages}
  {011103} (\bibinfo {year} {2006})},\ \Eprint
  {http://arxiv.org/abs/hep-ex/0605075} {arXiv:hep-ex/0605075 [hep-ex]}
  \BibitemShut {NoStop}%
%%CITATION = HEP-EX/0605075;%%
\bibitem [{\citenamefont {Kato}\ \emph {et~al.}(2014)\citenamefont {Kato} \emph
  {et~al.}}]{Kato:2013ynr}%
  \BibitemOpen
  \bibfield  {author} {\bibinfo {author} {\bibfnamefont {Y.}~\bibnamefont
  {Kato}} \emph {et~al.} (\bibinfo {collaboration} {Belle}),\ }\href {\doibase
  10.1103/PhysRevD.89.052003} {\bibfield  {journal} {\bibinfo  {journal} {Phys.
  Rev.}\ }\textbf {\bibinfo {volume} {D89}},\ \bibinfo {pages} {052003}
  (\bibinfo {year} {2014})},\ \Eprint {http://arxiv.org/abs/1312.1026}
  {arXiv:1312.1026 [hep-ex]} \BibitemShut {NoStop}%
%%CITATION = ARXIV:1312.1026;%%
\bibitem [{\citenamefont {Savage}\ and\ \citenamefont
  {Wise}(1990)}]{Savage:1990di}%
  \BibitemOpen
  \bibfield  {author} {\bibinfo {author} {\bibfnamefont {M.~J.}\ \bibnamefont
  {Savage}}\ and\ \bibinfo {author} {\bibfnamefont {M.~B.}\ \bibnamefont
  {Wise}},\ }\href {\doibase 10.1016/0370-2693(90)90035-5} {\bibfield
  {journal} {\bibinfo  {journal} {Phys. Lett.}\ }\textbf {\bibinfo {volume}
  {B248}},\ \bibinfo {pages} {177} (\bibinfo {year} {1990})}\BibitemShut
  {NoStop}%
%%CITATION = PHLTA,B248,177;%%
\bibitem [{\citenamefont {Brambilla}\ \emph {et~al.}(2005)\citenamefont
  {Brambilla}, \citenamefont {Vairo},\ and\ \citenamefont
  {Rosch}}]{Brambilla:2005yk}%
  \BibitemOpen
  \bibfield  {author} {\bibinfo {author} {\bibfnamefont {N.}~\bibnamefont
  {Brambilla}}, \bibinfo {author} {\bibfnamefont {A.}~\bibnamefont {Vairo}}, \
  and\ \bibinfo {author} {\bibfnamefont {T.}~\bibnamefont {Rosch}},\ }\href
  {\doibase 10.1103/PhysRevD.72.034021} {\bibfield  {journal} {\bibinfo
  {journal} {Phys. Rev.}\ }\textbf {\bibinfo {volume} {D72}},\ \bibinfo {pages}
  {034021} (\bibinfo {year} {2005})},\ \Eprint
  {http://arxiv.org/abs/hep-ph/0506065} {arXiv:hep-ph/0506065 [hep-ph]}
  \BibitemShut {NoStop}%
%%CITATION = HEP-PH/0506065;%%
\bibitem [{\citenamefont {Pineda}\ and\ \citenamefont
  {Soto}(1998)}]{Pineda:1997bj}%
  \BibitemOpen
  \bibfield  {author} {\bibinfo {author} {\bibfnamefont {A.}~\bibnamefont
  {Pineda}}\ and\ \bibinfo {author} {\bibfnamefont {J.}~\bibnamefont {Soto}},\
  }\bibfield  {booktitle} {\emph {\bibinfo {booktitle} {{Quantum
  chromodynamics. Proceedings, Conference, QCD'97, Montpellier, France, July
  3-9, 1997}}},\ }\href {\doibase 10.1016/S0920-5632(97)01102-X} {\bibfield
  {journal} {\bibinfo  {journal} {Nucl. Phys. Proc. Suppl.}\ }\textbf {\bibinfo
  {volume} {64}},\ \bibinfo {pages} {428} (\bibinfo {year} {1998})},\ \Eprint
  {http://arxiv.org/abs/hep-ph/9707481} {arXiv:hep-ph/9707481 [hep-ph]}
  \BibitemShut {NoStop}%
%%CITATION = HEP-PH/9707481;%%
\bibitem [{\citenamefont {Brambilla}\ \emph {et~al.}(2000)\citenamefont
  {Brambilla}, \citenamefont {Pineda}, \citenamefont {Soto},\ and\
  \citenamefont {Vairo}}]{Brambilla:1999xf}%
  \BibitemOpen
  \bibfield  {author} {\bibinfo {author} {\bibfnamefont {N.}~\bibnamefont
  {Brambilla}}, \bibinfo {author} {\bibfnamefont {A.}~\bibnamefont {Pineda}},
  \bibinfo {author} {\bibfnamefont {J.}~\bibnamefont {Soto}}, \ and\ \bibinfo
  {author} {\bibfnamefont {A.}~\bibnamefont {Vairo}},\ }\href {\doibase
  10.1016/S0550-3213(99)00693-8} {\bibfield  {journal} {\bibinfo  {journal}
  {Nucl. Phys.}\ }\textbf {\bibinfo {volume} {B566}},\ \bibinfo {pages} {275}
  (\bibinfo {year} {2000})},\ \Eprint {http://arxiv.org/abs/hep-ph/9907240}
  {arXiv:hep-ph/9907240 [hep-ph]} \BibitemShut {NoStop}%
%%CITATION = HEP-PH/9907240;%%
\bibitem [{\citenamefont {Soto}(2003)}]{Soto:2003ft}%
  \BibitemOpen
  \bibfield  {author} {\bibinfo {author} {\bibfnamefont {J.}~\bibnamefont
  {Soto}},\ }in\ \href@noop {} {\emph {\bibinfo {booktitle} {{Quark confinement
  and the hadron spectrum. Proceedings, 5th International Conference, Gargnano,
  Italy, September 10-14, 2002}}}}\ (\bibinfo {year} {2003})\ pp.\ \bibinfo
  {pages} {227--235},\ \Eprint {http://arxiv.org/abs/hep-ph/0301138}
  {arXiv:hep-ph/0301138 [hep-ph]} \BibitemShut {NoStop}%
%%CITATION = HEP-PH/0301138;%%
\bibitem [{\citenamefont {Soto}\ and\ \citenamefont
  {Tarr\'us~Castell\`a}(2020)}]{Soto:2020xpm}%
  \BibitemOpen
  \bibfield  {author} {\bibinfo {author} {\bibfnamefont {J.}~\bibnamefont
  {Soto}}\ and\ \bibinfo {author} {\bibfnamefont {J.}~\bibnamefont
  {Tarr\'us~Castell\`a}},\ }\href {\doibase 10.1103/PhysRevD.102.014012}
  {\bibfield  {journal} {\bibinfo  {journal} {Phys. Rev. D}\ }\textbf {\bibinfo
  {volume} {102}},\ \bibinfo {pages} {014012} (\bibinfo {year} {2020})},\
  \Eprint {http://arxiv.org/abs/2005.00552} {arXiv:2005.00552 [hep-ph]}
  \BibitemShut {NoStop}%
\bibitem [{\citenamefont {Najjar}\ and\ \citenamefont
  {Bali}(2009)}]{Najjar:2009da}%
  \BibitemOpen
  \bibfield  {author} {\bibinfo {author} {\bibfnamefont {J.}~\bibnamefont
  {Najjar}}\ and\ \bibinfo {author} {\bibfnamefont {G.}~\bibnamefont {Bali}},\
  }\bibfield  {booktitle} {\emph {\bibinfo {booktitle} {{Proceedings, 27th
  International Symposium on Lattice field theory (Lattice 2009): Beijing, P.R.
  China, July 26-31, 2009}}},\ }\href {\doibase 10.22323/1.091.0089} {\bibfield
   {journal} {\bibinfo  {journal} {PoS}\ }\textbf {\bibinfo {volume}
  {LAT2009}},\ \bibinfo {pages} {089} (\bibinfo {year} {2009})},\ \Eprint
  {http://arxiv.org/abs/0910.2824} {arXiv:0910.2824 [hep-lat]} \BibitemShut
  {NoStop}%
%%CITATION = ARXIV:0910.2824;%%
\bibitem [{\citenamefont {Najjar}(2009)}]{Najjarthesis}%
  \BibitemOpen
  \bibfield  {author} {\bibinfo {author} {\bibfnamefont {J.}~\bibnamefont
  {Najjar}},\ }\emph {\bibinfo {title} {Static-static-light Baryonic potentials
  from lattice QCD}},\ \href@noop {} {\bibinfo {type} {Diploma thesis}},\
  \bibinfo  {school} {Universit\"at Regensburg} (\bibinfo {year}
  {2009})\BibitemShut {NoStop}%
\bibitem [{\citenamefont {Alexandrou}\ \emph {et~al.}(1994)\citenamefont
  {Alexandrou}, \citenamefont {Borrelli}, \citenamefont {Gusken}, \citenamefont
  {Jegerlehner}, \citenamefont {Schilling}, \citenamefont {Siegert},\ and\
  \citenamefont {Sommer}}]{Alexandrou:1994dm}%
  \BibitemOpen
  \bibfield  {author} {\bibinfo {author} {\bibfnamefont {C.}~\bibnamefont
  {Alexandrou}}, \bibinfo {author} {\bibfnamefont {A.}~\bibnamefont
  {Borrelli}}, \bibinfo {author} {\bibfnamefont {S.}~\bibnamefont {Gusken}},
  \bibinfo {author} {\bibfnamefont {F.}~\bibnamefont {Jegerlehner}}, \bibinfo
  {author} {\bibfnamefont {K.}~\bibnamefont {Schilling}}, \bibinfo {author}
  {\bibfnamefont {G.}~\bibnamefont {Siegert}}, \ and\ \bibinfo {author}
  {\bibfnamefont {R.}~\bibnamefont {Sommer}},\ }\href {\doibase
  10.1016/0370-2693(94)90985-7} {\bibfield  {journal} {\bibinfo  {journal}
  {Phys. Lett.}\ }\textbf {\bibinfo {volume} {B337}},\ \bibinfo {pages} {340}
  (\bibinfo {year} {1994})},\ \Eprint {http://arxiv.org/abs/hep-lat/9407027}
  {arXiv:hep-lat/9407027 [hep-lat]} \BibitemShut {NoStop}%
%%CITATION = HEP-LAT/9407027;%%
\bibitem [{\citenamefont {Bowler}\ \emph {et~al.}(1996)\citenamefont {Bowler},
  \citenamefont {Kenway}, \citenamefont {Oliveira}, \citenamefont {Richards},
  \citenamefont {Uberholz}, \citenamefont {Lellouch}, \citenamefont {Nieves},
  \citenamefont {Sachrajda}, \citenamefont {Stella},\ and\ \citenamefont
  {Wittig}}]{Bowler:1996ws}%
  \BibitemOpen
  \bibfield  {author} {\bibinfo {author} {\bibfnamefont {K.~C.}\ \bibnamefont
  {Bowler}}, \bibinfo {author} {\bibfnamefont {R.~D.}\ \bibnamefont {Kenway}},
  \bibinfo {author} {\bibfnamefont {O.}~\bibnamefont {Oliveira}}, \bibinfo
  {author} {\bibfnamefont {D.~G.}\ \bibnamefont {Richards}}, \bibinfo {author}
  {\bibfnamefont {P.}~\bibnamefont {Uberholz}}, \bibinfo {author}
  {\bibfnamefont {L.}~\bibnamefont {Lellouch}}, \bibinfo {author}
  {\bibfnamefont {J.}~\bibnamefont {Nieves}}, \bibinfo {author} {\bibfnamefont
  {C.~T.}\ \bibnamefont {Sachrajda}}, \bibinfo {author} {\bibfnamefont
  {N.}~\bibnamefont {Stella}}, \ and\ \bibinfo {author} {\bibfnamefont
  {H.}~\bibnamefont {Wittig}} (\bibinfo {collaboration} {UKQCD}),\ }\href
  {\doibase 10.1103/PhysRevD.54.3619} {\bibfield  {journal} {\bibinfo
  {journal} {Phys. Rev.}\ }\textbf {\bibinfo {volume} {D54}},\ \bibinfo {pages}
  {3619} (\bibinfo {year} {1996})},\ \Eprint
  {http://arxiv.org/abs/hep-lat/9601022} {arXiv:hep-lat/9601022 [hep-lat]}
  \BibitemShut {NoStop}%
%%CITATION = HEP-LAT/9601022;%%
\bibitem [{\citenamefont {Ali~Khan}\ \emph {et~al.}(2000)\citenamefont
  {Ali~Khan}, \citenamefont {Bhattacharya}, \citenamefont {Collins},
  \citenamefont {Davies}, \citenamefont {Gupta}, \citenamefont {Morningstar},
  \citenamefont {Shigemitsu},\ and\ \citenamefont {Sloan}}]{AliKhan:1999yb}%
  \BibitemOpen
  \bibfield  {author} {\bibinfo {author} {\bibfnamefont {A.}~\bibnamefont
  {Ali~Khan}}, \bibinfo {author} {\bibfnamefont {T.}~\bibnamefont
  {Bhattacharya}}, \bibinfo {author} {\bibfnamefont {S.}~\bibnamefont
  {Collins}}, \bibinfo {author} {\bibfnamefont {C.~T.~H.}\ \bibnamefont
  {Davies}}, \bibinfo {author} {\bibfnamefont {R.}~\bibnamefont {Gupta}},
  \bibinfo {author} {\bibfnamefont {C.}~\bibnamefont {Morningstar}}, \bibinfo
  {author} {\bibfnamefont {J.}~\bibnamefont {Shigemitsu}}, \ and\ \bibinfo
  {author} {\bibfnamefont {J.~H.}\ \bibnamefont {Sloan}},\ }\href {\doibase
  10.1103/PhysRevD.62.054505} {\bibfield  {journal} {\bibinfo  {journal} {Phys.
  Rev.}\ }\textbf {\bibinfo {volume} {D62}},\ \bibinfo {pages} {054505}
  (\bibinfo {year} {2000})},\ \Eprint {http://arxiv.org/abs/hep-lat/9912034}
  {arXiv:hep-lat/9912034 [hep-lat]} \BibitemShut {NoStop}%
%%CITATION = HEP-LAT/9912034;%%
\bibitem [{\citenamefont {Mathur}\ \emph {et~al.}(2002)\citenamefont {Mathur},
  \citenamefont {Lewis},\ and\ \citenamefont {Woloshyn}}]{Mathur:2002ce}%
  \BibitemOpen
  \bibfield  {author} {\bibinfo {author} {\bibfnamefont {N.}~\bibnamefont
  {Mathur}}, \bibinfo {author} {\bibfnamefont {R.}~\bibnamefont {Lewis}}, \
  and\ \bibinfo {author} {\bibfnamefont {R.~M.}\ \bibnamefont {Woloshyn}},\
  }\href {\doibase 10.1103/PhysRevD.66.014502} {\bibfield  {journal} {\bibinfo
  {journal} {Phys. Rev.}\ }\textbf {\bibinfo {volume} {D66}},\ \bibinfo {pages}
  {014502} (\bibinfo {year} {2002})},\ \Eprint
  {http://arxiv.org/abs/hep-ph/0203253} {arXiv:hep-ph/0203253 [hep-ph]}
  \BibitemShut {NoStop}%
%%CITATION = HEP-PH/0203253;%%
\bibitem [{\citenamefont {Lewis}\ \emph {et~al.}(2001)\citenamefont {Lewis},
  \citenamefont {Mathur},\ and\ \citenamefont {Woloshyn}}]{Lewis:2001iz}%
  \BibitemOpen
  \bibfield  {author} {\bibinfo {author} {\bibfnamefont {R.}~\bibnamefont
  {Lewis}}, \bibinfo {author} {\bibfnamefont {N.}~\bibnamefont {Mathur}}, \
  and\ \bibinfo {author} {\bibfnamefont {R.~M.}\ \bibnamefont {Woloshyn}},\
  }\href {\doibase 10.1103/PhysRevD.64.094509} {\bibfield  {journal} {\bibinfo
  {journal} {Phys. Rev.}\ }\textbf {\bibinfo {volume} {D64}},\ \bibinfo {pages}
  {094509} (\bibinfo {year} {2001})},\ \Eprint
  {http://arxiv.org/abs/hep-ph/0107037} {arXiv:hep-ph/0107037 [hep-ph]}
  \BibitemShut {NoStop}%
%%CITATION = HEP-PH/0107037;%%
\bibitem [{\citenamefont {Flynn}\ \emph {et~al.}(2003)\citenamefont {Flynn},
  \citenamefont {Mescia},\ and\ \citenamefont {Tariq}}]{Flynn:2003vz}%
  \BibitemOpen
  \bibfield  {author} {\bibinfo {author} {\bibfnamefont {J.~M.}\ \bibnamefont
  {Flynn}}, \bibinfo {author} {\bibfnamefont {F.}~\bibnamefont {Mescia}}, \
  and\ \bibinfo {author} {\bibfnamefont {A.~S.~B.}\ \bibnamefont {Tariq}}
  (\bibinfo {collaboration} {UKQCD}),\ }\href {\doibase
  10.1088/1126-6708/2003/07/066} {\bibfield  {journal} {\bibinfo  {journal}
  {JHEP}\ }\textbf {\bibinfo {volume} {07}},\ \bibinfo {pages} {066} (\bibinfo
  {year} {2003})},\ \Eprint {http://arxiv.org/abs/hep-lat/0307025}
  {arXiv:hep-lat/0307025 [hep-lat]} \BibitemShut {NoStop}%
%%CITATION = HEP-LAT/0307025;%%
\bibitem [{\citenamefont {Chiu}\ and\ \citenamefont
  {Hsieh}(2005)}]{Chiu:2005zc}%
  \BibitemOpen
  \bibfield  {author} {\bibinfo {author} {\bibfnamefont {T.-W.}\ \bibnamefont
  {Chiu}}\ and\ \bibinfo {author} {\bibfnamefont {T.-H.}\ \bibnamefont
  {Hsieh}},\ }\bibfield  {booktitle} {\emph {\bibinfo {booktitle} {{The
  structure of baryons. Proceedings, 10th International Conference, Baryons'04,
  Palaiseau, France, October 25-29, 2004}}},\ }\href {\doibase
  10.1016/j.nuclphysa.2005.03.090} {\bibfield  {journal} {\bibinfo  {journal}
  {Nucl. Phys.}\ }\textbf {\bibinfo {volume} {A755}},\ \bibinfo {pages} {471}
  (\bibinfo {year} {2005})},\ \Eprint {http://arxiv.org/abs/hep-lat/0501021}
  {arXiv:hep-lat/0501021 [hep-lat]} \BibitemShut {NoStop}%
%%CITATION = HEP-LAT/0501021;%%
\bibitem [{\citenamefont {Na}\ and\ \citenamefont
  {Gottlieb}(2007)}]{Na:2007pv}%
  \BibitemOpen
  \bibfield  {author} {\bibinfo {author} {\bibfnamefont {H.}~\bibnamefont
  {Na}}\ and\ \bibinfo {author} {\bibfnamefont {S.~A.}\ \bibnamefont
  {Gottlieb}},\ }\bibfield  {booktitle} {\emph {\bibinfo {booktitle}
  {{Proceedings, 25th International Symposium on Lattice field theory (Lattice
  2007): Regensburg, Germany, July 30-August 4, 2007}}},\ }\href {\doibase
  10.22323/1.042.0124} {\bibfield  {journal} {\bibinfo  {journal} {PoS}\
  }\textbf {\bibinfo {volume} {LATTICE2007}},\ \bibinfo {pages} {124} (\bibinfo
  {year} {2007})},\ \Eprint {http://arxiv.org/abs/0710.1422} {arXiv:0710.1422
  [hep-lat]} \BibitemShut {NoStop}%
%%CITATION = ARXIV:0710.1422;%%
\bibitem [{\citenamefont {Liu}\ \emph {et~al.}(2010)\citenamefont {Liu},
  \citenamefont {Lin}, \citenamefont {Orginos},\ and\ \citenamefont
  {Walker-Loud}}]{Liu:2009jc}%
  \BibitemOpen
  \bibfield  {author} {\bibinfo {author} {\bibfnamefont {L.}~\bibnamefont
  {Liu}}, \bibinfo {author} {\bibfnamefont {H.-W.}\ \bibnamefont {Lin}},
  \bibinfo {author} {\bibfnamefont {K.}~\bibnamefont {Orginos}}, \ and\
  \bibinfo {author} {\bibfnamefont {A.}~\bibnamefont {Walker-Loud}},\ }\href
  {\doibase 10.1103/PhysRevD.81.094505} {\bibfield  {journal} {\bibinfo
  {journal} {Phys. Rev.}\ }\textbf {\bibinfo {volume} {D81}},\ \bibinfo {pages}
  {094505} (\bibinfo {year} {2010})},\ \Eprint {http://arxiv.org/abs/0909.3294}
  {arXiv:0909.3294 [hep-lat]} \BibitemShut {NoStop}%
%%CITATION = ARXIV:0909.3294;%%
\bibitem [{\citenamefont {Lin}\ \emph {et~al.}(2011)\citenamefont {Lin},
  \citenamefont {Cohen}, \citenamefont {Liu}, \citenamefont {Mathur},
  \citenamefont {Orginos},\ and\ \citenamefont {Walker-Loud}}]{Lin:2010wb}%
  \BibitemOpen
  \bibfield  {author} {\bibinfo {author} {\bibfnamefont {H.-W.}\ \bibnamefont
  {Lin}}, \bibinfo {author} {\bibfnamefont {S.~D.}\ \bibnamefont {Cohen}},
  \bibinfo {author} {\bibfnamefont {L.}~\bibnamefont {Liu}}, \bibinfo {author}
  {\bibfnamefont {N.}~\bibnamefont {Mathur}}, \bibinfo {author} {\bibfnamefont
  {K.}~\bibnamefont {Orginos}}, \ and\ \bibinfo {author} {\bibfnamefont
  {A.}~\bibnamefont {Walker-Loud}},\ }\href {\doibase
  10.1016/j.cpc.2010.07.004} {\bibfield  {journal} {\bibinfo  {journal}
  {Comput. Phys. Commun.}\ }\textbf {\bibinfo {volume} {182}},\ \bibinfo
  {pages} {24} (\bibinfo {year} {2011})},\ \Eprint
  {http://arxiv.org/abs/1002.4710} {arXiv:1002.4710 [hep-lat]} \BibitemShut
  {NoStop}%
%%CITATION = ARXIV:1002.4710;%%
\bibitem [{\citenamefont {Briceno}\ \emph {et~al.}(2012)\citenamefont
  {Briceno}, \citenamefont {Lin},\ and\ \citenamefont
  {Bolton}}]{Briceno:2012wt}%
  \BibitemOpen
  \bibfield  {author} {\bibinfo {author} {\bibfnamefont {R.~A.}\ \bibnamefont
  {Briceno}}, \bibinfo {author} {\bibfnamefont {H.-W.}\ \bibnamefont {Lin}}, \
  and\ \bibinfo {author} {\bibfnamefont {D.~R.}\ \bibnamefont {Bolton}},\
  }\href {\doibase 10.1103/PhysRevD.86.094504} {\bibfield  {journal} {\bibinfo
  {journal} {Phys. Rev.}\ }\textbf {\bibinfo {volume} {D86}},\ \bibinfo {pages}
  {094504} (\bibinfo {year} {2012})},\ \Eprint {http://arxiv.org/abs/1207.3536}
  {arXiv:1207.3536 [hep-lat]} \BibitemShut {NoStop}%
%%CITATION = ARXIV:1207.3536;%%
\bibitem [{\citenamefont {Alexandrou}\ \emph {et~al.}(2012)\citenamefont
  {Alexandrou}, \citenamefont {Carbonell}, \citenamefont {Christaras},
  \citenamefont {Drach}, \citenamefont {Gravina},\ and\ \citenamefont
  {Papinutto}}]{Alexandrou:2012xk}%
  \BibitemOpen
  \bibfield  {author} {\bibinfo {author} {\bibfnamefont {C.}~\bibnamefont
  {Alexandrou}}, \bibinfo {author} {\bibfnamefont {J.}~\bibnamefont
  {Carbonell}}, \bibinfo {author} {\bibfnamefont {D.}~\bibnamefont
  {Christaras}}, \bibinfo {author} {\bibfnamefont {V.}~\bibnamefont {Drach}},
  \bibinfo {author} {\bibfnamefont {M.}~\bibnamefont {Gravina}}, \ and\
  \bibinfo {author} {\bibfnamefont {M.}~\bibnamefont {Papinutto}},\ }\href
  {\doibase 10.1103/PhysRevD.86.114501} {\bibfield  {journal} {\bibinfo
  {journal} {Phys. Rev.}\ }\textbf {\bibinfo {volume} {D86}},\ \bibinfo {pages}
  {114501} (\bibinfo {year} {2012})},\ \Eprint {http://arxiv.org/abs/1205.6856}
  {arXiv:1205.6856 [hep-lat]} \BibitemShut {NoStop}%
%%CITATION = ARXIV:1205.6856;%%
\bibitem [{\citenamefont {Basak}\ \emph {et~al.}(2012)\citenamefont {Basak},
  \citenamefont {Datta}, \citenamefont {Padmanath}, \citenamefont {Majumdar},\
  and\ \citenamefont {Mathur}}]{Basak:2012py}%
  \BibitemOpen
  \bibfield  {author} {\bibinfo {author} {\bibfnamefont {S.}~\bibnamefont
  {Basak}}, \bibinfo {author} {\bibfnamefont {S.}~\bibnamefont {Datta}},
  \bibinfo {author} {\bibfnamefont {M.}~\bibnamefont {Padmanath}}, \bibinfo
  {author} {\bibfnamefont {P.}~\bibnamefont {Majumdar}}, \ and\ \bibinfo
  {author} {\bibfnamefont {N.}~\bibnamefont {Mathur}},\ }\bibfield  {booktitle}
  {\emph {\bibinfo {booktitle} {{Proceedings, 30th International Symposium on
  Lattice Field Theory (Lattice 2012): Cairns, Australia, June 24-29, 2012}}},\
  }\href {\doibase 10.22323/1.164.0141} {\bibfield  {journal} {\bibinfo
  {journal} {PoS}\ }\textbf {\bibinfo {volume} {LATTICE2012}},\ \bibinfo
  {pages} {141} (\bibinfo {year} {2012})},\ \Eprint
  {http://arxiv.org/abs/1211.6277} {arXiv:1211.6277 [hep-lat]} \BibitemShut
  {NoStop}%
%%CITATION = ARXIV:1211.6277;%%
\bibitem [{\citenamefont {Bali}\ \emph {et~al.}(2013)\citenamefont {Bali},
  \citenamefont {Collins},\ and\ \citenamefont {Perez-Rubio}}]{Bali:2012ua}%
  \BibitemOpen
  \bibfield  {author} {\bibinfo {author} {\bibfnamefont {G.}~\bibnamefont
  {Bali}}, \bibinfo {author} {\bibfnamefont {S.}~\bibnamefont {Collins}}, \
  and\ \bibinfo {author} {\bibfnamefont {P.}~\bibnamefont {Perez-Rubio}},\
  }\bibfield  {booktitle} {\emph {\bibinfo {booktitle} {{Proceedings, FAIR NExt
  generation ScientistS 2012 (FAIRNESS 2012): Hersonissos, Greece, September
  3-8, 2012}}},\ }\href {\doibase 10.1088/1742-6596/426/1/012017} {\bibfield
  {journal} {\bibinfo  {journal} {J. Phys. Conf. Ser.}\ }\textbf {\bibinfo
  {volume} {426}},\ \bibinfo {pages} {012017} (\bibinfo {year} {2013})},\
  \Eprint {http://arxiv.org/abs/1212.0565} {arXiv:1212.0565 [hep-lat]}
  \BibitemShut {NoStop}%
%%CITATION = ARXIV:1212.0565;%%
\bibitem [{\citenamefont {Namekawa}\ \emph {et~al.}(2013)\citenamefont
  {Namekawa} \emph {et~al.}}]{Namekawa:2013vu}%
  \BibitemOpen
  \bibfield  {author} {\bibinfo {author} {\bibfnamefont {Y.}~\bibnamefont
  {Namekawa}} \emph {et~al.} (\bibinfo {collaboration} {PACS-CS}),\ }\href
  {\doibase 10.1103/PhysRevD.87.094512} {\bibfield  {journal} {\bibinfo
  {journal} {Phys. Rev.}\ }\textbf {\bibinfo {volume} {D87}},\ \bibinfo {pages}
  {094512} (\bibinfo {year} {2013})},\ \Eprint {http://arxiv.org/abs/1301.4743}
  {arXiv:1301.4743 [hep-lat]} \BibitemShut {NoStop}%
%%CITATION = ARXIV:1301.4743;%%
\bibitem [{\citenamefont {Brown}\ \emph {et~al.}(2014)\citenamefont {Brown},
  \citenamefont {Detmold}, \citenamefont {Meinel},\ and\ \citenamefont
  {Orginos}}]{Brown:2014ena}%
  \BibitemOpen
  \bibfield  {author} {\bibinfo {author} {\bibfnamefont {Z.~S.}\ \bibnamefont
  {Brown}}, \bibinfo {author} {\bibfnamefont {W.}~\bibnamefont {Detmold}},
  \bibinfo {author} {\bibfnamefont {S.}~\bibnamefont {Meinel}}, \ and\ \bibinfo
  {author} {\bibfnamefont {K.}~\bibnamefont {Orginos}},\ }\href {\doibase
  10.1103/PhysRevD.90.094507} {\bibfield  {journal} {\bibinfo  {journal} {Phys.
  Rev.}\ }\textbf {\bibinfo {volume} {D90}},\ \bibinfo {pages} {094507}
  (\bibinfo {year} {2014})},\ \Eprint {http://arxiv.org/abs/1409.0497}
  {arXiv:1409.0497 [hep-lat]} \BibitemShut {NoStop}%
%%CITATION = ARXIV:1409.0497;%%
\bibitem [{\citenamefont {P\'erez-Rubio}\ \emph {et~al.}(2015)\citenamefont
  {P\'erez-Rubio}, \citenamefont {Collins},\ and\ \citenamefont
  {Bali}}]{Bali:2015lka}%
  \BibitemOpen
  \bibfield  {author} {\bibinfo {author} {\bibfnamefont {P.}~\bibnamefont
  {P\'erez-Rubio}}, \bibinfo {author} {\bibfnamefont {S.}~\bibnamefont
  {Collins}}, \ and\ \bibinfo {author} {\bibfnamefont {G.~S.}\ \bibnamefont
  {Bali}},\ }\href {\doibase 10.1103/PhysRevD.92.034504} {\bibfield  {journal}
  {\bibinfo  {journal} {Phys. Rev.}\ }\textbf {\bibinfo {volume} {D92}},\
  \bibinfo {pages} {034504} (\bibinfo {year} {2015})},\ \Eprint
  {http://arxiv.org/abs/1503.08440} {arXiv:1503.08440 [hep-lat]} \BibitemShut
  {NoStop}%
%%CITATION = ARXIV:1503.08440;%%
\bibitem [{\citenamefont {Alexandrou}\ and\ \citenamefont
  {Kallidonis}(2017)}]{Alexandrou:2017xwd}%
  \BibitemOpen
  \bibfield  {author} {\bibinfo {author} {\bibfnamefont {C.}~\bibnamefont
  {Alexandrou}}\ and\ \bibinfo {author} {\bibfnamefont {C.}~\bibnamefont
  {Kallidonis}},\ }\href {\doibase 10.1103/PhysRevD.96.034511} {\bibfield
  {journal} {\bibinfo  {journal} {Phys. Rev.}\ }\textbf {\bibinfo {volume}
  {D96}},\ \bibinfo {pages} {034511} (\bibinfo {year} {2017})},\ \Eprint
  {http://arxiv.org/abs/1704.02647} {arXiv:1704.02647 [hep-lat]} \BibitemShut
  {NoStop}%
%%CITATION = ARXIV:1704.02647;%%
\bibitem [{\citenamefont {Can}\ \emph {et~al.}(2019)\citenamefont {Can},
  \citenamefont {Bahtiyar}, \citenamefont {Erkol}, \citenamefont {Gubler},
  \citenamefont {Oka},\ and\ \citenamefont {Takahashi}}]{Can:2019wts}%
  \BibitemOpen
  \bibfield  {author} {\bibinfo {author} {\bibfnamefont {K.~U.}\ \bibnamefont
  {Can}}, \bibinfo {author} {\bibfnamefont {H.}~\bibnamefont {Bahtiyar}},
  \bibinfo {author} {\bibfnamefont {G.}~\bibnamefont {Erkol}}, \bibinfo
  {author} {\bibfnamefont {P.}~\bibnamefont {Gubler}}, \bibinfo {author}
  {\bibfnamefont {M.}~\bibnamefont {Oka}}, \ and\ \bibinfo {author}
  {\bibfnamefont {T.~T.}\ \bibnamefont {Takahashi}},\ }\bibfield  {booktitle}
  {\emph {\bibinfo {booktitle} {{Proceedings, 8th International Conference on
  Quarks and Nuclear Physics (QNP2018): Tsukuba, Japan, November 13-17,
  2018}}},\ }\href {\doibase 10.7566/JPSCP.26.022028} {\bibfield  {journal}
  {\bibinfo  {journal} {JPS Conf. Proc.}\ }\textbf {\bibinfo {volume} {26}},\
  \bibinfo {pages} {022028} (\bibinfo {year} {2019})}\BibitemShut {NoStop}%
%%CITATION = INSPIRE-1764545;%%
\bibitem [{\citenamefont {Padmanath}\ \emph {et~al.}(2015)\citenamefont
  {Padmanath}, \citenamefont {Edwards}, \citenamefont {Mathur},\ and\
  \citenamefont {Peardon}}]{Padmanath:2015jea}%
  \BibitemOpen
  \bibfield  {author} {\bibinfo {author} {\bibfnamefont {M.}~\bibnamefont
  {Padmanath}}, \bibinfo {author} {\bibfnamefont {R.~G.}\ \bibnamefont
  {Edwards}}, \bibinfo {author} {\bibfnamefont {N.}~\bibnamefont {Mathur}}, \
  and\ \bibinfo {author} {\bibfnamefont {M.}~\bibnamefont {Peardon}},\ }\href
  {\doibase 10.1103/PhysRevD.91.094502} {\bibfield  {journal} {\bibinfo
  {journal} {Phys. Rev.}\ }\textbf {\bibinfo {volume} {D91}},\ \bibinfo {pages}
  {094502} (\bibinfo {year} {2015})},\ \Eprint
  {http://arxiv.org/abs/1502.01845} {arXiv:1502.01845 [hep-lat]} \BibitemShut
  {NoStop}%
%%CITATION = ARXIV:1502.01845;%%
\bibitem [{\citenamefont {Mathur}\ and\ \citenamefont
  {Padmanath}(2019)}]{Mathur:2018rwu}%
  \BibitemOpen
  \bibfield  {author} {\bibinfo {author} {\bibfnamefont {N.}~\bibnamefont
  {Mathur}}\ and\ \bibinfo {author} {\bibfnamefont {M.}~\bibnamefont
  {Padmanath}},\ }\href {\doibase 10.1103/PhysRevD.99.031501} {\bibfield
  {journal} {\bibinfo  {journal} {Phys. Rev.}\ }\textbf {\bibinfo {volume}
  {D99}},\ \bibinfo {pages} {031501} (\bibinfo {year} {2019})},\ \Eprint
  {http://arxiv.org/abs/1807.00174} {arXiv:1807.00174 [hep-lat]} \BibitemShut
  {NoStop}%
%%CITATION = ARXIV:1807.00174;%%
\bibitem [{\citenamefont {Lewis}\ and\ \citenamefont
  {Woloshyn}(2009)}]{Lewis:2008fu}%
  \BibitemOpen
  \bibfield  {author} {\bibinfo {author} {\bibfnamefont {R.}~\bibnamefont
  {Lewis}}\ and\ \bibinfo {author} {\bibfnamefont {R.~M.}\ \bibnamefont
  {Woloshyn}},\ }\href {\doibase 10.1103/PhysRevD.79.014502} {\bibfield
  {journal} {\bibinfo  {journal} {Phys. Rev.}\ }\textbf {\bibinfo {volume}
  {D79}},\ \bibinfo {pages} {014502} (\bibinfo {year} {2009})},\ \Eprint
  {http://arxiv.org/abs/0806.4783} {arXiv:0806.4783 [hep-lat]} \BibitemShut
  {NoStop}%
%%CITATION = ARXIV:0806.4783;%%
\bibitem [{\citenamefont {Mohanta}\ and\ \citenamefont
  {Basak}(2019)}]{Mohanta:2019mxo}%
  \BibitemOpen
  \bibfield  {author} {\bibinfo {author} {\bibfnamefont {P.}~\bibnamefont
  {Mohanta}}\ and\ \bibinfo {author} {\bibfnamefont {S.}~\bibnamefont
  {Basak}},\ }\href@noop {} {\  (\bibinfo {year} {2019})},\ \Eprint
  {http://arxiv.org/abs/1911.03741} {arXiv:1911.03741 [hep-lat]} \BibitemShut
  {NoStop}%
%%CITATION = ARXIV:1911.03741;%%
\bibitem [{\citenamefont {Caswell}\ and\ \citenamefont
  {Lepage}(1986)}]{Caswell:1985ui}%
  \BibitemOpen
  \bibfield  {author} {\bibinfo {author} {\bibfnamefont {W.~E.}\ \bibnamefont
  {Caswell}}\ and\ \bibinfo {author} {\bibfnamefont {G.~P.}\ \bibnamefont
  {Lepage}},\ }\href {\doibase 10.1016/0370-2693(86)91297-9} {\bibfield
  {journal} {\bibinfo  {journal} {Phys. Lett.}\ }\textbf {\bibinfo {volume}
  {167B}},\ \bibinfo {pages} {437} (\bibinfo {year} {1986})}\BibitemShut
  {NoStop}%
%%CITATION = PHLTA,167B,437;%%
\bibitem [{\citenamefont {Bodwin}\ \emph {et~al.}(1995)\citenamefont {Bodwin},
  \citenamefont {Braaten},\ and\ \citenamefont {Lepage}}]{Bodwin:1994jh}%
  \BibitemOpen
  \bibfield  {author} {\bibinfo {author} {\bibfnamefont {G.~T.}\ \bibnamefont
  {Bodwin}}, \bibinfo {author} {\bibfnamefont {E.}~\bibnamefont {Braaten}}, \
  and\ \bibinfo {author} {\bibfnamefont {G.~P.}\ \bibnamefont {Lepage}},\
  }\href {\doibase 10.1103/PhysRevD.55.5853, 10.1103/PhysRevD.51.1125}
  {\bibfield  {journal} {\bibinfo  {journal} {Phys. Rev.}\ }\textbf {\bibinfo
  {volume} {D51}},\ \bibinfo {pages} {1125} (\bibinfo {year} {1995})},\
  \bibinfo {note} {[Erratum: Phys. Rev.D55,5853(1997)]},\ \Eprint
  {http://arxiv.org/abs/hep-ph/9407339} {arXiv:hep-ph/9407339 [hep-ph]}
  \BibitemShut {NoStop}%
%%CITATION = HEP-PH/9407339;%%
\bibitem [{\citenamefont {Manohar}(1997)}]{Manohar:1997qy}%
  \BibitemOpen
  \bibfield  {author} {\bibinfo {author} {\bibfnamefont {A.~V.}\ \bibnamefont
  {Manohar}},\ }\href {\doibase 10.1103/PhysRevD.56.230} {\bibfield  {journal}
  {\bibinfo  {journal} {Phys. Rev.}\ }\textbf {\bibinfo {volume} {D56}},\
  \bibinfo {pages} {230} (\bibinfo {year} {1997})},\ \Eprint
  {http://arxiv.org/abs/hep-ph/9701294} {arXiv:hep-ph/9701294 [hep-ph]}
  \BibitemShut {NoStop}%
%%CITATION = HEP-PH/9701294;%%
\bibitem [{\citenamefont {Hu}\ and\ \citenamefont {Mehen}(2006)}]{Hu:2005gf}%
  \BibitemOpen
  \bibfield  {author} {\bibinfo {author} {\bibfnamefont {J.}~\bibnamefont
  {Hu}}\ and\ \bibinfo {author} {\bibfnamefont {T.}~\bibnamefont {Mehen}},\
  }\href {\doibase 10.1103/PhysRevD.73.054003} {\bibfield  {journal} {\bibinfo
  {journal} {Phys. Rev.}\ }\textbf {\bibinfo {volume} {D73}},\ \bibinfo {pages}
  {054003} (\bibinfo {year} {2006})},\ \Eprint
  {http://arxiv.org/abs/hep-ph/0511321} {arXiv:hep-ph/0511321 [hep-ph]}
  \BibitemShut {NoStop}%
%%CITATION = HEP-PH/0511321;%%
\bibitem [{\citenamefont {Mehen}(2017)}]{Mehen:2017nrh}%
  \BibitemOpen
  \bibfield  {author} {\bibinfo {author} {\bibfnamefont {T.}~\bibnamefont
  {Mehen}},\ }\href {\doibase 10.1103/PhysRevD.96.094028} {\bibfield  {journal}
  {\bibinfo  {journal} {Phys. Rev.}\ }\textbf {\bibinfo {volume} {D96}},\
  \bibinfo {pages} {094028} (\bibinfo {year} {2017})},\ \Eprint
  {http://arxiv.org/abs/1708.05020} {arXiv:1708.05020 [hep-ph]} \BibitemShut
  {NoStop}%
%%CITATION = ARXIV:1708.05020;%%
\bibitem [{\citenamefont {Mehen}\ and\ \citenamefont
  {Mohapatra}(2019)}]{Mehen:2019cxn}%
  \BibitemOpen
  \bibfield  {author} {\bibinfo {author} {\bibfnamefont {T.~C.}\ \bibnamefont
  {Mehen}}\ and\ \bibinfo {author} {\bibfnamefont {A.}~\bibnamefont
  {Mohapatra}},\ }\href {\doibase 10.1103/PhysRevD.100.076014} {\bibfield
  {journal} {\bibinfo  {journal} {Phys. Rev.}\ }\textbf {\bibinfo {volume}
  {D100}},\ \bibinfo {pages} {076014} (\bibinfo {year} {2019})},\ \Eprint
  {http://arxiv.org/abs/1905.06965} {arXiv:1905.06965 [hep-ph]} \BibitemShut
  {NoStop}%
%%CITATION = ARXIV:1905.06965;%%
\bibitem [{\citenamefont {Brambilla}\ \emph {et~al.}(2018)\citenamefont
  {Brambilla}, \citenamefont {Krein}, \citenamefont {Tarr\'us~Castell\`a},\
  and\ \citenamefont {Vairo}}]{Brambilla:2017uyf}%
  \BibitemOpen
  \bibfield  {author} {\bibinfo {author} {\bibfnamefont {N.}~\bibnamefont
  {Brambilla}}, \bibinfo {author} {\bibfnamefont {G.}~\bibnamefont {Krein}},
  \bibinfo {author} {\bibfnamefont {J.}~\bibnamefont {Tarr\'us~Castell\`a}}, \
  and\ \bibinfo {author} {\bibfnamefont {A.}~\bibnamefont {Vairo}},\ }\href
  {\doibase 10.1103/PhysRevD.97.016016} {\bibfield  {journal} {\bibinfo
  {journal} {Phys. Rev. D}\ }\textbf {\bibinfo {volume} {97}},\ \bibinfo
  {pages} {016016} (\bibinfo {year} {2018})},\ \Eprint
  {http://arxiv.org/abs/1707.09647} {arXiv:1707.09647 [hep-ph]} \BibitemShut
  {NoStop}%
\bibitem [{\citenamefont {Pineda}(2001)}]{Pineda:2001zq}%
  \BibitemOpen
  \bibfield  {author} {\bibinfo {author} {\bibfnamefont {A.}~\bibnamefont
  {Pineda}},\ }\href {\doibase 10.1088/1126-6708/2001/06/022} {\bibfield
  {journal} {\bibinfo  {journal} {JHEP}\ }\textbf {\bibinfo {volume} {06}},\
  \bibinfo {pages} {022} (\bibinfo {year} {2001})},\ \Eprint
  {http://arxiv.org/abs/hep-ph/0105008} {arXiv:hep-ph/0105008 [hep-ph]}
  \BibitemShut {NoStop}%
%%CITATION = HEP-PH/0105008;%%
\bibitem [{\citenamefont {Bazavov}\ \emph {et~al.}(2018)\citenamefont {Bazavov}
  \emph {et~al.}}]{Bazavov:2018omf}%
  \BibitemOpen
  \bibfield  {author} {\bibinfo {author} {\bibfnamefont {A.}~\bibnamefont
  {Bazavov}} \emph {et~al.} (\bibinfo {collaboration} {Fermilab Lattice, MILC,
  TUMQCD}),\ }\href {\doibase 10.1103/PhysRevD.98.054517} {\bibfield  {journal}
  {\bibinfo  {journal} {Phys. Rev.}\ }\textbf {\bibinfo {volume} {D98}},\
  \bibinfo {pages} {054517} (\bibinfo {year} {2018})},\ \Eprint
  {http://arxiv.org/abs/1802.04248} {arXiv:1802.04248 [hep-lat]} \BibitemShut
  {NoStop}%
%%CITATION = ARXIV:1802.04248;%%
\bibitem [{\citenamefont {Juge}\ \emph {et~al.}(2003)\citenamefont {Juge},
  \citenamefont {Kuti},\ and\ \citenamefont {Morningstar}}]{Juge:2002br}%
  \BibitemOpen
  \bibfield  {author} {\bibinfo {author} {\bibfnamefont {K.~J.}\ \bibnamefont
  {Juge}}, \bibinfo {author} {\bibfnamefont {J.}~\bibnamefont {Kuti}}, \ and\
  \bibinfo {author} {\bibfnamefont {C.}~\bibnamefont {Morningstar}},\ }\href
  {\doibase 10.1103/PhysRevLett.90.161601} {\bibfield  {journal} {\bibinfo
  {journal} {Phys. Rev. Lett.}\ }\textbf {\bibinfo {volume} {90}},\ \bibinfo
  {pages} {161601} (\bibinfo {year} {2003})},\ \Eprint
  {http://arxiv.org/abs/hep-lat/0207004} {arXiv:hep-lat/0207004 [hep-lat]}
  \BibitemShut {NoStop}%
%%CITATION = HEP-LAT/0207004;%%
\bibitem [{\citenamefont {Bali}\ \emph {et~al.}(2005)\citenamefont {Bali},
  \citenamefont {Neff}, \citenamefont {Duessel}, \citenamefont {Lippert},\ and\
  \citenamefont {Schilling}}]{Bali:2005fu}%
  \BibitemOpen
  \bibfield  {author} {\bibinfo {author} {\bibfnamefont {G.~S.}\ \bibnamefont
  {Bali}}, \bibinfo {author} {\bibfnamefont {H.}~\bibnamefont {Neff}}, \bibinfo
  {author} {\bibfnamefont {T.}~\bibnamefont {Duessel}}, \bibinfo {author}
  {\bibfnamefont {T.}~\bibnamefont {Lippert}}, \ and\ \bibinfo {author}
  {\bibfnamefont {K.}~\bibnamefont {Schilling}} (\bibinfo {collaboration}
  {SESAM}),\ }\href {\doibase 10.1103/PhysRevD.71.114513} {\bibfield  {journal}
  {\bibinfo  {journal} {Phys. Rev.}\ }\textbf {\bibinfo {volume} {D71}},\
  \bibinfo {pages} {114513} (\bibinfo {year} {2005})},\ \Eprint
  {http://arxiv.org/abs/hep-lat/0505012} {arXiv:hep-lat/0505012 [hep-lat]}
  \BibitemShut {NoStop}%
%%CITATION = HEP-LAT/0505012;%%
\bibitem [{\citenamefont {M{\"u}ller}\ \emph {et~al.}(2019)\citenamefont
  {M{\"u}ller}, \citenamefont {Philipsen}, \citenamefont {Reisinger},\ and\
  \citenamefont {Wagner}}]{Mueller:2019mkh}%
  \BibitemOpen
  \bibfield  {author} {\bibinfo {author} {\bibfnamefont {L.}~\bibnamefont
  {M{\"u}ller}}, \bibinfo {author} {\bibfnamefont {O.}~\bibnamefont
  {Philipsen}}, \bibinfo {author} {\bibfnamefont {C.}~\bibnamefont
  {Reisinger}}, \ and\ \bibinfo {author} {\bibfnamefont {M.}~\bibnamefont
  {Wagner}},\ }\href {\doibase 10.1103/PhysRevD.100.054503} {\bibfield
  {journal} {\bibinfo  {journal} {Phys. Rev.}\ }\textbf {\bibinfo {volume}
  {D100}},\ \bibinfo {pages} {054503} (\bibinfo {year} {2019})},\ \Eprint
  {http://arxiv.org/abs/1907.01482} {arXiv:1907.01482 [hep-lat]} \BibitemShut
  {NoStop}%
%%CITATION = ARXIV:1907.01482;%%
\bibitem [{\citenamefont {Yamamoto}\ \emph {et~al.}(2008)\citenamefont
  {Yamamoto}, \citenamefont {Suganuma},\ and\ \citenamefont
  {Iida}}]{Yamamoto:2008jz}%
  \BibitemOpen
  \bibfield  {author} {\bibinfo {author} {\bibfnamefont {A.}~\bibnamefont
  {Yamamoto}}, \bibinfo {author} {\bibfnamefont {H.}~\bibnamefont {Suganuma}},
  \ and\ \bibinfo {author} {\bibfnamefont {H.}~\bibnamefont {Iida}},\ }\href
  {\doibase 10.1103/PhysRevD.78.014513} {\bibfield  {journal} {\bibinfo
  {journal} {Phys. Rev.}\ }\textbf {\bibinfo {volume} {D78}},\ \bibinfo {pages}
  {014513} (\bibinfo {year} {2008})},\ \Eprint {http://arxiv.org/abs/0806.3554}
  {arXiv:0806.3554 [hep-lat]} \BibitemShut {NoStop}%
%%CITATION = ARXIV:0806.3554;%%
\bibitem [{\citenamefont {Luscher}\ and\ \citenamefont
  {Weisz}(2002)}]{Luscher:2002qv}%
  \BibitemOpen
  \bibfield  {author} {\bibinfo {author} {\bibfnamefont {M.}~\bibnamefont
  {Luscher}}\ and\ \bibinfo {author} {\bibfnamefont {P.}~\bibnamefont
  {Weisz}},\ }\href {\doibase 10.1088/1126-6708/2002/07/049} {\bibfield
  {journal} {\bibinfo  {journal} {JHEP}\ }\textbf {\bibinfo {volume} {07}},\
  \bibinfo {pages} {049} (\bibinfo {year} {2002})},\ \Eprint
  {http://arxiv.org/abs/hep-lat/0207003} {arXiv:hep-lat/0207003 [hep-lat]}
  \BibitemShut {NoStop}%
%%CITATION = HEP-LAT/0207003;%%
\bibitem [{\citenamefont {Tanabashi}\ \emph {et~al.}(2018)\citenamefont
  {Tanabashi} \emph {et~al.}}]{Tanabashi:2018oca}%
  \BibitemOpen
  \bibfield  {author} {\bibinfo {author} {\bibfnamefont {M.}~\bibnamefont
  {Tanabashi}} \emph {et~al.} (\bibinfo {collaboration} {Particle Data
  Group}),\ }\href {\doibase 10.1103/PhysRevD.98.030001} {\bibfield  {journal}
  {\bibinfo  {journal} {Phys. Rev.}\ }\textbf {\bibinfo {volume} {D98}},\
  \bibinfo {pages} {030001} (\bibinfo {year} {2018})}\BibitemShut {NoStop}%
%%CITATION = PHRVA,D98,030001;%%
\bibitem [{\citenamefont {Fleming}\ and\ \citenamefont
  {Mehen}(2006)}]{Fleming:2005pd}%
  \BibitemOpen
  \bibfield  {author} {\bibinfo {author} {\bibfnamefont {S.}~\bibnamefont
  {Fleming}}\ and\ \bibinfo {author} {\bibfnamefont {T.}~\bibnamefont
  {Mehen}},\ }\href {\doibase 10.1103/PhysRevD.73.034502} {\bibfield  {journal}
  {\bibinfo  {journal} {Phys. Rev.}\ }\textbf {\bibinfo {volume} {D73}},\
  \bibinfo {pages} {034502} (\bibinfo {year} {2006})},\ \Eprint
  {http://arxiv.org/abs/hep-ph/0509313} {arXiv:hep-ph/0509313 [hep-ph]}
  \BibitemShut {NoStop}%
%%CITATION = HEP-PH/0509313;%%
\bibitem [{\citenamefont {Alexandrou}\ \emph {et~al.}(2014)\citenamefont
  {Alexandrou}, \citenamefont {Drach}, \citenamefont {Jansen}, \citenamefont
  {Kallidonis},\ and\ \citenamefont {Koutsou}}]{Alexandrou:2014sha}%
  \BibitemOpen
  \bibfield  {author} {\bibinfo {author} {\bibfnamefont {C.}~\bibnamefont
  {Alexandrou}}, \bibinfo {author} {\bibfnamefont {V.}~\bibnamefont {Drach}},
  \bibinfo {author} {\bibfnamefont {K.}~\bibnamefont {Jansen}}, \bibinfo
  {author} {\bibfnamefont {C.}~\bibnamefont {Kallidonis}}, \ and\ \bibinfo
  {author} {\bibfnamefont {G.}~\bibnamefont {Koutsou}},\ }\href {\doibase
  10.1103/PhysRevD.90.074501} {\bibfield  {journal} {\bibinfo  {journal} {Phys.
  Rev.}\ }\textbf {\bibinfo {volume} {D90}},\ \bibinfo {pages} {074501}
  (\bibinfo {year} {2014})},\ \Eprint {http://arxiv.org/abs/1406.4310}
  {arXiv:1406.4310 [hep-lat]} \BibitemShut {NoStop}%
%%CITATION = ARXIV:1406.4310;%%
\bibitem [{\citenamefont {Soto}(2018)}]{Soto:2017one}%
  \BibitemOpen
  \bibfield  {author} {\bibinfo {author} {\bibfnamefont {J.}~\bibnamefont
  {Soto}},\ }\href {\doibase 10.1016/j.nuclphysbps.2018.03.020} {\bibfield
  {journal} {\bibinfo  {journal} {Nucl.\ Part.\ Phys.\ Proc.}\ }\textbf
  {\bibinfo {volume} {294-296}},\ \bibinfo {pages} {87} (\bibinfo {year}
  {2018})},\ \Eprint {http://arxiv.org/abs/1709.08038} {arXiv:1709.08038
  [hep-ph]} \BibitemShut {NoStop}%
\bibitem [{\citenamefont {Brambilla}\ \emph {et~al.}(2019)\citenamefont
  {Brambilla}, \citenamefont {Lai}, \citenamefont {Segovia}, \citenamefont
  {Tarr\'us~Castell\`a},\ and\ \citenamefont {Vairo}}]{Brambilla:2018pyn}%
  \BibitemOpen
  \bibfield  {author} {\bibinfo {author} {\bibfnamefont {N.}~\bibnamefont
  {Brambilla}}, \bibinfo {author} {\bibfnamefont {W.~K.}\ \bibnamefont {Lai}},
  \bibinfo {author} {\bibfnamefont {J.}~\bibnamefont {Segovia}}, \bibinfo
  {author} {\bibfnamefont {J.}~\bibnamefont {Tarr\'us~Castell\`a}}, \ and\
  \bibinfo {author} {\bibfnamefont {A.}~\bibnamefont {Vairo}},\ }\href
  {\doibase 10.1103/PhysRevD.99.014017} {\bibfield  {journal} {\bibinfo
  {journal} {Phys.\ Rev.\ D}\ }\textbf {\bibinfo {volume} {99}},\ \bibinfo
  {pages} {014017} (\bibinfo {year} {2019})},\ \Eprint
  {http://arxiv.org/abs/1805.07713} {arXiv:1805.07713 [hep-ph]} \BibitemShut
  {NoStop}%
\bibitem [{\citenamefont {Brambilla}\ \emph {et~al.}(2020)\citenamefont
  {Brambilla}, \citenamefont {Lai}, \citenamefont {Segovia},\ and\
  \citenamefont {Tarr\'us~Castell\`a}}]{Brambilla:2019jfi}%
  \BibitemOpen
  \bibfield  {author} {\bibinfo {author} {\bibfnamefont {N.}~\bibnamefont
  {Brambilla}}, \bibinfo {author} {\bibfnamefont {W.~K.}\ \bibnamefont {Lai}},
  \bibinfo {author} {\bibfnamefont {J.}~\bibnamefont {Segovia}}, \ and\
  \bibinfo {author} {\bibfnamefont {J.}~\bibnamefont {Tarr\'us~Castell\`a}},\
  }\href {\doibase 10.1103/PhysRevD.101.054040} {\bibfield  {journal} {\bibinfo
   {journal} {Phys. Rev.}\ }\textbf {\bibinfo {volume} {D101}},\ \bibinfo
  {pages} {054040} (\bibinfo {year} {2020})},\ \Eprint
  {http://arxiv.org/abs/1908.11699} {arXiv:1908.11699 [hep-ph]} \BibitemShut
  {NoStop}%
%%CITATION = ARXIV:1908.11699;%%
\bibitem [{\citenamefont {Burdman}\ and\ \citenamefont
  {Donoghue}(1992)}]{Burdman:1992gh}%
  \BibitemOpen
  \bibfield  {author} {\bibinfo {author} {\bibfnamefont {G.}~\bibnamefont
  {Burdman}}\ and\ \bibinfo {author} {\bibfnamefont {J.~F.}\ \bibnamefont
  {Donoghue}},\ }\href {\doibase 10.1016/0370-2693(92)90068-F} {\bibfield
  {journal} {\bibinfo  {journal} {Phys.\ Lett.\ B}\ }\textbf {\bibinfo {volume}
  {280}},\ \bibinfo {pages} {287} (\bibinfo {year} {1992})}\BibitemShut
  {NoStop}%
\bibitem [{\citenamefont {Shi}\ \emph {et~al.}(2020)\citenamefont {Shi},
  \citenamefont {Wang}, \citenamefont {Zhao},\ and\ \citenamefont
  {Meißner}}]{Shi:2020qde}%
  \BibitemOpen
  \bibfield  {author} {\bibinfo {author} {\bibfnamefont {Y.-J.}\ \bibnamefont
  {Shi}}, \bibinfo {author} {\bibfnamefont {W.}~\bibnamefont {Wang}}, \bibinfo
  {author} {\bibfnamefont {Z.-X.}\ \bibnamefont {Zhao}}, \ and\ \bibinfo
  {author} {\bibfnamefont {U.-G.}\ \bibnamefont {Meißner}},\ }\href@noop {} {\
   (\bibinfo {year} {2020})},\ \Eprint {http://arxiv.org/abs/2002.02785}
  {arXiv:2002.02785 [hep-ph]} \BibitemShut {NoStop}%
\bibitem [{\citenamefont {Berwein}\ \emph {et~al.}(2015)\citenamefont
  {Berwein}, \citenamefont {Brambilla}, \citenamefont {Tarr\'us~Castell\`a},\
  and\ \citenamefont {Vairo}}]{Berwein:2015vca}%
  \BibitemOpen
  \bibfield  {author} {\bibinfo {author} {\bibfnamefont {M.}~\bibnamefont
  {Berwein}}, \bibinfo {author} {\bibfnamefont {N.}~\bibnamefont {Brambilla}},
  \bibinfo {author} {\bibfnamefont {J.}~\bibnamefont {Tarr\'us~Castell\`a}}, \
  and\ \bibinfo {author} {\bibfnamefont {A.}~\bibnamefont {Vairo}},\ }\href
  {\doibase 10.1103/PhysRevD.92.114019} {\bibfield  {journal} {\bibinfo
  {journal} {Phys. Rev.}\ }\textbf {\bibinfo {volume} {D92}},\ \bibinfo {pages}
  {114019} (\bibinfo {year} {2015})},\ \Eprint
  {http://arxiv.org/abs/1510.04299} {arXiv:1510.04299 [hep-ph]} \BibitemShut
  {NoStop}%
%%CITATION = ARXIV:1510.04299;%%
\end{thebibliography}%

\end{document}